\documentclass[5p, sort&compress, number]{elsarticle}

\makeatletter

\def\@author#1{\g@addto@macro\elsauthors{\normalsize%
    \def\baselinestretch{1}%
    \upshape\authorsep#1\unskip\textsuperscript{%
      \ifx\@fnmark\@empty\else\unskip\sep\@fnmark\let\sep=,\fi
      \ifx\@corref\@empty\else\unskip\sep\@corref\let\sep=,\fi
      }%
    \def\authorsep{\unskip,\space}%
    \global\let\@fnmark\@empty
    \global\let\@corref\@empty  
    \global\let\sep\@empty}%
    \@eadauthor={#1}
}

\def\ps@pprintTitle{%
 \let\@oddhead\@empty
 \let\@evenhead\@empty
 \def\@oddfoot{}%
 \let\@evenfoot\@oddfoot}
\makeatother

\usepackage{placeins}
\usepackage[switch,modulo]{lineno}
\usepackage{float}
\usepackage{graphicx}
\usepackage{lipsum}
\usepackage{flushend}
\usepackage{amssymb}
\usepackage{amsmath}
\usepackage{color}
\usepackage{url}
\usepackage{multirow}
\usepackage{isotope}
\usepackage{verbatim}
\usepackage{hyperref}

\begin{document}

\begin{frontmatter}

\title{The MIGDAL experiment: \\
Measuring a rare atomic process to aid the search for dark matter}

\author[IMP]{H.~M.~Ara\'ujo\corref{correspondingauthor1}}
\address[IMP]{Department of Physics, Imperial College London, UK}
\cortext[correspondingauthor1]{h.araujo@imperial.ac.uk}
\author[PPD]{S.~N.~Balashov}
\address[PPD]{Particle Physics Department, STFC Rutherford Appleton Laboratory, UK}
\author[IMP]{J.~E.~Borg}
\author[CERN]{F.~M.~Brunbauer}
\address[CERN]{CERN -- European Organization for Nuclear Research, Geneva, Switzerland}
\author[ISIS]{C.~Cazzaniga}
\address[ISIS]{ISIS Neutron and Muon Beams, STFC Rutherford Appleton Laboratory, UK}
\author[ISIS]{C.~D.~Frost}
\author[HIP]{F.~Garcia}
\address[HIP]{Helsinki Institute of Physics, University of Helsinki, Finland}
\author[RHUL]{A.~C.~Kaboth}
\address[RHUL]{Department of Physics, Royal Holloway University of London, UK}
\author[ISIS]{M.~Kastriotou}
\author[BIR]{I.~Katsioulas}
\address[BIR]{School of Physics and Astronomy, University of Birmingham, UK}
\author[PPD]{A.~Khazov}
\author[OXF]{H.~Kraus}
\address[OXF]{Department of Physics, University of Oxford, UK}
\author[SHE]{V.~A.~Kudryavtsev}
\address[SHE]{Department of Physics \& Astronomy, University of Sheffield, UK}
\author[ISIS]{S.~Lilley}
\author[LIP]{A.~Lindote}
\address[LIP]{LIP--Coimbra \& Department of Physics, University of Coimbra, Portugal}
\author[UNM]{D.~Loomba}
\address[UNM]{Department of Physics \& Astronomy, University of New Mexico, USA}
\author[LIP]{M.~I.~Lopes}
\author[LIP,UAM]{E.~Lopez~Asamar}
\address[UAM]{Department of Physics, Universidad Aut\'onoma de Madrid, Spain}
\author[ISIS]{P.~Luna~Dapica}
\author[PPD]{P.~A.~Majewski\corref{correspondingauthor2}}
\cortext[correspondingauthor2]{pawel.majewski@stfc.ac.uk}
\author[IMP,PPD]{T.~Marley}
\author[KCL]{C.~McCabe}
\address[KCL]{Department of Physics, King's College London, UK}
\author[UNM]{A.~F.~Mills}
\author[IMP,PPD]{M.~Nakhostin}
\author[BIR]{T.~Neep}
\author[LIP]{F.~Neves}
\author[BIR]{K.~Nikolopoulos}
\author[CERN]{E.~Oliveri}
\author[CERN]{L.~Ropelewski}
\author[UNM]{E.~Tilly}
\author[LIP]{V.~N.~Solovov}
\author[IMP]{T.~J.~Sumner}
\author[RTD]{J.~Tarrant}
\address[RTD]{Technology Department, STFC Rutherford Appleton Laboratory, UK}
\author[ISIS]{R.~Turnley}
\author[PPD]{M.~G.~D.~van~der~Grinten}
\author[CERN]{R.~Veenhof}

\begin{abstract}
We present the Migdal In Galactic Dark mAtter expLoration (MIGDAL) experiment aiming at the unambiguous observation and study of the so-called Migdal effect induced by fast-neutron scattering. It is hoped that this elusive atomic process can be exploited to enhance the reach of direct dark matter search experiments to lower masses, but it is still lacking experimental confirmation. Our goal is to detect the predicted atomic electron emission which is thought to accompany nuclear scattering with low, but calculable, probability, by deploying an Optical Time Projection Chamber filled with a low-pressure gas based on CF$_4$. Initially, pure CF$_4$ will be used, and then in mixtures containing other elements employed by leading dark matter search technologies -- including noble species, plus Si and Ge. High resolution track images generated by a Gas Electron Multiplier stack, together with timing information from scintillation and ionisation readout, will be used for 3D reconstruction of the characteristic event topology expected for this process -- an arrangement of two tracks sharing a common vertex, with one belonging to a Migdal electron and the other to a nuclear recoil. Different energy-loss rate distributions along both tracks will be used as a powerful discrimination tool against background events. In this article we present the design of the experiment, informed by extensive particle and track simulations and detailed estimations of signal and background rates. In pure CF$_4$ we expect to observe 8.9 (29.3) Migdal events per calendar day of exposure to an intense D-D (D-T) neutron generator beam at the NILE facility located at the Rutherford Appleton Laboratory (UK). With our nominal assumptions, 5$\sigma$ median discovery significance can be achieved in under one day with either generator.
\end{abstract}


\end{frontmatter}



\setcounter{tocdepth}{2}
\tableofcontents


\section{Introduction}
\label{S:Introduction}

Most dark matter (DM) direct detection experiments search for rare nuclear recoils from the elastic scattering of DM particles off nuclei in ordinary matter. Below some small nuclear recoil (NR) threshold -- typically $\mathcal{O}$(keV) -- this signature is not detectable, defining a DM mass threshold which depends on the mass of the target nucleus. This simple model assumes a recoiling nucleus moving together with its electrons, inducing ionisation and excitation of the neighbouring atoms. However, it has long been recognised that the sudden acceleration of the nucleus can lead to direct ionisation of the atomic electrons -- both in nuclear scattering~\cite{Mig1939,Migdal1977} as well as in $\alpha$ and $\beta$ radioactive decay~\cite{Migdal1941,Feinberg1941,Levinger1953}. This phenomenon, the so-called `Migdal effect', predicts a small but non-zero probability for atomic ionisation if the timescale for the nuclear `jolt' is much shorter than the electronic orbital periods. In these cases the nucleus initially moves relative to its electrons without `carrying' them, which may lead to the ionisation of the recoiling atom -- producing an electronic recoil (ER) signal in the detector.

Several early studies recognised that this effect may provide an alternative signature for the direct detection of DM~\cite{Vergados:2004bm,Moustakidis:2005gx,Bernabei:2007jz,Sharma2017b}, but it was only recently that Migdal’s approach has been reformulated to yield a relation between NR and ER energies and ionisation probabilities as a function of ER energy~\cite{Ibe:2017yqa}, highlighting this as an attractive process to search for sub-GeV mass DM particles: for a given mass, the maximum available ER energy exceeds that in the NR channel -- and ER signals are easier to detect as their responses are `unquenched' -- effectively making sub-threshold NR interactions detectable indirectly, albeit with low probability.

The Migdal effect has been observed in nuclear decay processes in decades past.  In the 1950s, a measurement was performed of K- and L-shell ionisation accompanying $\beta$ decay from \isotope[147]{Pm} and \isotope[210]{Bi}~\cite{Boehm1954}. Later, the effect was measured from K-, L- and M-shell electron shake-off accompanying the $\alpha$ decay in \isotope[238]{Pu} and \isotope[210]{Po}~\cite{Rapaport1974,Rapaport1975,Rapaport1975-2}, and in mono-cetylphosphate with a large content of \isotope[32]P~\cite{Berlovich1965}. More recently, electron ionisation has been observed following the $\beta$ decay of trapped \isotope[6]He$^+$ ions and $\beta^+$ decay in heavier \isotope[19]Ne$^+$ and \isotope[35]Ar$^+$ ions~\cite{Couratin2012,Fabian2018}.

Apart from the observation of the effect in both light and heavy elements, these measurements confirm that the phenomenon is not restricted to isolated atoms, appearing also in molecular compounds and in solids. Additionally, although we will refer to this process exclusively as the `Migdal effect', these and other references show that this phenomenon has been described by various names through the decades, including, but not limited to, `electron shake-off'\footnote{Following the ionisation of an atom or molecule (e.g., through photoionisation), it is possible for additional electron ionisation to occur, a phenomenon also known as shake-off (e.g.,~\protect{\cite{Bristow1982,Persson2001,Kochur2006}}). This should not be confused with the effect that we consider here, where ionisation follows from a sudden perturbation to the {\it nucleus} of the atom or molecule.} `neutron-impact ionisation'~\cite{Liertzer2012,Pindzola2014}, `electron excitation by neutron-nucleus scattering'~\cite{Lovesey1982}, or `atom excitation by jolting'~\cite[p.~149]{Landau1981}.

Despite the lack of experimental confirmation in nuclear scattering -- where the change in the motion of the nucleus is brought about by scattering with an electrically neutral projectile -- this process has been invoked to extend the science reach of several DM search technologies to sub-GeV particles~\cite{Dolan2018,Baxter:2019pnz,Essig:2019xkx} -- with those with very low energy thresholds and ER backgrounds benefiting the most. Several collaborations have now published DM results exploiting the reformulation of the Migdal effect in Ref.~\cite{Ibe:2017yqa}: LUX applied it first, decreasing the DM mass threshold from 4~GeV to 400~MeV~\cite{Akerib:2018hck}, followed by searches from several other experiments~\cite{Armengaud:2019kfj,Liu:2019kzq,XENON:2019zpr,SENSEI:2020dpa,Adhikari2022,Armengaud2022,Agnes2022}. However, to our knowledge no measurement has confirmed the theory to date, even for isolated atoms -- and there could be important departures for molecular species, liquids and solids (see, e.g.,~\cite{Lovesey1987,Liang:2019nnx,Knapen:2020aky}).

The MIGDAL (Migdal In Galactic Dark mAtter expLoration) collaboration aims to achieve the unambiguous detection of the Migdal effect under the most favourable conditions. We will use energetic neutrons as projectiles and a low-pressure gas detector so that ionisation tracks from NR and ER can be imaged and traced to a common vertex, which is the tell-tale signature of the Migdal effect. Although our measurements will probe an energy regime well above that being exploited by DM experiments, the systematic study of Migdal probabilities in various atomic and molecular species will allow us to establish if the theoretical predictions are sound over a wide energy regime. Work to develop the theoretical calculations at relevant energies has progressed in parallel~\cite{MIGDALth}.

Mature technology to achieve an unambiguous observation already exists, developed partly by the collaborating groups~\cite{Phan:2015pda,Phan:2017sep,Brunbauer2018}, among others -- especially in the directional DM detection community~\cite{Vahsen2020,Vahsen2021}.

\begin{figure}[t]
\centerline{\includegraphics[width=1.0\linewidth]{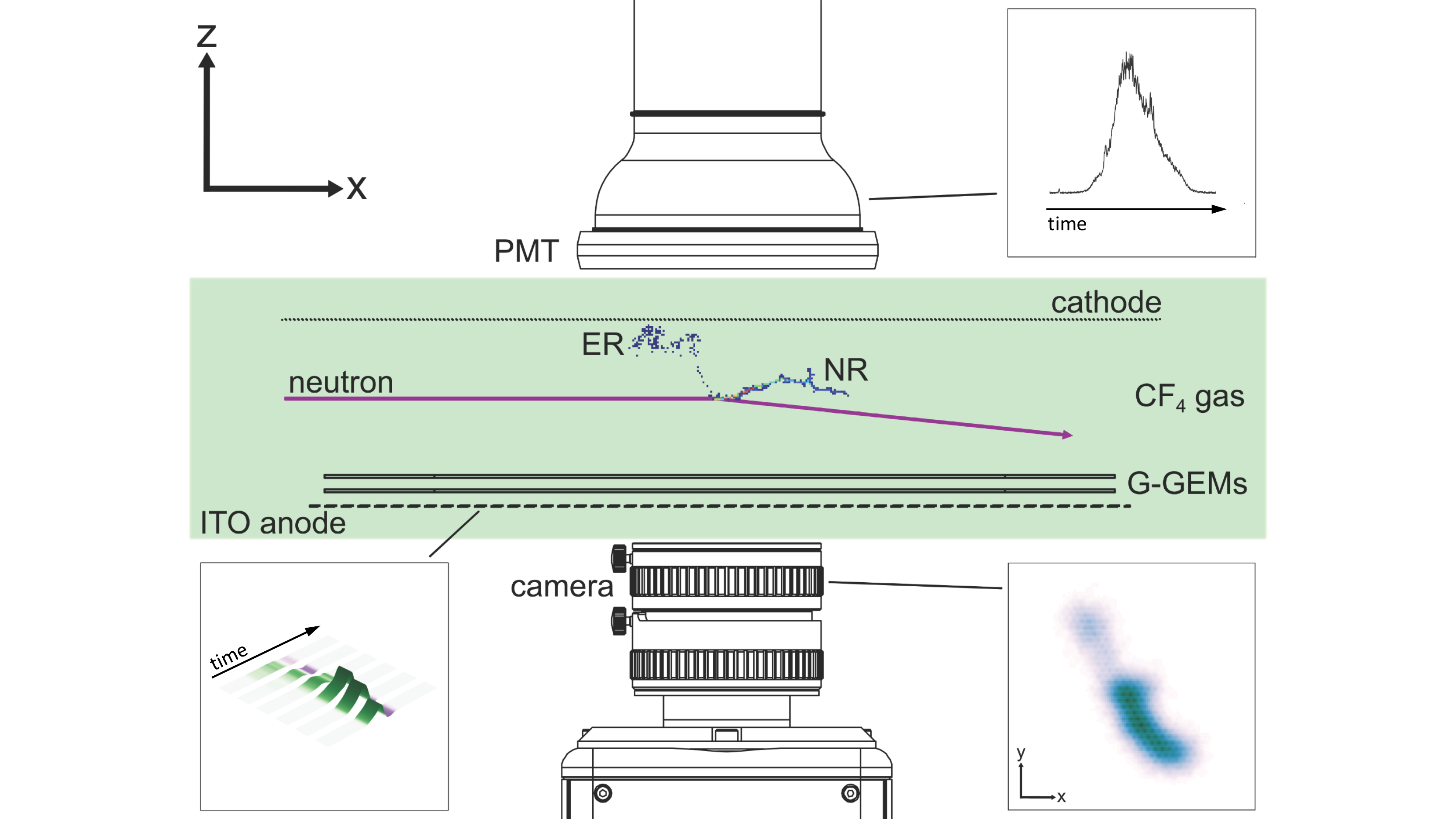}}
\caption{Schematic representation of the MIGDAL experiment showing the OTPC exposed to neutrons, with interactions in the low-pressure CF$_4$ gas amplified by a double glass-GEM system; the active volume of the OTPC is approximately 10$\times$10~cm$^2$ by 3~cm in the drift direction; neutron interactions take place in a volume $\approx 9\!\times\!9$~cm$^2$ by 1.3~cm. Optical signals are recorded by an external camera and a photomultiplier tube, while amplified track ionisation is detected by an ITO-strip anode. An example 2D-projected Migdal event (scaled 10$\times$) is shown, featuring a 5-keV electron and a 150-keV fluorine nuclear recoil originating from the same vertex -- simulated using Degrad~\protect{\cite{Degrad}} and SRIM~\protect{\cite{SRIM}} plus Garfield++\protect{\cite{Garfield}}, respectively. Illustrative signals in the various detector systems are also shown.}
\label{Fig:ExperimentIllustration}
\end{figure}

The measurement is illustrated schematically in Fig.~\ref{Fig:ExperimentIllustration}. Our choice of base gas is CF$_4$ -- for its high scintillation yield and emission spectrum compatible with complementary metal oxide semiconductor (CMOS) camera readout -- as the active (working) medium in an Optical Time Projection Chamber (OTPC). The detector allows three-dimensional (3D) track reconstruction through the following detector sub-systems: $i)$ track ionisation is drifted to a double glass Gas Electron Multiplier (GEM) system and converted to an optical signal which is imaged by a CMOS camera; $ii)$ the amplified charge is collected at an Indium Tin Oxide (ITO) anode plane segmented into readout strips to obtain the perpendicular coordinate; $iii)$ a photomultiplier tube (PMT) detects both the primary and secondary scintillation light to provide the absolute `depth' coordinate. The detector is exposed to high-flux D-D (2.47~MeV) and D-T (14.7~MeV) neutron generators, with significant shielding and collimation providing background mitigation and a controlled scattering environment. 

Our initial goal is to observe clearly the Migdal effect in pure CF$_4$; subsequently, this will be mixed with other gases, including the noble elements and other gases based on Si and Ge. This article focuses on the initial pure-CF$_4$ deployment, but some discussion is offered on future measurements with other gas mixtures. There have been other proposals to achieve a measurement in argon and xenon gas at higher pressures~\cite{Nakamura:2020kex} and in the condensed phase~\cite{Bell2022,Adams2022}.

In this article we describe the design of the experiment, informed by simulations and preliminary test data from ancillary systems. The paper is organised as follows. In Section~\ref{S:SignalsBackgrounds} we describe the behaviour of NR and ER tracks in low-pressure gas, and introduce the neutron-induced processes responsible for signal and background interactions; we summarise calculated Migdal rates, giving context to the experimental challenge. An appendix reviews pertinent aspects of neutron scattering kinematics and neutron cross sections. In Section~\ref{S:Experiment} an overview of the experiment is given, including the main design drivers for each detector subsystem and key design choices. In Section~\ref{S:NeutronSource} we describe the neutron beam and host facility, the design of the collimator and shield elements, and mention key beam-induced backgrounds. In Section~\ref{S:TrackSimulations} we detail the modelling of particle tracks in the gas and their detection by the optical and charge readout systems. Section~\ref{S:Sensitivity} discusses the expected sensitivity of the experiment to Migdal events. Section~\ref{S:Gases} addresses the extension of the measurement to other gas mixtures. We conclude by discussing the outlook for our programme in Section~\ref{S:Conclusion}.

\section{Signal and backgrounds}
\label{S:SignalsBackgrounds}

Detection of the rare Migdal event topology, consisting of two short tracks with a common vertex, using a low-pressure OTPC detector requires optimisation based mostly around the gas composition and density (pressure). The latter operational parameter governs almost the whole experimental approach by impacting two key physical parameters which are in tension with each other: the neutron scattering rate in the active volume increases with pressure, but the length of the resulting ER and NR tracks decreases. The electron track length is particularly important as this determines the ability to discriminate between ER and NR tracks, as well as the detection threshold for Migdal events. In this section we discuss the behaviour of NR and ER tracks in the low-pressure gas, as well as the signal and background rates after a simple threshold based on track length has been defined. Other experimental parameters that depend on the gas composition and density will be discussed in Section~\ref{S:Experiment}, where a more detailed account of the experiment is given.

\subsection{Tracks in low pressure gas}

\begin{figure*}[t!]
\centerline{
    \includegraphics[width=0.47\linewidth]{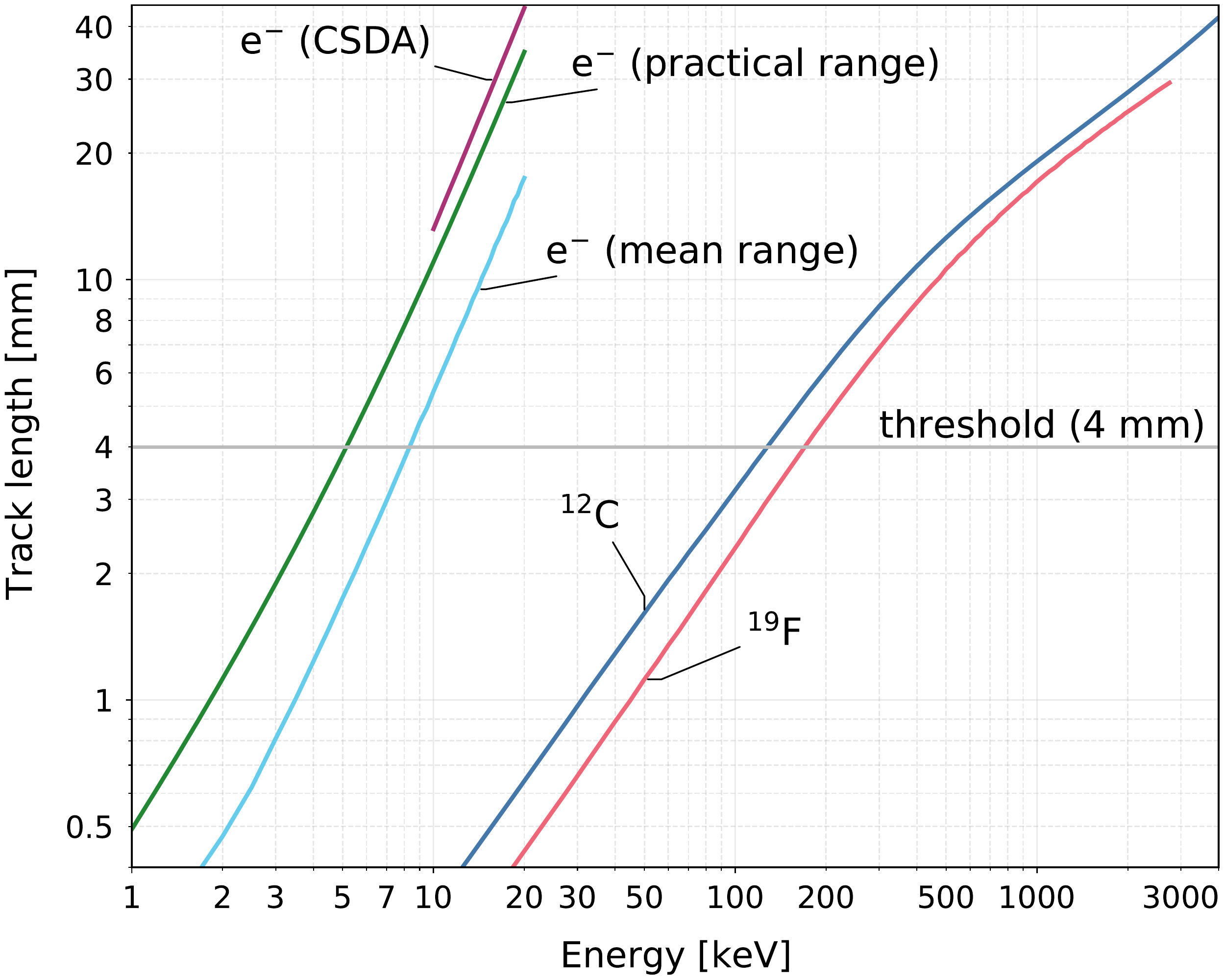}\qquad
    \includegraphics[width=0.47\linewidth]{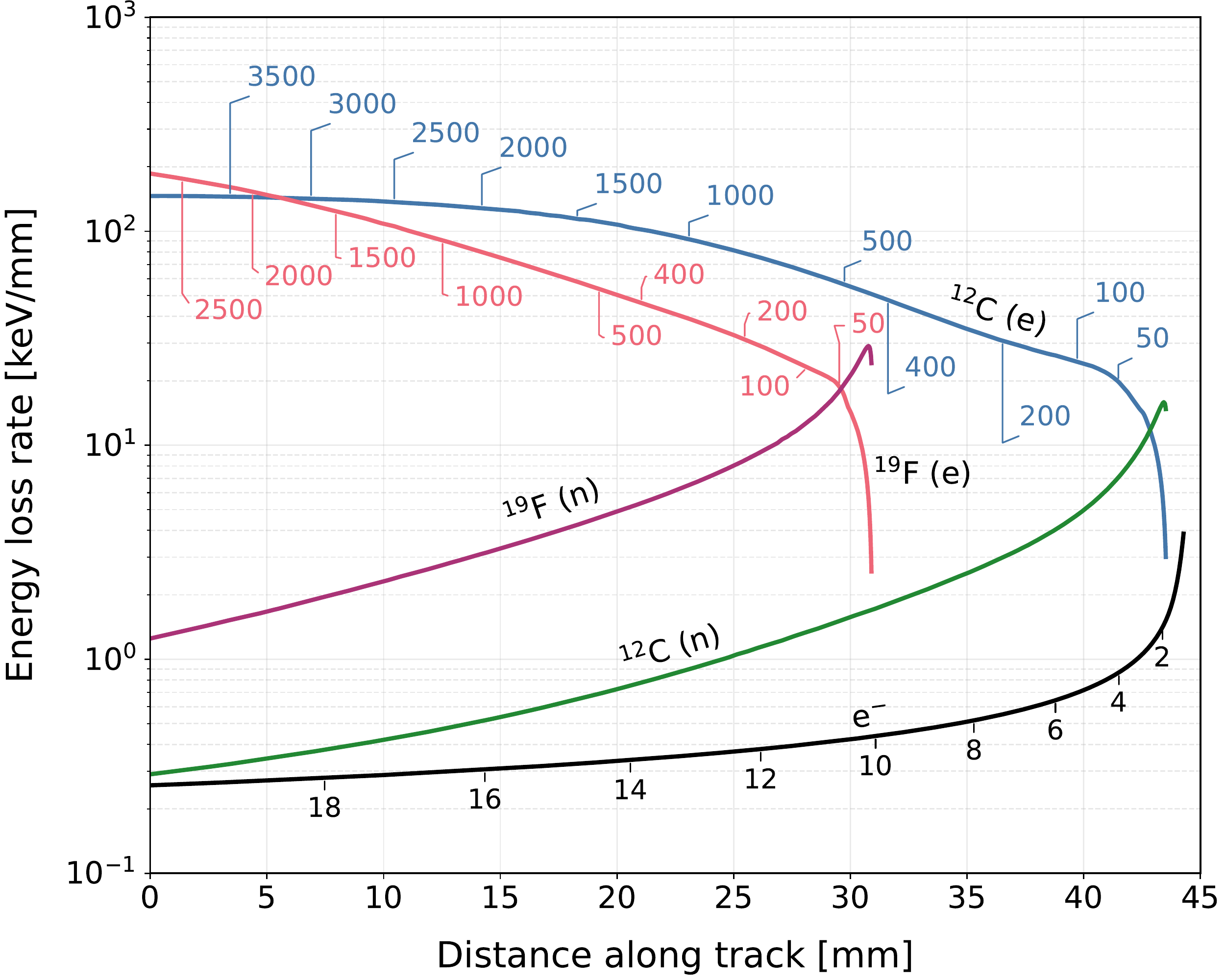}}
\caption{Left -- Track length in CF$_4$ at 50~Torr for electrons (mean projected range calculated with Degrad~\protect{\cite{Degrad}}, CSDA range with ESTAR~\protect{\cite{ESTAR}}, and the practical range formula from Ref.~\protect{\cite{Kobetich1968}}), and mean projected range for carbon and fluorine ions from SRIM~\protect{\cite{SRIM}}). Right -- Electronic and nuclear energy loss rates (CSDA) along carbon and fluorine ion tracks in CF$_4$ at 50~Torr, calculated with SRIM and electronic energy loss for 20~keV electrons obtained with ESTAR; called out values are interim particle energies (in keV) remaining at that point in the track.}
\label{Fig:dedx_and_track_length}
\end{figure*}

Low-energy electrons have convoluted tracks, and their spatial extent can be characterised by various metrics of `range' -- with some of the most common depicted in Fig.~\ref{Fig:dedx_and_track_length} (left). For the purpose of design optimisation we adopted the `practical range' for electrons~\cite{Kobetich1968}. This is longer than the `mean projected range' along the direction of incidence (which is the mean of a wide distribution), but not quite as long as the `Continuous-Slowing-Down Approximation (CSDA) range', which corresponds to the full track length in 3D. Simulations show that the practical range corresponds approximately to the 99$^\mathrm{th}$ percentile of the projected range distribution in our conditions. We adopt a 4-mm practical range to motivate an energy threshold of 5~keV for electrons as a nominal design value, which is conveniently calibrated in CF$_4$ by 5.9~keV X-rays from $^{55}$Fe. Other estimators for electron range will be used when analysing data, both real and simulated, but for the design process we opted for this simple metric; the sensitivity of the experiment does not change dramatically for electron threshold ranges $\pm$1~mm around this value.

At our energies of interest NR tracks are straighter than those from low-energy electrons and hence their mean projected range is closer to the CSDA range. The mean projected range for $^{12}$C and $^{19}$F recoils is also shown in Fig.~\ref{Fig:dedx_and_track_length} (left), obtained from SRIM~\cite{SRIM}. In this case, the 4-mm threshold translates to 130~keV and 170~keV for the two recoil species, respectively.

This point is illustrated in Fig.~\ref{Fig:dedx_and_track_length} (right), which shows the energy loss rates as a function of distance along the track for 20~keV electrons and for $^{19}$F and $^{12}$C ions with end-point energies resulting from D-T neutrons (2.8~MeV and 4.2~MeV). Electronic and nuclear stopping powers are shown for the ions, which are due to inelastic collisions with bound electrons and with nuclei, respectively. The numerical values indicated at various points along the electronic energy loss curves are the interim particle energies in keV during the slowing down process. For example, it can be inferred that the CSDA range for 170~keV $^{19}$F ions is 4--5~mm, which is close to the value seen on the left panel of the same figure.

Significantly, a distinctive `head-tail' difference can be observed in the energy loss rate along the tracks between electrons and the heavier ion recoils. The former lose more energy at the end of the track, while for the latter the opposite is true. This feature gives unique power to identify Migdal events against some backgrounds.

We note that the ratio between electronic and nuclear losses decreases towards low recoil energies; the fraction of the energy available in the electronic channel (the so-called `quenching factor') has been derived for binary gases including CF$_4$~\cite{Hitachi2008}. A measurement of this parameter is one of the goals of the experiment, exploiting the directional capability of the OTPC. As discussed later, the NR spectrum leading to a Migdal detection is not critically dependent on this parameter.\footnote{Hereafter, we use `keV$_{\mathrm{ee}}$' units to denote the electron-equivalent energy of nuclear recoil interactions which can be measured via scintillation or ionisation post-quenching, and reserve `keV' to denote particle energy.}

The opposing trend in energy loss rate along ER and NR tracks also poses a technical challenge for track reconstruction in the immediate vicinity of the Migdal vertex. As Fig.~\ref{Fig:dedx_and_track_length} (right) confirms, the two recoil types have very different magnitudes in `dE/dx' at production: a 5~keV electron loses energy at its minimum rate of 0.74~keV/mm, while a 2.7~MeV fluorine recoil does so at its maximum rate of 186~keV/mm. An accurate measurement of both recoils requires some 3~orders of magnitude in GEM dynamic range (specifically, in ionisation density in the GEM holes between the maximum signal and the noise floor for optical detection), which brings the OPTC close to the Raether limit~\cite{Raether1964} -- a challenging regime for achieving stable operation.

\label{SS:dedx}

\subsection{Signal rates}
\label{SS:Signal}

For the purpose of signal estimation we define an observable Migdal event as an interaction containing an ER track protruding out of the penumbra of an NR track in the 2D image, such that the Migdal topology can be unambiguously detected. The extent of the NR penumbra is determined by the large energy deposition near the vertex and diffusion of its ionisation cloud. We adopt the nominal 5~keV electron threshold, noting that lower energies will nonetheless be recorded in our data and available for analysis, depending on whether these can be detected past the NR penumbra. Here we caution that Migdal emission necessarily leads to atomic deexcitation~\cite{Ibe:2017yqa}, and for light atoms this is dominated by Auger (or Coster-Kronig) emission rather than X-ray fluorescence. However, in most of this article we assume that, for nominal operating conditions, the sub-mm Auger electron tracks ($\lesssim$0.65~keV~\cite{Larkins1987}) will not be resolvable near the vertex. We return to this topic in Section~\ref{S:Gases}, since some interesting conclusions can be drawn for heavier elements.

The Migdal signal rate induced by D-D or D-T neutrons depends on the properties of the neutron beam, the volume and pressure of the OTPC, the neutron interaction cross section with atomic nuclei, and on the Migdal electron emission probability -- how likely it is for an electron to be ionised during the sudden-collision process between the neutron and an atomic nucleus. The total neutron flux and that entering the active region of the OTPC are given in Table~\ref{Tab:nrates}. The two generator yields differ by an order of magnitude, but the D-D collimator is shorter giving a comparable flux through the detector, as described in Section~\ref{S:NeutronSource} where the neutron systems are detailed.

\begin{table}[t!]
\caption{Nominal neutron beam parameters and estimated event rates from the D-D and D-T generators in CF$_4$ gas at 50~Torr. Beam widths are indicated at FWHM and for 99\% of flux (`halo'). Interaction rates are given for all NR tracks originating in the active region and also for those fully contained within an 8$\times$8~cm$^2$ fiducial region and track length greater than 4~mm. Migdal interaction rates (no efficiencies considered) are given for tracks contained in the fiducial region, at track thresholds of 3~mm, 4~mm and 5~mm. The baseline scenario (4~mm, in bold) integrates the NR spectrum from 130~keV for carbon and 170~keV for fluorine, and electron energies in the range 5--15~keV; rates are indicated also for a low electron energy threshold (0.5~keV).}
\vspace{5pt}
\centering
{\small
\begin{tabular}{l | r r }
\hline
Generator & D-D  & D-T \\
Nominal neutron energy (MeV) & 2.47 & 14.7 \\
\hline
Neutron intensity (n/s)\\
\quad Emitted ($4\pi$) & $1\!\times\!10^{9}$ & $1\!\times\!10^{10}$ \\
\quad Active region    & $2.6\!\times\!10^{5}$ & $4.7\!\times\!10^{5}$ \\
Beam width/halo (cm)\\
\quad Vertical     &  9.0/9.0 & 9.2/9.2 \\
\quad Horizontal   &  1.4/1.8 & 1.3/1.5 \\
Recoil spectrum (MeV) & & \\
\quad $^{12}$C mean/end-point & 0.37/0.71 & 0.97/4.2 \\
\quad $^{19}$F mean/end-point & 0.17/0.47 & 0.52/2.8 \\
Interaction rates (evt/s) \\
\quad Total            &    53  &  68  \\
\quad Signal-inducing  &    53  &  48  \\
\quad Elastic          &    40  &  37  \\
Contained tracks$^\dagger$ (evt/s)\\
\quad Total            &    15  &  22  \\
\quad Signal-inducing  &    15  &  18  \\
\quad Elastic          &    11  &  14  \\
Migdal rates$^\dagger$ (interactions/day) \\
\quad NR track$\,\geq\,$4~mm, ER$>$0.5~keV & 7,250 & 26,430  \\
\quad Tracks$\,\geq\,$3~mm (ER\,$\in$\,4--15~keV)  &    89 &  255  \\
\quad Tracks$\,\geq\,$4~mm (ER\,$\in$\,5--15~keV) &    {\bf 43} &  {\bf 131}  \\
\quad Tracks$\,\geq\,$5~mm (ER\,$\in$\,6--15~keV)  &    22 &  74  \\
\hline
\end{tabular}
}
\begin{flushleft}
{\footnotesize $^\dagger$ In 8$\times$8~cm$^2$ fiducial region, NR track $\ge\,$4~mm unless indicated.}
\end{flushleft}
\label{Tab:nrates}
\end{table}

The experiment is designed such that neutrons interact minimally with the detector structure around the low-pressure CF$_4$ gas in the active region, where they scatter off $^{12}$C or $^{19}$F nuclei. The mean and end-point of the respective recoil spectra are given in Table~\ref{Tab:nrates}. The neutron scattering rates on CF$_4$ gas at 50~Torr in the active region of the OTPC are also listed, calculated with GEANT4~\cite{Agostinelli2002}~v10.5.1 (G4NDL~4.5); event rates of 50--70~per second are expected. Although we will analyse all tracks starting inside the TPC, only a fraction of those will be above the NR threshold, and only a fraction of those are fully contained (head and tail) within a more restricted fiducial region (8$\times$8~cm$^2$), and the rates for these tracks are also listed in the table.

Neutron scattering cross sections at D-D and D-T energies are listed in Table~\ref{Tab:nxs}, taken from the ENDF/B-VIII.0~\cite{Brown2018} database which was used to validate our Monte Carlo simulations. This table lists processes that produce events featuring `bare’ nuclear recoils, i.e.~those with no accompanying charged tracks. In addition to elastic scattering, these include inelastic scattering and neutron capture (where the $\gamma$-rays can easily escape the detector) as well as $(n,2n)$ reactions: these can all contribute to the Migdal signal rate (we label this sum $\sigma_s$). In general, the dominant interaction in CF$_4$ is with $^{19}$F mostly due to molecular composition. In particular, elastic scattering on $^{19}$F dominates the interaction rate in CF$_4$ for D-D neutrons, and yields 62\% of all signal-inducing NR events; $^{12}$C provides a further 12\%. Inelastic scattering on $^{19}$F contributes the remaining 26\%, with around half coming from scattering via the $n=2$ level at 197~keV. With D-T neutrons, elastic scattering on $^{19}$F contributes to $\sigma_s$ in similar proportions to the D-D case -- but more inelastic levels from both isotopes are accessible in this case (adding up to 21\% of the signal-inducing recoil events). This discussion highlights another advantage of CF$_4$ as the target gas: the natural abundance of carbon and fluorine is dominated by a single isotope of each species, so the neutron inelastic interactions are relatively simple.

\begin{table}[t]
\caption{Neutron cross sections at 2.47~MeV and 14.7~MeV from ENDF/B-VIII.0~\protect{\cite{Brown2018}} (all values in mb); $\sigma_0$ denotes the total cross section, and the four signal-inducing processes ($\sigma_s$) include partial cross sections for elastic scattering $(n,n)$, inelastic scattering $(n,n')$, the $(n,2n)$ threshold reaction, and radiative capture $(n,\gamma)$.}
\vspace{5pt}
\centering
{\small
\setlength{\tabcolsep}{8pt}
\renewcommand{\arraystretch}{1.1}
\begin{tabular}{l | r r | r r }
\hline
& \multicolumn{2}{c|}{ 2.47~MeV (D-D)} & \multicolumn{2}{c}{14.7~MeV (D-T)} \\
\hline
nucleus             & $^{12}$C  & $^{19}$F  
                    & $^{12}$C  & $^{19}$F \\
\hline
$\sigma_0$          & 1,613      & 3,038     
                    & 1,379      & 1,786     \\
$(n,n)$             & 1,613      & 2,131     
                    &   895      &   985     \\
$(n,n')$            &    --      &   907     
                    &   426      &   235     \\
$(n,2n)$            &    --      &    --     
                    &    --      &    52     \\
$(n,\gamma)$        &  0.05      &  0.09     
                    &  0.15      &  0.03     \\
$\sigma_s/\sigma_0$ & 100\%      & 100\%
                    &  96\%      &  71\%     \\
\hline
\end{tabular}
}
\label{Tab:nxs}
\end{table}

Each signal-inducing NR has a small probability of emitting a detectable electron via the Migdal effect. In calculating the Migdal event rates, we use the `semi-inclusive’ probabilities as a function of the electron kinetic energy from Ref.~\cite{MIGDALth}. This improves on the calculations by Ibe and co-workers~\cite{Ibe:2017yqa} in two key ways. Firstly, it does not employ the dipole approximation, which breaks down in our NR energy regime. Secondly, the semi-inclusive rate accounts for the possibility of ionisation together with excitation of other electrons in the atom, as well as multiple-ionisation in the Migdal process in which one electron is above the ER threshold while additional electrons are below threshold. The semi-inclusive process therefore leads to the same characteristic Migdal event topology of an ER and NR track with a common vertex since any additional electrons will not be observable with our detector. When we refer to electron energy in the ensuing discussion, we exclusively refer to the kinetic energy of the above-threshold Migdal electron.

We note that the calculations in both Ref.~\cite{MIGDALth} and Ref.~\cite{Ibe:2017yqa} are for isolated atoms rather than for the nuclei in a CF$_4$ molecule. Corrections to these probabilities, and hence to the Migdal event rates that we present here, are therefore expected, although we anticipate that these will be relatively small. This is because, over most of our ROI, electrons emitted with D-D and D-T neutrons are the inner-most ones, where the deviation from atomic wave-functions are small. A more extensive discussion of the theoretical rates is given in Ref.~\cite{MIGDALth}.

\begin{figure*}[htb!]
\centerline{
    \includegraphics[width=0.47\linewidth]{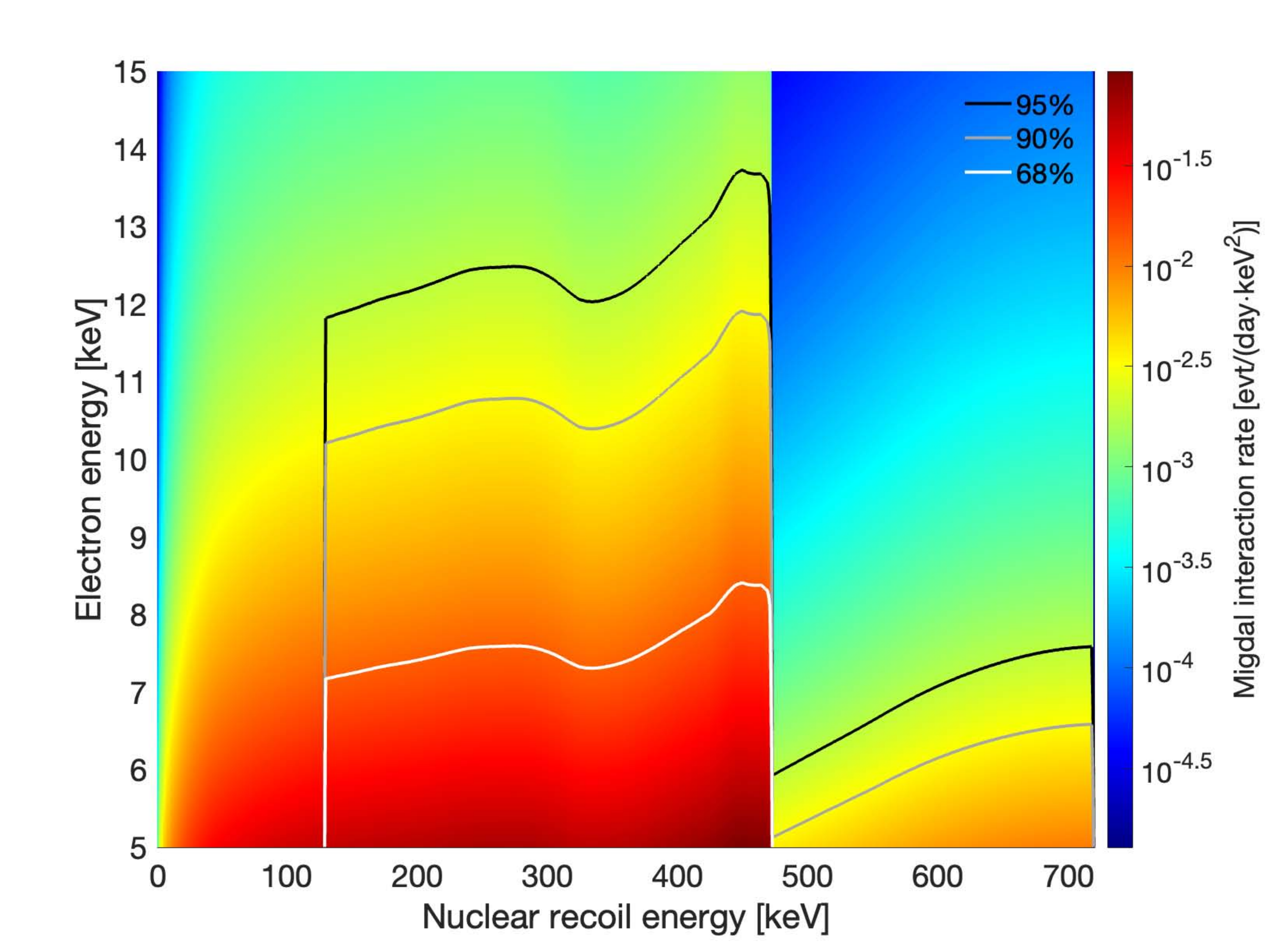}\qquad
    \includegraphics[width=0.47\linewidth]{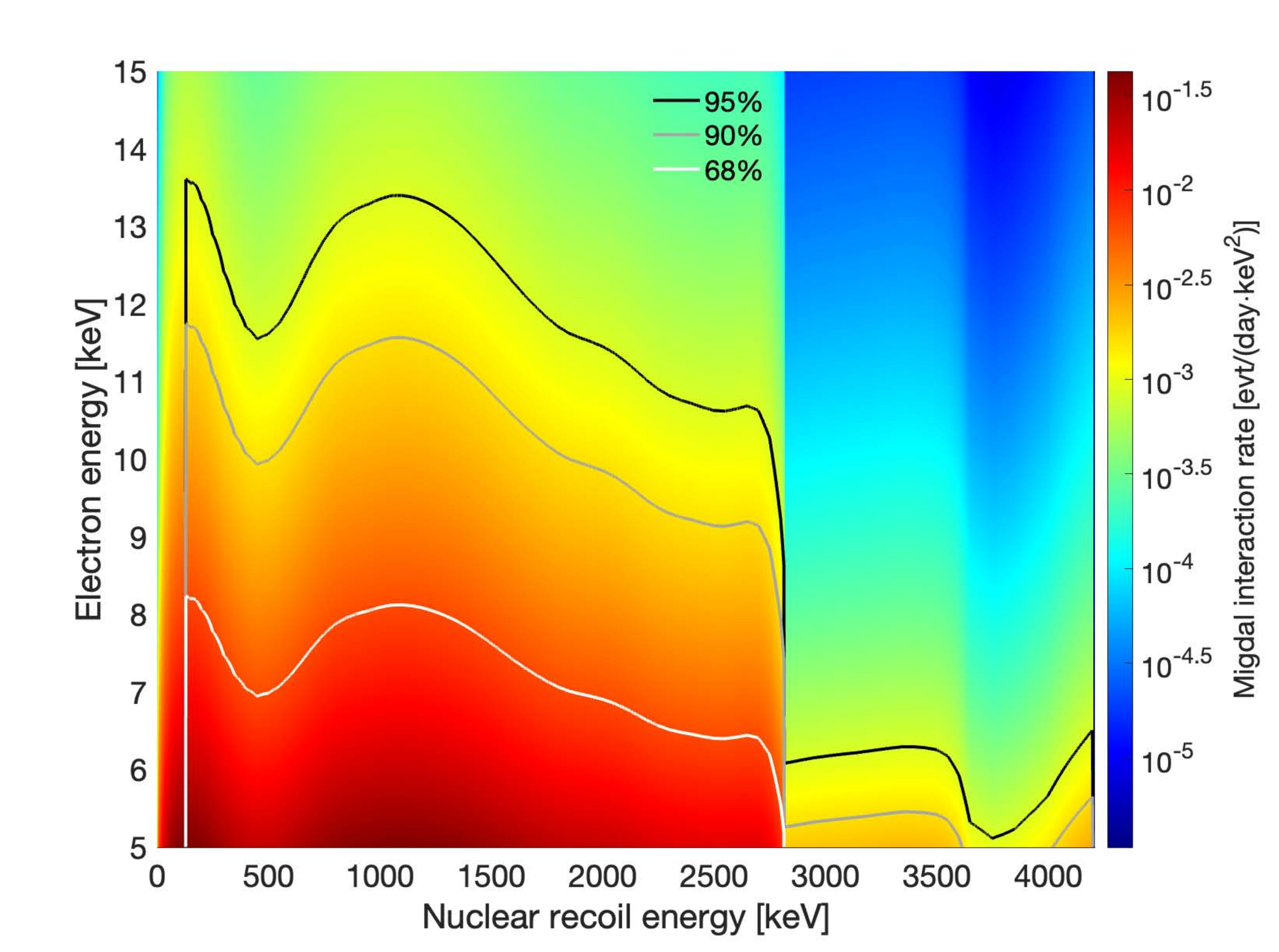}}
    \caption{Double-differential Migdal rates for tracks contained in the OTPC from D-D (left) and D-T (right) generators. The contours are based on the NR thresholds of 130~keV and 170~keV for C and F, respectively. The area bound by the contours encompasses 68\%, 90\% and 95\% of the signal; we note that the $y$-axis refers to the energy carried by the most energetic electron in the semi-inclusive calculation~\cite{MIGDALth}.}
\label{Fig:Signal}
\end{figure*}

Double-differential Migdal event rates with respect to NR and ER energies are shown in Fig.~\ref{Fig:Signal} for D-D (left) and D-T (right) neutrons. The plots consider only tracks contained in the 8$\times$8~cm$^2$ fiducial region; this selection softens the NR spectrum in the search sample, especially removing carbon recoils from D-T neutrons. The sharp discontinuities at 470~keV (left) and 2,815~keV (right) NR energy occur at the $^{19}$F end-point energies and highlight that the Migdal rates in CF$_4$ are dominated by fluorine. The log-scale colour maps illustrate that, in general, the rate drops off exponentially with increasing electron energy; therefore, a low ER threshold is essential for this experiment. The differential rate depends on the product of the electron emission probability with the differential neutron cross section. The interplay between these leads to features in the Migdal rate, e.g.~the rise at $E_r \sim 400$~keV (left) and the broad peak at $E_r \sim 1000$~keV (right).

As described previously, we define our nominal region of interest (ROI) to include NR and ER tracks each longer than 4~mm sharing a common vertex. This corresponds to NR energies greater than 130~keV and 170~keV for C and F ions, and ER tracks with energy greater than 5~keV. At the upper end of the ROI we extend to the NR end-point energies, while for ER tracks we accept up to 15~keV. As the Migdal rate drops off steeply with ER energy, this upper value has little effect on the signal rate -- but it will play a role in limiting background rates (cf.~Section~\ref{SS:Backgrounds}).

Migdal rates per live day are given in the last few rows of Table~\ref{Tab:nrates} for the nominal ROI plus additional threshold scenarios. These are obtained by integrating the differential rates exemplified in Fig.~\ref{Fig:Signal} as indicated in the table. With both neutron sources the Migdal yield originates mostly in fluorine, with the carbon contribution being modest: 10\% for D-D and 14\% for D-T. The first entry utilises a very low ER threshold (0.5~keV) to highlight the large number of events with small track lengths hidden below the detection threshold. Other scenarios correspond to a smaller ($\pm1$~mm) deviation from the nominal value of 4~mm. We conclude that the number of potentially detectable Migdal events per live day is always significant, especially for the D-T generator, and that small departures from the nominal electron threshold are reasonably inconsequential: a decrease to 3~mm approximately doubles the event rate, while an increase to 5~mm threshold halves the rate.

\subsection{Angular distribution of nuclear recoils}

\begin{figure}[t!]
\centering
\includegraphics[width=0.47\linewidth]{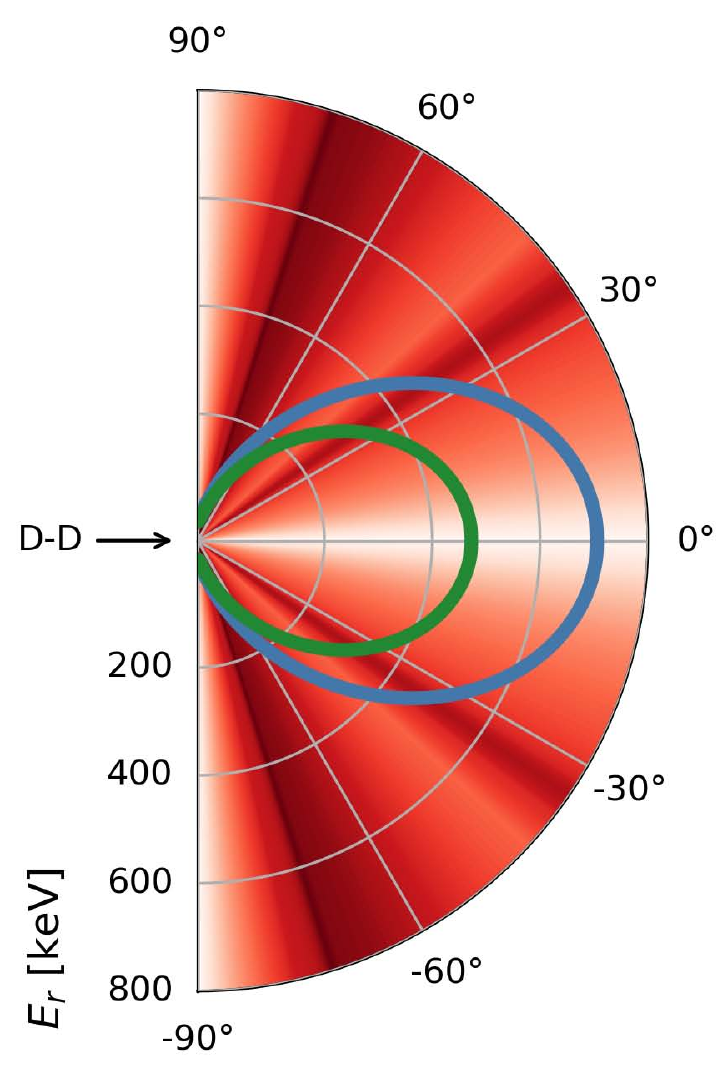}\quad
\includegraphics[width=0.47\linewidth]{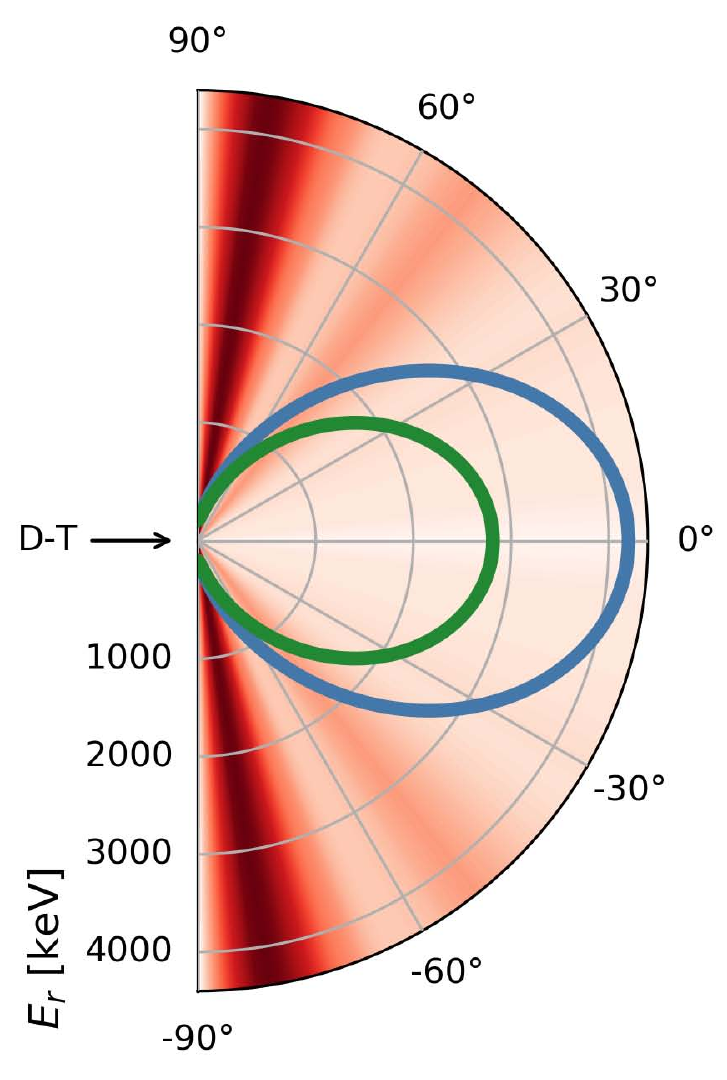}
\caption{Polar representations of nuclear recoil angle distributions from D-D (left) and D-T (right) neutrons. The black arrows indicate the direction of the neutron beam. The colour maps represent the neutron cross section as a function of NR angle in the laboratory frame, with darker red corresponding to increasing cross section on a linear scale; these include the signal-generating processes listed in Table~\ref{Tab:nxs} for CF$_4$. A small contribution from $^{19}$F($n$,$2n$) in the backward direction is omitted in the D-T plot. The solid lines show calculated recoil energies for elastic scattering only, with $^{12}$C in blue and $^{19}$F in green. The energy scale refers to this energy plot only, which is independent of the colour map.}
\label{Fig:polar}
\end{figure}

In addition to event rates, we may also consider the angular distributions of the recoiling nuclei in the laboratory frame, as this will impact the design of the OTPC and inform how data are analysed. Figure~\ref{Fig:polar} illustrates the neutron angular cross sections for CF$_4$, as well as energy spectra for the two atomic species recoiling elastically. In the elastic case the recoil energy can be uniquely determined from the scattering angle, while in inelastic scattering the energy (not shown) is double-valued for a given recoil angle, up to a maximum angle which is lower than 90$^\circ$ in the laboratory frame. Inelastic scattering does contribute to some features observed in the figure. An extensive discussion of the nuclear recoil distributions induced by the scattering of fast neutrons is given in~\ref{A:NeutronScattering}.

As expected, most elastic recoils are emitted with low energies at large angles with respect to the beam and, conversely, the most energetic recoils are emitted in the forward ($+x$) direction, and the same is true for the inelastic case; this determines the optimal orientation between the beam and the OTPC, including the optimisation of its geometry, accounting for the maximum NR range as a function of angle and the orientation of the charge readout strips as discussed later.

For D-D neutrons, both elastic and inelastic scattering contribute to the broad peak observed near 72$^\circ$ in the overall angular cross section, with a smaller enhancement around 34$^\circ$ which is mostly due to inelastic scattering off the $n=3,4,5$ levels in $^{19}$F. At this neutron energy, fluorine recoils are more numerous at all angles: from a factor of approximately four in the forward direction, to greater than 10 above 70$^\circ$.

The polar distribution for recoils from D-T neutrons is strongly peaked around 80$^\circ$, with a modest enhancement near 52$^\circ$. Around 3\% of the signal-inducing events come from $^{19}$F($n$,$2n$) reactions, with half of those emitted in the backward direction (these are not represented in the figure). In the D-T experiment, $^{12}$C recoils dominate below 15$^\circ$, while the fluorine-to-carbon ratio peaks sharply (13$\times$) at around 57$^\circ$: this is due both to an enhancement of the fluorine cross sections and to a significant dip in the carbon cross sections near those angles.

A good understanding of these distributions will aid with data analysis: $i$) to make robust `head-tail' determinations for NR interactions, since the only processes that can yield `backward' tracks have low probability; $ii$) by enabling NR energy reconstruction from scattering kinematics: for quenching factor measurements, for analysis of tracks which are not fully contained in the OTPC, and possibly to circumvent NR response saturation near the start of the NR track; and $iii$) by allowing us to focus on angular regions with higher signal yields and, possibly, to help distinguish between the contribution of each atomic species -- see example in Section~\ref{S:Gases}.

\subsection{Background mitigation strategy}
\label{SS:BackgroundStrategy}

The exposure of a sensitive detector to a very high flux of energetic neutrons will inevitably lead to the production of background topologies which can bear close resemblance to the signature presented by the Migdal signal. These backgrounds include various atomic processes leading to particle emission directly from the neutron-induced NR track, as well as random associations between an NR track and unrelated low-energy electrons -- either beam-coincident or purely accidental.

A quantitative discussion of background processes is postponed until Section~\ref{SS:Backgrounds}, as this must follow a full description of the experiment. Here we introduce the main strategy for background mitigation, which is central to the experimental design: the operation of the detector using a low pressure gas. This enables two powerful avenues for background discrimination: $i)$ the reconstruction of extended tracks allows selection of the distinct Migdal event topology; and $ii)$ by reducing the interaction probability for energetic photons (of both internal and external origin) near the NR track, an important class of potential backgrounds is mitigated effectively. We do not anticipate this measurement to be background limited.

\begin{figure}[t]
\centerline{\includegraphics[width=0.95\linewidth]{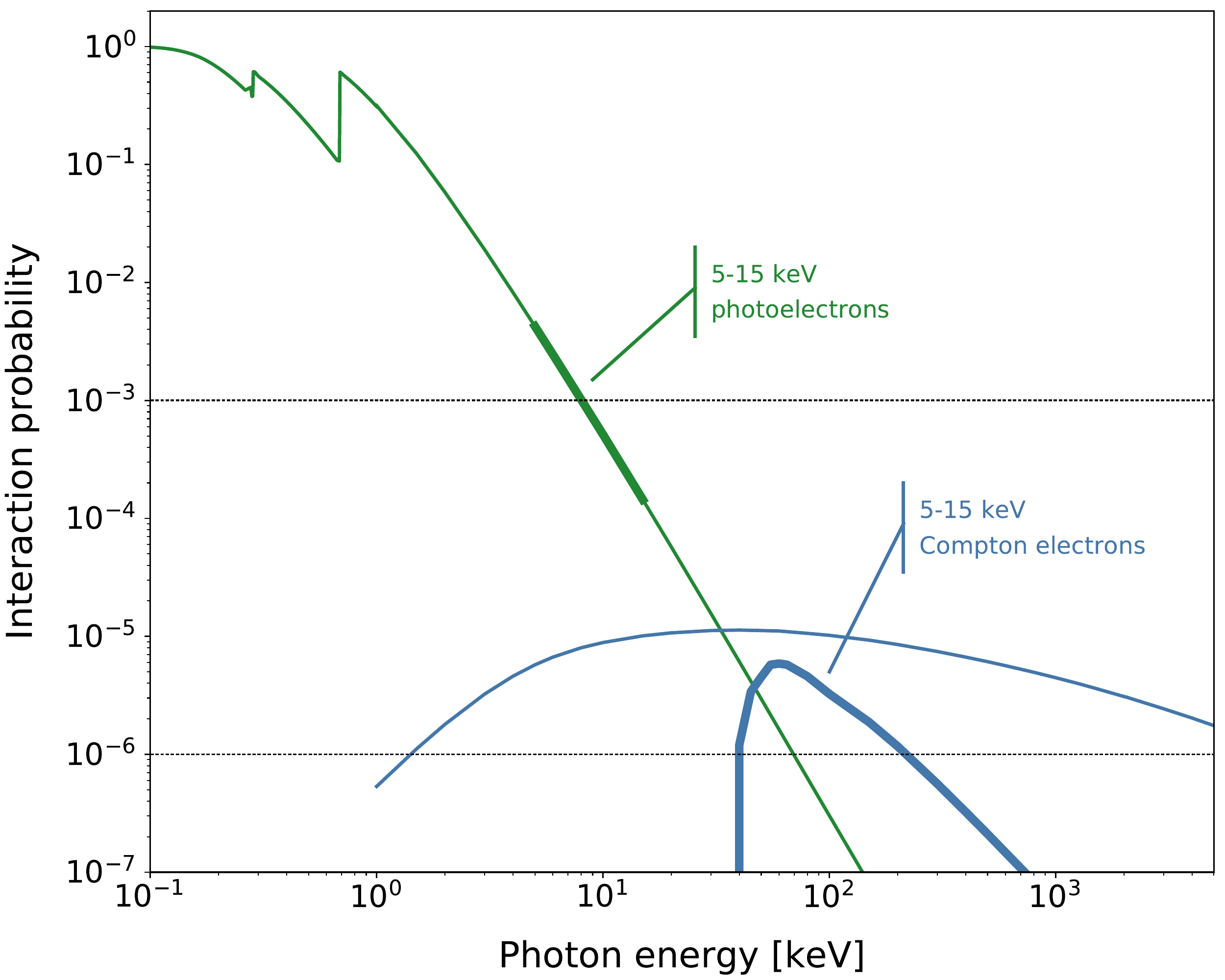}}
\caption{Photon interaction probabilities within a 3~mm distance for CF$_4$ at 50~Torr, using data from Refs.~\protect{\cite{FFAST,XCOM}}. Photoelectric data are shown in green, with the 5--15~keV photoelectron energy range highlighted by the thicker green line. Compton interaction probabilities are shown in blue, with the 5--15~keV Compton electron region of interest shown by the thicker blue line; the latter was obtained by integrating the Klein-Nishina cross section formula between the appropriate angles.}
\label{Fig:PhotonsInCF4}
\end{figure}

Photon interaction probabilities for 50~Torr of CF$_4$ are shown in Fig.~\ref{Fig:PhotonsInCF4}. The thicker lines highlight the probability that 5--15~keV photoelectrons or Compton electrons are produced up to a distance of 3~mm from the photon origin, which we consider here to coincide with the NR track vertex. A low-energy electron track in this (generously sized) region may give rise to a background topology (note that the spatial resolution of the OTPC is a fraction of this value in all three dimensions). Photoelectric absorption causing ROI electrons within 3~mm occurs with $\mathcal{O}(10^{-3})$ probability; this is applicable to X-ray emission from various atomic processes. In turn, Compton scattering of $\gamma$-rays is down at $\mathcal{O}(10^{-6})$ probability; this applies, for example, to nuclear radiation from inelastic neutron scattering. Both types of photon-emitting process would otherwise be important sources of background.

The use of a light-element gas in the first stage of the experiment is also important for the success of the initial measurement. Electron and X-ray emission from atomic deexcitation anywhere along the NR track are in principle problematic, and as long as atomic shell energies lie below the electron ROI these will not produce a background. K-shell binding energies for carbon and fluorine are 284~eV and 697~eV, respectively. One notable example already highlighted above is the Auger emission accompanying the Migdal effect -- although we expect some Auger electrons to be resolvable in $^{55}$Fe calibration, they should not be visible past the NR track penumbra for light-element recoils.

\section{The MIGDAL detector}
\label{S:Experiment}

\subsection{The Optical Time Projection Chamber}
\label{SS:OTPC}

\begin{figure*}[ht]
\centerline{\includegraphics[width=0.9\textwidth]{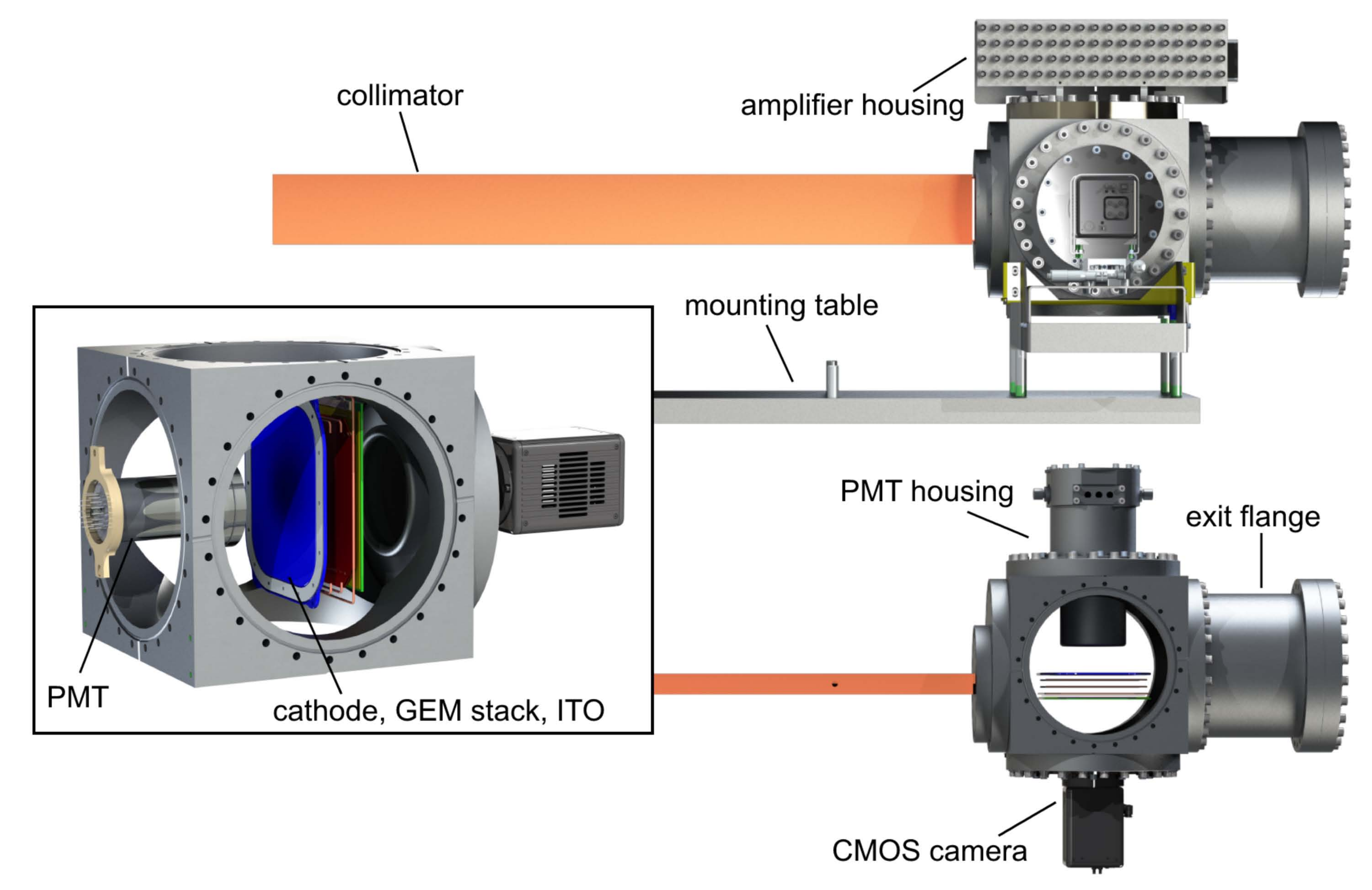}}
  \caption{Schematic representation of the MIGDAL detector, showing a side view (upper panel) and the top view of the setup (lower panel). The inset details the main chamber volume, with the OTPC viewed by a CMOS camera and a photomultiplier tube, both located behind reentrant viewport flanges outside of the gas space. The collimator shown is for the D-T experiment; the D-D collimator is just under half the length.}
\label{Fig:Experiment}
\end{figure*}

TPCs are excellent radiation detectors, providing a complete, three-dimensional (3D) image of the ionisation released in an active volume filled with a gas or a liquid. A key technical advance was the introduction of gas scintillation with optical readout for 2D imaging~\cite{Policarpo1977}, later in combination with the Gas Electron Multiplier (GEM)~\cite{Sauli2016,Fraga2003}. The OTPC technology has been successfully used, for example, in imaging of exotic nuclear decays~\cite{Cwiok2005}. Modern detection techniques deploying TPCs with imaging capability are reviewed in Ref.~\cite{Diaz2018}.

The MIGDAL experiment will benefit from these advances to deploy a combination of 2D-projection imaging using a CMOS camera, signal amplification with the cascade of two thick glass-GEMs~\cite{Fujiwara2016}, and electronic readout in the drift direction providing timing information for depth-coordinate reconstruction. Operation with a low pressure scintillating gas (50~Torr of CF$_4$, 0.24~mg/cm$^3$) will allow the reconstruction of low-energy tracks in 3D.

The setup is illustrated in Fig.~\ref{Fig:Experiment} -- with the longer collimator designed for the D-T generator. The TPC structure will be installed vertically in a 10-inch cubic vacuum chamber made from aluminium alloy, sealed by 6 aluminium Conflat flanges (DN200CF). Flanges at the front and rear incorporate 150~$\mu$m and 250~$\mu$m-thick aluminium windows to minimise neutron interactions near the active volume; the rear window is located at the end of a 19.5~cm long exit flange. These windows are epoxy-sealed and must be able to endure stresses from vacuum/pressure cycling. Flanges at either side of the cube support reentrant optical viewports to allow prompt and secondary scintillation light to be detected by a photomultiplier and a camera, which view the TPC from opposite sides. The two remaining flanges provide electrical connections (bias and readout), gas inlet/outlet ports, and two small 50-$\mu$m-thick aluminium windows for calibration with external sources.

The D-T neutron beam is coupled to the active region of the TPC using a 1-m long air-filled collimator made from pure copper, installed between the neutron generator head and the front of the cube. Together with a front shield composed of various materials, the collimator creates a well-defined neutron beam with minimum halo passing through the detector. These elements -- and the shorter D-D collimator -- are detailed in Section~\ref{S:NeutronSource}.\footnote{In an earlier design this was a vacuum collimator with entrance window near the neutron generator, to mitigate secondary particle production in the window material near the OTPC. After careful evaluation of these backgrounds we have opted for a less challenging design using an air-filled collimator, moving the detector window nearer the chamber. The exit window arrangement was similarly retracted (as shown in Fig.~\ref{Fig:Experiment}), although the actual window flange can also be attached directly to the chamber.}

\subsubsection{Active region}

Particle tracks will be created in the 3.0-cm drift region defined between a cathode mesh and the first GEM, as exemplified in Fig.~\ref{Fig:ExperimentIllustration}; the active area defined by the GEMs measures 10$\times$10~cm$^2$. The short drift gap will allow full development of most tracks while minimising diffusion of the primary ionisation. The cathode mesh (15.0$\times$12.6~cm$^2$) is woven from 280~$\mathrm{\mu}$m aluminium wire and has 66\% optical transparency, transmitting the CF$_4$ luminescence through to the photomultiplier. Its support ring (18.6$\times$17.4~cm$^2$) is made from two aluminium alloy halves, with the mesh clamped in between.

Three field-shaping rings, made of 2-mm diameter pure copper wire and spaced by 10~mm, will maintain a uniform electric field in the active volume. The two upper field-shaping electrodes are interrupted at the entrance and exit of the neutron beam to prevent spurious interactions that could generate background. The third is located at the level of the top electrode of the first GEM, and is wider to match the cathode ring dimensions.

\begin{figure*}[htb]
\centerline{
\includegraphics[height=7.0cm]{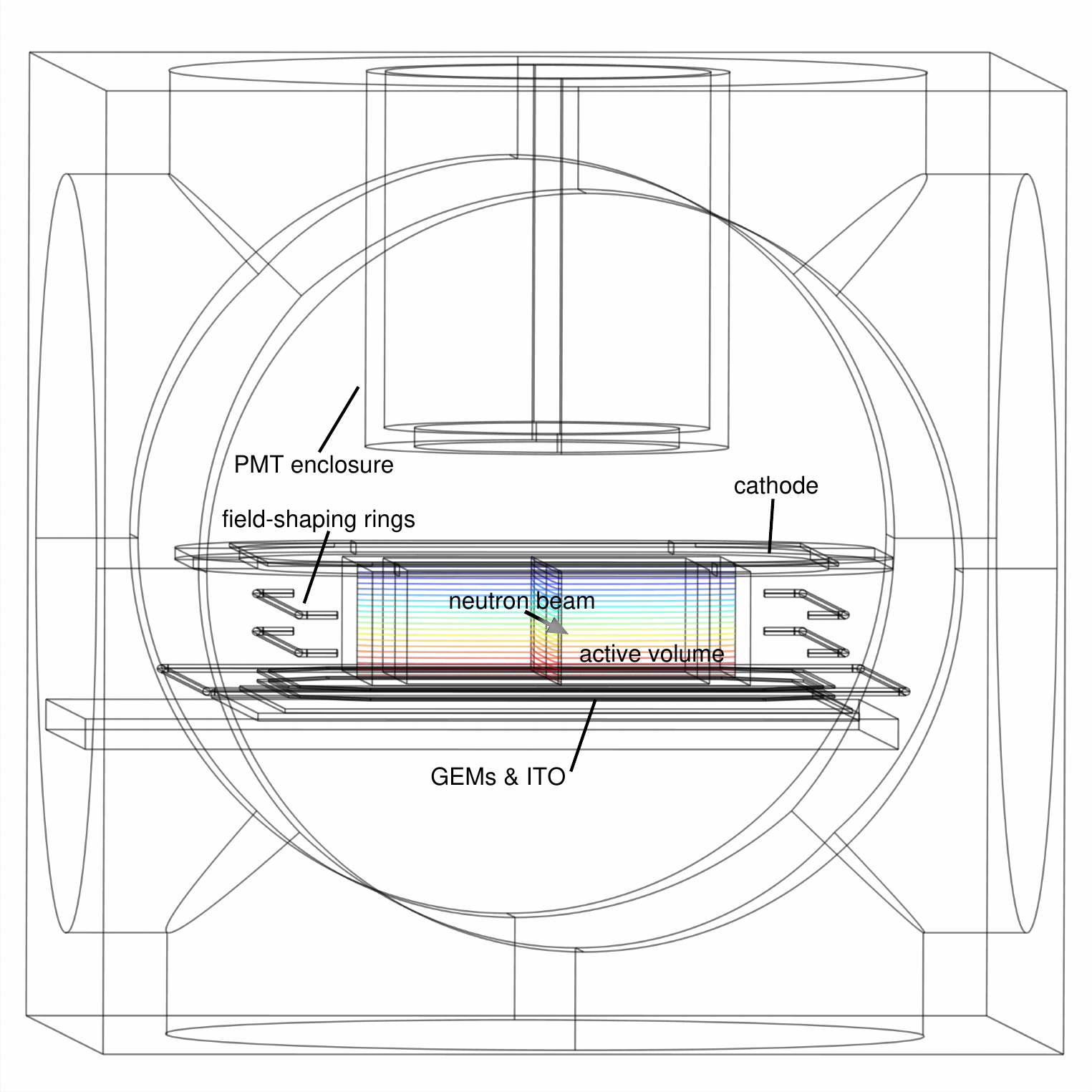}\qquad
\includegraphics[height=7.0cm]{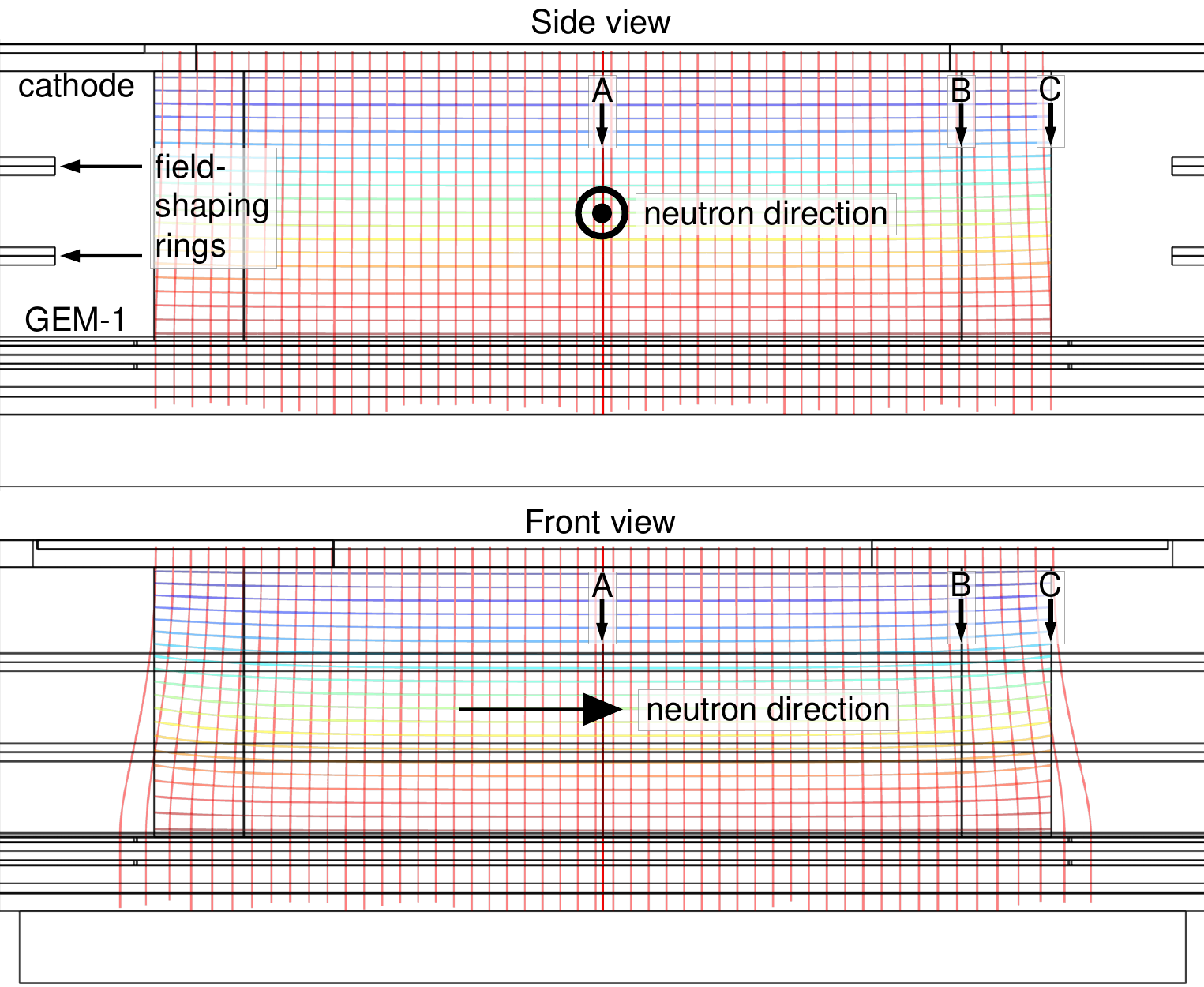}}
  \caption{Electrostatic design of the TPC. Left -- Detector model used in COMSOL; the active volume between the cathode and the first GEM lies between the two sets of field-shaping electrodes. Right -- Distribution of electric field lines between the cathode and the first GEM in the direction perpendicular (upper map) and parallel to the beam (lower map).}
\label{Fig:Fields}
\end{figure*}

The discontinuity of the field-shaping rings for the neutron beam means that the field is distorted in the drift region, which will require the fiducialisation of the active volume during analysis. To optimise the field in the active volume, electrostatic finite-element analyses have been developed using the COMSOL Multiphysics software with the AC/DC module. A full model of the detector including all TPC elements is shown in Fig.~\ref{Fig:Fields} (left). The field has been studied in two volumes with transverse dimensions of 10.0$\times$10.0~cm$^2$ and 8.0$\times$8.0~cm$^2$. The electric fields in the three key regions of the detector -- the drift, transfer and induction regions -- are 200, 600 and 400~V/cm, respectively. In the drift region the field is constant along the TPC axis -- direction `A' in Fig.~\ref{Fig:Fields} (right) -- while there is a strong non-uniformity of $\approx$40\% in field strength at the corners of the larger area (along C); this decreases to $\approx$10\% at the corners of our more restricted 8.0$\times$8.0~cm$^2$ signal-search area (along B).

The drift field in the restricted 8$\times$8~cm$^2$ region will be 200~V/cm, at which the transverse electron diffusion in pure CF$_4$ at 50~Torr has a minimum value of 260~$\mathrm{\mu}$m/$\smash{\sqrt{\mathrm{cm}}}$; in these conditions the drift velocity is $\approx$13~cm/$\mu$s. The electric field non-uniformity due to discontinuity of the field-shaping rings will create a variation of the electron diffusion and drift velocity at the level of 5\%, which is acceptable for our experiment.

The two glass-GEMs with 14.5$\times$14.5~cm$^2$ overall dimension and 10.0$\times$10.0~cm$^2$ active area will be cascaded for signal amplification, with a combined gain of $\sim$10$^5$; these are manufactured by Radiment Lab Inc., Japan. The charge-transfer region between them is only 2~mm long in order to minimise transverse diffusion. The glass-GEMs are 570~$\mathrm{\mu}$m thick, with 2-$\mathrm{\mu}$m copper cladding on both sides, and a dense pattern of holes with $\approx$170~$\mathrm{\mu}$m diameter and a pitch of 280~$\mathrm{\mu}$m. Beyond the second GEM electrons will drift until they are collected by a segmented anode made from a transparent ITO layer patterned into readout strips; the length of this so-called induction region is also 2~mm to ensure the short duration of the induced pulses.

\subsubsection{Gas system}

\begin{figure}[ht]
\centerline{\includegraphics[width=\columnwidth]{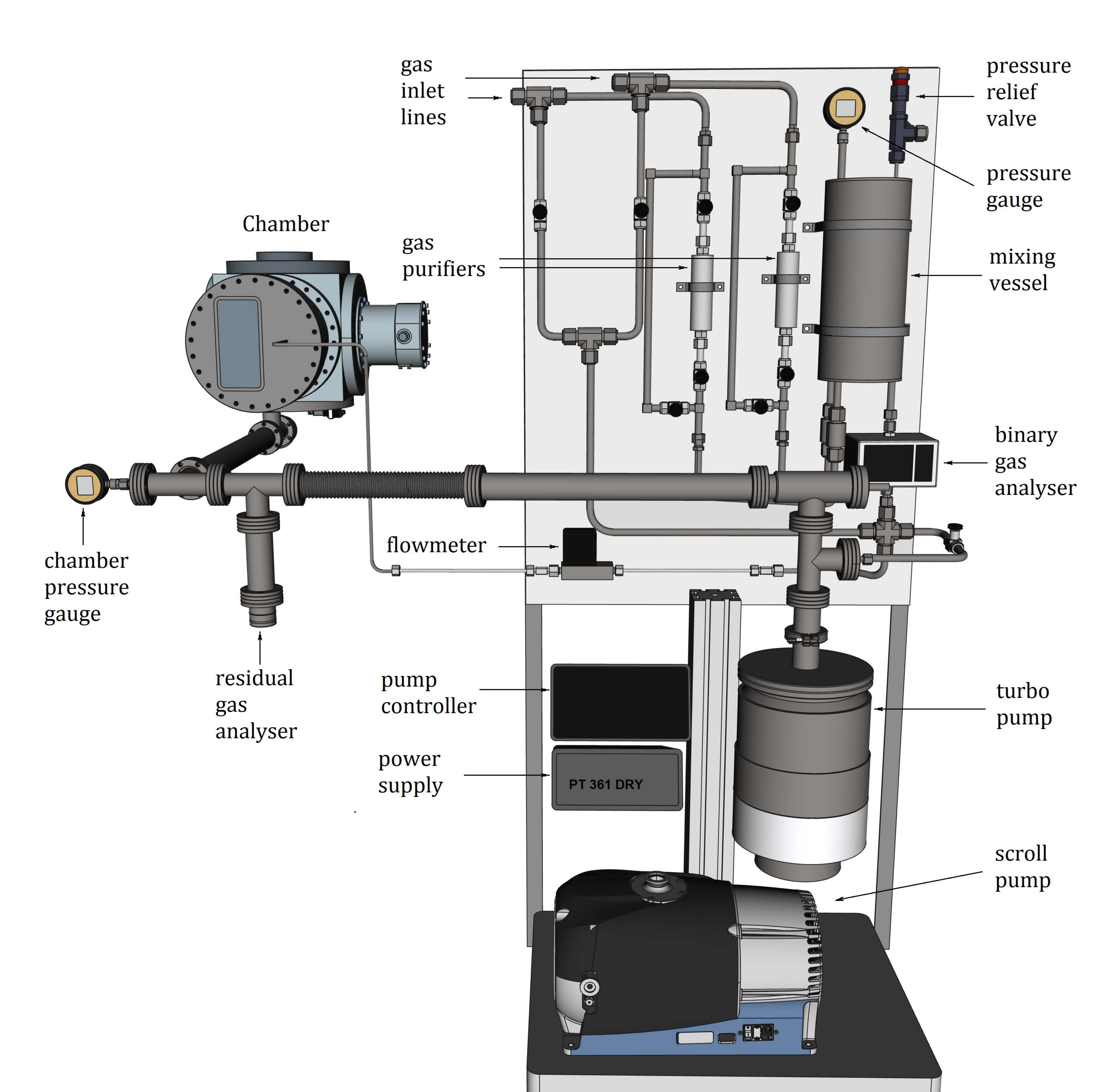}}
  \caption{The gas system design, including pumping station and gas mixing/assaying and delivery hardware.} 
\label{Fig:Gas}
\end{figure}

The gas system consists of a vacuum pumping station and a gas filling station, as illustrated in Fig.~\ref{Fig:Gas}. It is located inside the neutron `bunker' (described in Section~\ref{S:NeutronSource}), and supplied by gas cylinders (CF$_4$ plus noble gases) located outside of this area. The performance of the detector is influenced by gas purity, especially the presence of electronegative contaminants, and the goal is to be able to take high-quality data for over one day on a single gas load. Therefore, the design involves only metal-gasket seals and other low-outgassing materials. To achieve high purity initially, the chamber will be pumped to $\sim$10$^{-5}$~Torr, with the quality of the vacuum monitored by a Pfeiffer QMC-200 residual gas analyser (RGA); this is followed by purging the system with pure argon. Entegris GateKeeper GPU-80 gas purifiers are used to purify both types of gas. The mixing chamber will be filled to prepare the required gas mixture prior to insertion into the detector, with the composition measured precisely with a Lambda~BGA244 Binary Gas Analyser (BGA). When the required ratio is achieved, the mixing vessel is isolated and the gas is introduced to the detector through a needle valve to the desired pressure. The gas flow rate is monitored with a Teledyne Hastings HFM300 flowmeter. The pressure in the chamber is measured by a Keller LEO5 digital manometer to a precision of $<$1~Torr at 50~Torr.

\subsection{Optical readout}
\label{SS:OpticalReadout}

The optical readout of the detector involves two sensors: a photomultiplier tube (PMT) to detect the prompt and secondary scintillation signals from the CF$_4$ gas; and a camera system to image the particle tracks in the $x,y$ plane after amplification by the GEMs. We describe these in turn.

\subsubsection{Camera system}
\label{SSS:Camera}

A scientific CMOS camera (Hamamatsu ORCA-Fusion, C14440) will be used to image the secondary scintillation light produced by particle tracks as their ionisation is amplified by the GEMs. The camera is mounted inside a light-tight enclosure on the side of the chamber, viewing the second GEM through a Kodial glass viewport and the ITO anode plate -- as shown previously in Fig.~\ref{Fig:Experiment}. The optical system is focused on the GEM surface using a fast lens (EHD‐25085‐C~F0.85) with focal length of 25.6~mm. The optical system can image the 10$\times$10~cm$^2$ active area at distances of $\sim$12~cm. Precise alignment of the optical axis is achieved using a 2D linear stage.

The 15$\times$15~mm$^2$ CMOS sensor has 2,304$\times$2,304 pixels, with a  spectral sensitivity extending from $\sim$320~nm to $\sim$1,000~nm. This is well-matched to the secondary scintillation spectrum of CF$_4$, with the camera reaching a quantum efficiency (QE) of $\approx$75\% at the $\sim$620~nm peak wavelength~\cite{Morozov2012}. The Peltier-cooled CMOS sensor operates at $-15^{\circ}$C in combination with liquid cooling, offering a read-out noise of $\sim$1.4~electrons (rms) and dark current of $\sim$0.2~electrons/pixel/second. The camera reads the contents of the pixels with 16-bit analog-to-digital converters. A 2$\times$2 digital binning will be utilised for reduced data volume and improved signal-to-noise ratio at the expense of lower image resolution.

The camera will be operated in a free-running (internal) trigger mode: it will continuously image the GEM system at the maximum rate of 89.1~frames/s, corresponding to an exposure of 11.2~ms per frame. The camera readout uses a rolling shutter which allows for zero dead-time at the expense of an 11.2~ms time separation between the top and bottom rows of each image. This time separation is challenging when trying to maintain synchronicity with the photomultiplier tube. If a scintillation event occurs when the readout is located at the middle row of the image, there is an equal probability of the event appearing on the current frame or the following frame, and some probability that the event is split across two frames. In the latter case the images can be combined, at the expense of signal-to-noise ratio.

The maximum f/0.85 aperture setting allows for maximal optical sensitivity, at the expense of a shallow depth of field and greater vignetting. The shallow depth of field is expected to introduce a slight blur at the edges of the image, but as the resolution is limited by the GEM pitch this should be a sub-dominant effect. Vignetting can be accounted for via a flat-fielding correction based on images of uniformly distributed tracks from an $^{55}$Fe calibration source, although signal-to-noise may be reduced towards the edges of the image where the intensity is compromised.

We estimate a detection efficiency for GEM photons integrated across the secondary scintillation emission spectrum of $\sim$0.1\%, including geometrical light collection, optical transmission coefficients and camera QE. Combined with a nominal charge gain of 10$^5$ and a photon-to-electron ratio of 0.34~\cite{Kaboth2008} (all angles), this yields just over 30 detected photons per electron at the output of the GEMs.

\subsubsection{Photomultiplier readout}

A VUV-sensitive, 3-inch Hamamatsu R11410 PMT is a distinct part of the optical readout system: it views the active region of the OTPC through its cathode mesh, from a reentrant enclosure attached to the main chamber and located opposite to the camera position -- as shown in Fig.~\ref{Fig:Experiment}. Within this hermetic enclosure, the PMT is secured against a MgF$_2$ viewport, 50~mm away from the cathode. The external location avoids operation at low pressure and proximity to the neutron beam, but adequate optical coupling to the active region is still achieved with the reentrant design. The PMT is negatively biased to deliver a gain of $5\!\times\!10^6$, and the near-ground signal is digitised by the DAQ system along with the ITO-strip signals.

The main function of the PMT readout is to detect both the primary and the secondary scintillation signals from the active volume, which we denote by `S1' and `S2', respectively. The S2 pulse provides a convenient trigger for the acquisition, while the S1 signal (recorded in the pre-trigger region of the waveforms) determines the interaction time, and hence the absolute depth coordinate ($z$). In addition, for tracks oriented towards the cathode or the first GEM, the time difference between these two pulses also will help identify tracks only partially contained in the drift region.

The primary scintillation of CF$_4$ has two main continua, one centred at 160~nm in the VUV region and a UV/visible component extending between 200~nm and $\sim$400~nm~\cite{Hesser1967,Pansky1995,Morozov2011}; a further continuum exists above 550~nm which is particularly strong in the case of secondary scintillation from electron avalanching~\cite{Morozov2012}. The PMT quantum efficiency is significant down to its 160~nm cutoff (e.g.~23\% at 165~nm), and the viewport transmission is 85\% at those wavelengths; the PMT enclosure is purged with argon gas to prevent significant VUV absorption. In summary, the PMT system has reasonable sensitivity to the VUV scintillation, it is fully sensitive to the UV/visible component, and is partially sensitive to the red/infrared emission.

The prompt CF$_4$ scintillation yield is affected noticeably by pressure, with the shorter wavelength components increasing and the longer wavelength ones progressively disappearing as the pressure decreases. Below 1~bar the emission from $\alpha$-particles in the range 220--500~nm increases to $>$2~photons/keV~\cite{Morozov2010}; we use this value to estimate the yield for NR tracks. Light collection simulations using the ANTS2 package~\cite{Morozov2016} provide an estimate of light collection efficiency averaged over the active volume of $\sim$1\%, peaking at 2\% at the centre of the OTPC.

An S1 analysis threshold of a single photoelectron (phe) is likely possible, enabled by the short drift length of the OTPC and high drift speed of the gas, combined with the low dark count rate of the PMT ($\sim$500~c/s). This threshold translates to an average NR energy of around 50~keV for interactions near the optical axis of the OTPC, rising to 100~keV for those near the edge of the active region. Clearly, such a low S1 threshold brings significant stochastic fluctuations from counting statistics to the NR threshold, but this detection efficiency curve is easily calculable.

Secondary light generated in the GEMs will also be detected by the PMT. The smaller response generated within the first GEM should be detectable, as the direct light collection efficiency is significant; however, photons from the second GEM -- mostly reflected inside the chamber -- still dominate the overall response due to the additional gain. The transit time between GEMs is too small for the two optical pulses to be fully resolved over the longitudinal diffusion of the electron cloud even for parallel tracks, and so the signals from the two GEMs combine into a single S2 pulse containing a few hundred phe for NR signals at threshold.

The absolute $z$ coordinate can be determined from the time delay between the S1 and S2 pulses. A timing resolution of 10~ns or better is achieved by the fast scintillation decay time of CF$_4$ (6.2~ns~\cite{Lehaut2015}, see also \cite{Morozov2011}), the 9~ns (FWHM) transit time spread for this PMT model, and a 2-ns sampling time at the DAQ digitisers. A spatial resolution of $\sim$1~mm or better should therefore be within reach for the smallest NR signals.

\subsection{Charge readout}
\label{SS:ChargeReadout}

\begin{figure}[htb]
\centerline{\includegraphics[width=0.85\columnwidth]{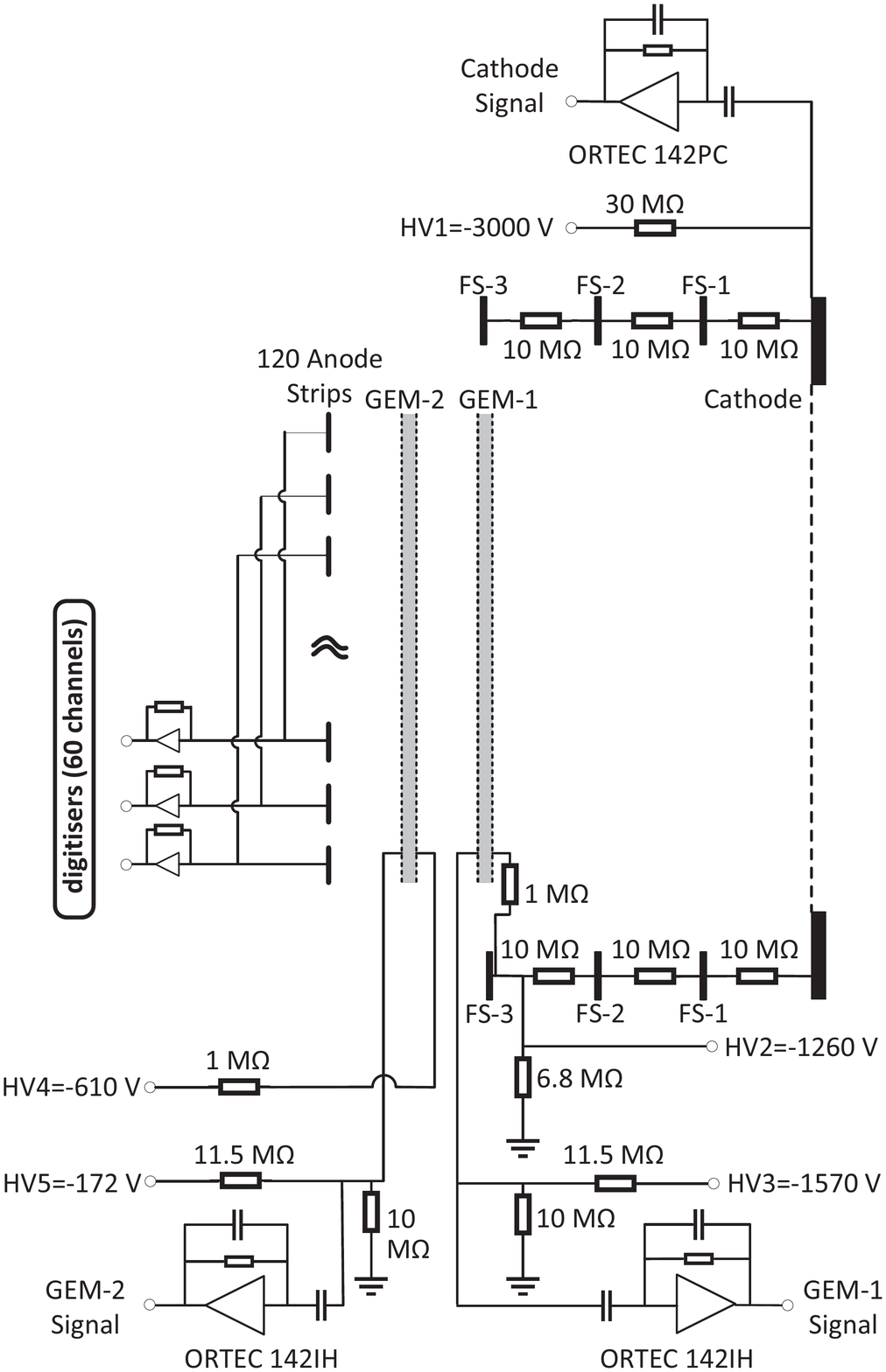}}
  \caption{Electrical circuit showing the OTPC bias and readout.} 
\label{Fig:TPC2}
\end{figure}

The OTPC bias and ionisation readout circuits are depicted in Fig.~\ref{Fig:TPC2}; the five bias voltages (HV1...5) set up the three electric fields and deliver 530~V across each GEM; the values are indicative only, they will be adjusted for the GEM resistance  (100--300~M$\Omega$ expected).

Depth resolution will be achieved by reading out the charge deposited on the transparent anode-strip plate at the end of the induction gap, as proposed in Ref.~\cite{Brunbauer2018}, using transimpedance amplifiers. The anode strips are made from a low-resistivity film (4~$\Omega$/square) of indium tin oxide (ITO) deposited on a 1.1-mm thick glass plate, with the ITO patterned into 120~strips on a pitch of 833~$\mu$m. This enables not only some lateral resolution, which helps reduce ambiguity and thus simplifies event reconstruction, but also reduces the capacitance seen by each amplifier, thus lowering the equivalent input charge noise. The camera readout drives the transparency requirement for this plate. The ends of the ITO strips are covered with an Al-on-Cr coating to enable wire-bonding to a Kapton flexible PCB/cable for connection to the vacuum feedthrough.

It is advantageous to orient the ITO strips perpendicularly to the beam axis for D-D neutron-induced recoil energies above $\sim$230~keV and D-T neutron-induced recoil energies above $\sim$750~keV, so this orientation was selected. Figure~\ref{Fig:Signal} does not suggest that most Migdal events will involve nuclear recoils above these energies, but that figure refers only to contained tracks, which tend to have lower energies. Eventually, we aim to analyse all tracks with origin within the fiducial region.

\begin{figure}[ht]
\centerline{\includegraphics[width=0.85\linewidth]{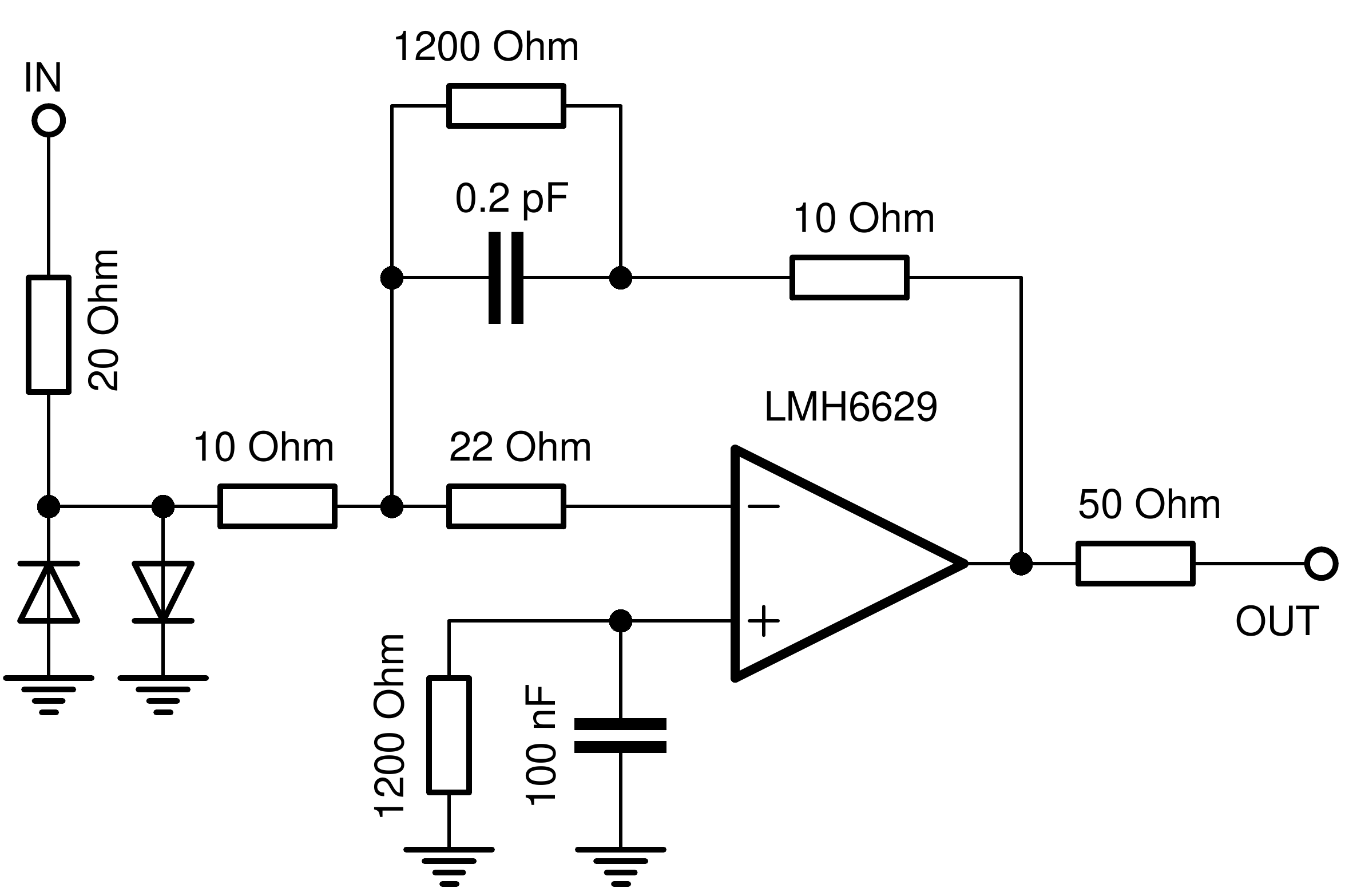}}
\caption{Circuit representing each of the 60 electronic channels. It utilises fast and low-noise commercial amplifiers (LMH6629), configured for high speed and low gain. The inputs are protected using D5V0F4U6SO low-capacitance ESD diodes. Various extra resistors have been added to ensure stability when operated with low gain and a capacitive source impedance.}
\label{Fig:Amplifier}
\end{figure}

Pairs of strips (60~strips apart) are connected to each DAQ readout channel to reduce channel count. To minimise capacitance and the risk of interference from external signal sources, the amplifiers will be located on two PCBs connected directly to the external connector of the vacuum feedthrough. These circuits, shown in Fig.~\ref{Fig:Amplifier}, are built from commercial parts, and include arc protection and amplifiers/buffers for driving the (10~m) cables to the DAQ in the control room outside of the neutron bunker.

Charge-sensitive preamplifiers (Ortec~142IH) equipped with spark protection circuits will be used to read out the charge signals induced on the GEMs. Those induced on the second GEM would allow us to measure the overall gas gain of the system for 5.9~keV X-rays, which is important for monitoring the detector gain stability during the experiment. Despite the large dynamic range of the preamplifier (100~MeV energy deposition in Si), it is expected that the preamplifier attached to the second GEM will saturate for energetic NRs ($\gtrsim$100~keV) and, therefore, the output of the preamplifier attached to the first GEM would be used for the energy measurement from $\sim$100~keV to 4~MeV.

Charge induced on the cathode is due to the motion of all charge carriers -- i.e.~electrons and positive ions -- in the drift region. A cathode pulse is composed of a fast electron component from electron drift towards the first GEM, and a slow ion component due to the drift of positive ions toward the cathode. The amplitude of the fast electron component carries very useful information on the depth of interaction along the $z$-axis and on the orientation of the NR track, as demonstrated in Ref.~\cite{Couturier2017}; eventually, we aim to make use of this information.

\subsection{Data acquisition}
\label{SS:DataAcquisition}

\begin{figure}[htp]
    \centering
    \includegraphics[width=\linewidth]{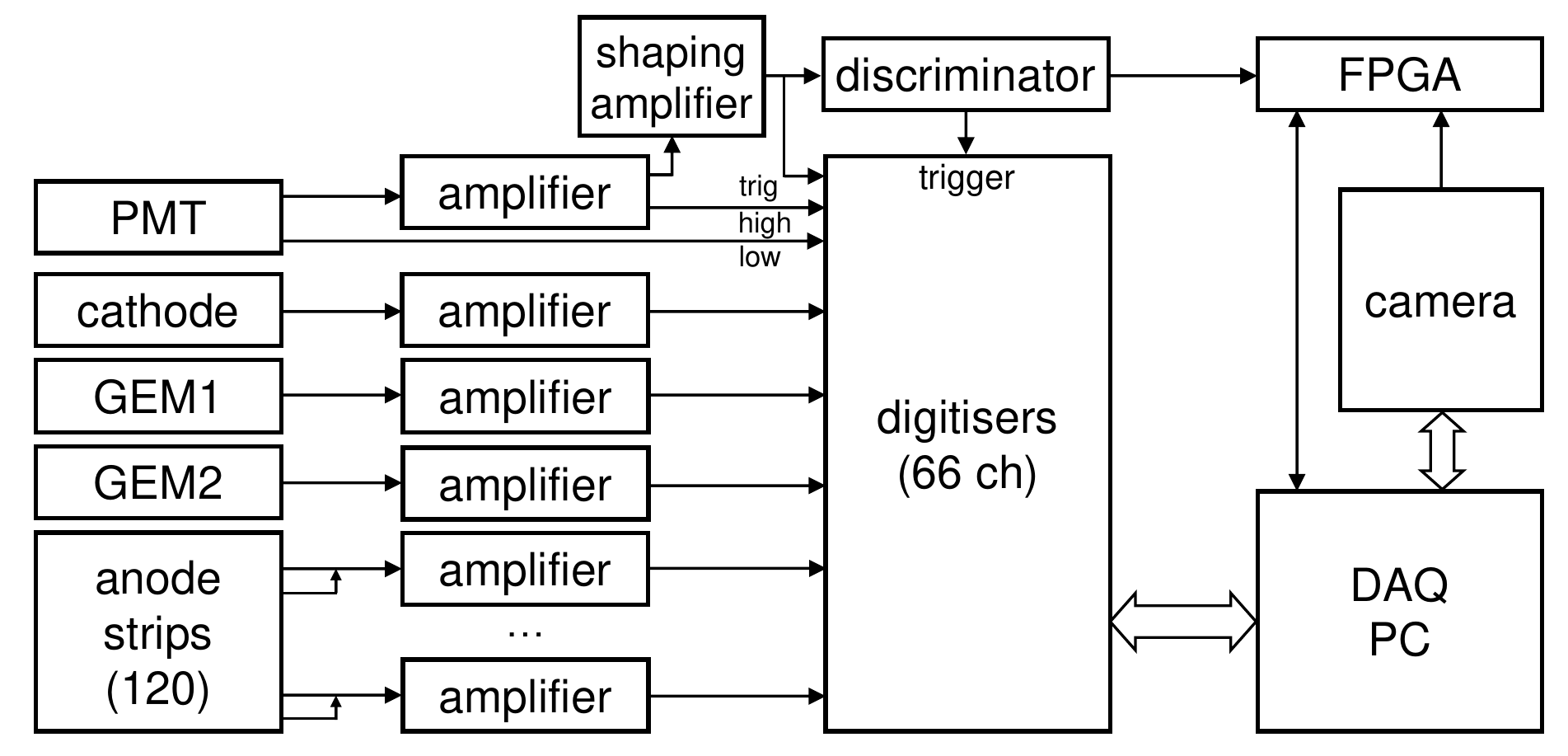}
    \caption{Box diagram highlighting the three main data acquisition elements: digitisers, camera and FPGA -- controlled and read out by a common PC.}
    \label{Fig:DAQ-diagram}
\end{figure}

At the heart of the data acquisition (DAQ) system is an Acqiris CC121 crate with 17~DC265 modules, for a total of 66~digitiser channels with 8-bit vertical resolution and maximum sampling frequency of 500~MS/s. This setup is used to record waveforms from the PMT anode (in dual-range mode), the higher-potential GEM electrodes, the cathode, and the 60~anode strip pairs. The trigger signal is derived from the PMT and fed to the external trigger input of the digitisers to start the acquisition of a new event, and is also recorded in the data stream as a separate channel. The camera images are stored using a separate data stream. Figure~\ref{Fig:DAQ-diagram} shows the DAQ block diagram.

A purpose-developed, Linux-based DAQ software (MiDAQ) controls the overall data acquisition workflow and is responsible for the interface with the Acqiris crate and the Hamamatsu camera. It allows configuration of the sampling rate, waveform and pre-trigger duration, and the full-scale of each channel, as well as to set the camera's exposure time, image binning and scan mode.

All signals of interest lie in a short time window of $\lesssim$0.5~$\mu$s, to which we add at least 0.5~$\mu$s of pre-trigger waveform for baseline parameter estimation. We plan to digitise at 500~MS/s (2~ns sampling) but may reduce this to 250~MS/s in some datasets to minimise deadtime and data volume. The camera is operated independently in free-running mode, and images can be directly synchronised with the recorded waveforms during offline analysis via microsecond-precision timestamps which are generated by the software for each DAQ event and each image, and recorded in both data streams.

\begin{figure}[t]
    \centering
    \includegraphics[width=\linewidth]{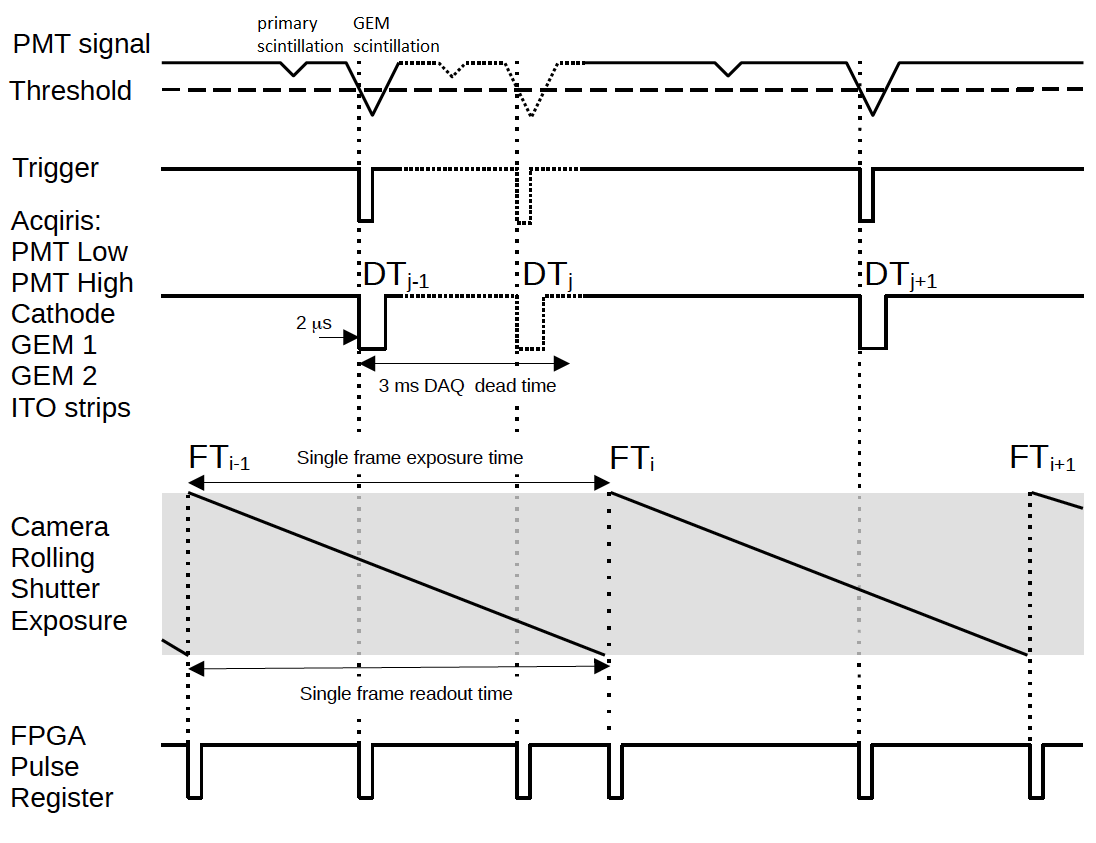}
    \caption{Timing diagram of the DAQ output electronic signals. A secondary scintillation pulse from the GEMs triggers the Acqiris digitisers to record waveforms of the induced charge pulses. The camera runs in a free continuous mode with a rolling shutter and a minimum exposure time of 11.2~ms. All the triggers are recorded by the FPGA pulse register.}
    \label{Fig:DAQ-timing}
\end{figure}

The timing diagram in Fig.~\ref{Fig:DAQ-timing} shows example timelines for three events recorded by the three main DAQ elements. When an interaction takes place in the active volume of the detector it produces a primary scintillation signal (S1) which is typically below our trigger threshold. The ensuing large secondary scintillation (S2) pulse produced by the GEMs triggers the Acqiris digitiser, which records also pulses induced on the ITO strips as well as on the cathode and the two GEMs.

Although the length of the recorded waveforms is only a few $\mu$s, the deadtime caused by the data-transfer bottleneck is around 3~ms. Due to this, the following event (depicted by a dotted line), will not be recorded by the digitiser, and signals from the PMT and the electrodes will be missing. However, it will still be captured by the camera, and hence it delivers only partial information about this event -- causing a potential background from random coincidences. To record this kind of occurrence and avoid event confusion in the image analysis an FPGA pulse counter with 1~ns timing resolution was developed; this will help MiDAQ record the timing of all triggers (DT) as well as the camera’s output pulses indicating the start time (FT) of each frame. The timing information of the recorded waveforms (DT$_j$) and individual exposures (FT$_i$) will be embedded into the digitiser’s data stream and in the frame’s metadata, respectively, and used later to synchronise between the j-th waveform and i-th frame. A future development will include the full digitisation of all PMT waveforms by the FPGA.

Ancillary sensors and other equipment required for operation of the experiment are queried each minute and their information is recorded by MiDAQ with each event, allowing its use during offline analysis. This includes the pressure and composition of the gas in the chamber, the ambient temperature, the bias voltages on the TPC electrodes and the position of the $^{55}$Fe calibration source.

Data from the Acqiris digitisers will be stored in binary format. Each event includes the 66~Acqiris waveforms, the unique trigger timestamp plus the ancillary sensors -- taking up around 12~kB per event after compression. In addition, each binary file includes a global header with the relevant information on the channel and trigger settings used in the run. The images from the CMOS camera are stored in MTIFF files, with each 16-bit TIFF image in $2\times2$ binning mode, occupying 1.2~MB after compression. The image timestamp and the FPGA trigger timing information are recorded as metadata. Significant local storage ensures continuous operation while saved data are transferred to a computing cluster for offline analysis.

A new software framework -- MiDAS -- was developed for data reduction and extraction of the information of interest from each triggered event. This will be carried out offline using dedicated C++ algorithms. Besides pulse identification and parameterisation functions, MiDAS provides also event display.

\subsection{Calibration}
\label{SS:Calibration}

The planned calibration measurements will deliver the necessary information to optimise the operation of the detector for each gas composition and pressure, and the 3D reconstruction of the ionisation track images. The various OTPC electrodes will be biased to provide high gas gain, good energy resolution, large dynamic range, minimum electron diffusion, and stable operation free from destructive sparks. Optimisation of these parameters will be carried out using low energy X-rays from $^{55}$Fe (5.9~keV) and $^{241}$Am (13.8~keV), $\alpha$-particles from $^{241}$Am, and highly-ionising fission fragments from $^{252}$Cf.

Two dedicated X-ray calibration windows are provided on one of the flanges, sealed with 50~$\mathrm{\mu}$m Al foil; these have around 5~mm wide entrance aperture and a pyramidal shape that allows exposure of the drift and transfer regions separately. The $^{55}$Fe source is permanently attached to the chamber on a movable shaft for remote deployment on demand. The 5.9~keV $^{55}$Fe X-rays are a key calibration tool, producing photoelectrons just above our 5~keV ER threshold uniformly across the active volume; sub-keV Auger electrons should be visible in some cases too. A 80~MBq source will be used for this purpose. Each X-ray interaction creates an average of 172~electron-ion pairs for a W-value of 34.2~eV for CF$_4$~\cite{Reinking1986}. From the number of primary pairs the gas gain of the GEM system will be determined from a charge-to-voltage calibration of the charge-sensitive preamplifier connected to the second GEM (cf.~Fig.~\ref{Fig:TPC2}). A second calibration point is provided by the 13.8~keV X-rays from a 7.7~MBq $^{241}$Am source.

The response of the camera to low-energy electrons (range, energy and uniformity) will be calibrated with the same X-ray sources. After subtracting the mean noise using dark frames, the electron energy is extracted by integrating the pixel values after applying a noise smoothing image filter. A previous energy measurement of $^{55}$Fe interactions with our CMOS camera yielded a good energy resolution of 28\% (FWHM). The electron range will also be studied carefully at these energies. These datasets will be essential for flat-fielding the camera image and the PMT response.

The transverse diffusion of electrons during their drift towards the anode is another critical parameter affecting the amount of detail that can be extracted from the particle tracks, i.e.~the spatial resolution in the camera images. This will be quantified by measuring the width of $\alpha$-particle tracks fired at known locations parallel to the anode plane. The $^{241}$Am $\alpha$-source can be deployed inside the chamber, behind a 1-cm long PEEK collimator with 1-mm aperture, which is attached to the cathode frame and pointing from the side of the active region. These tracks will also help develop the 3D reconstruction algorithms.

A $^{252}$Cf fission source will be used to evaluate the detector response to highly-ionising particles that can potentially limit the dynamic range of the detector by causing destructive sparks. Fission fragments can have ionisation densities much larger than those of C and F recoils from D-T neutrons, and hence the dynamic range of the detector can be probed and safe operating voltages determined before the actual neutron beam experiment. A 37~Bq $^{252}$Cf fission source has been acquired for this purpose. The average energies of the fission fragments are 103~MeV and 78~MeV for the average light and heavy fragments, respectively, leading to primary ionisation densities approximately ten times larger than the maximum expected from the C and F recoils.The source will be mounted in two locations inside the chamber: behind the cathode mesh and on the side of the TPC, with the particle beam oriented perpendicularly to the ITO strips. 

The uniformity and cross-talk of the anode-strip response will be measured using a precision pulse generator with a fast rise time (2~ns) using test inputs in the preamplifiers. A charge-injection probe was used prior to assembly to study the ionisation response at various points along a strip, to inform the electronics response model.

Blue LED pulses will be used periodically to monitor the single photoelectron response of the PMT and the timing synchronisation between the PMT and the camera. One LED is fibre-coupled into the space in front of the camera (also visible to the PMT) and another is installed in the PMT enclosure.

Calibration of the camera response to NR interactions is needed to evaluate their energy and hence enable the nuclear scattering kinematics: this requires first the discrimination between C and F recoils, which we aim to achieve by comparing their stopping power against total energy. We intend to measure the quenching factor (QF) of CF$_4$ and other gases in the $\sim$0.1--1~MeV energy range. In addition to supporting the Migdal analysis, these measurements are of interest to the dark matter community as such data are scarce. The well-defined neutron direction in concert with a measurement of the recoil angle -- within some angular resolution -- can potentially provide a precise estimate of the recoil energy -- cf.~\ref{A:NeutronScattering}. The recoiling species and recoil angle can be determined from the 3D reconstructed tracks. The estimated recoil kinetic energy can then be compared to the energy deposited in the drift region by the energetic ions to provide the electron-equivalent deposited energy and hence a QF measurement. We plan to conduct such measurements also at lower GEM gain (to avoid saturation) and at several drift electric fields to check for charge recombination effects.

\begin{figure}[t]
\centerline{\includegraphics[width=0.95\linewidth]{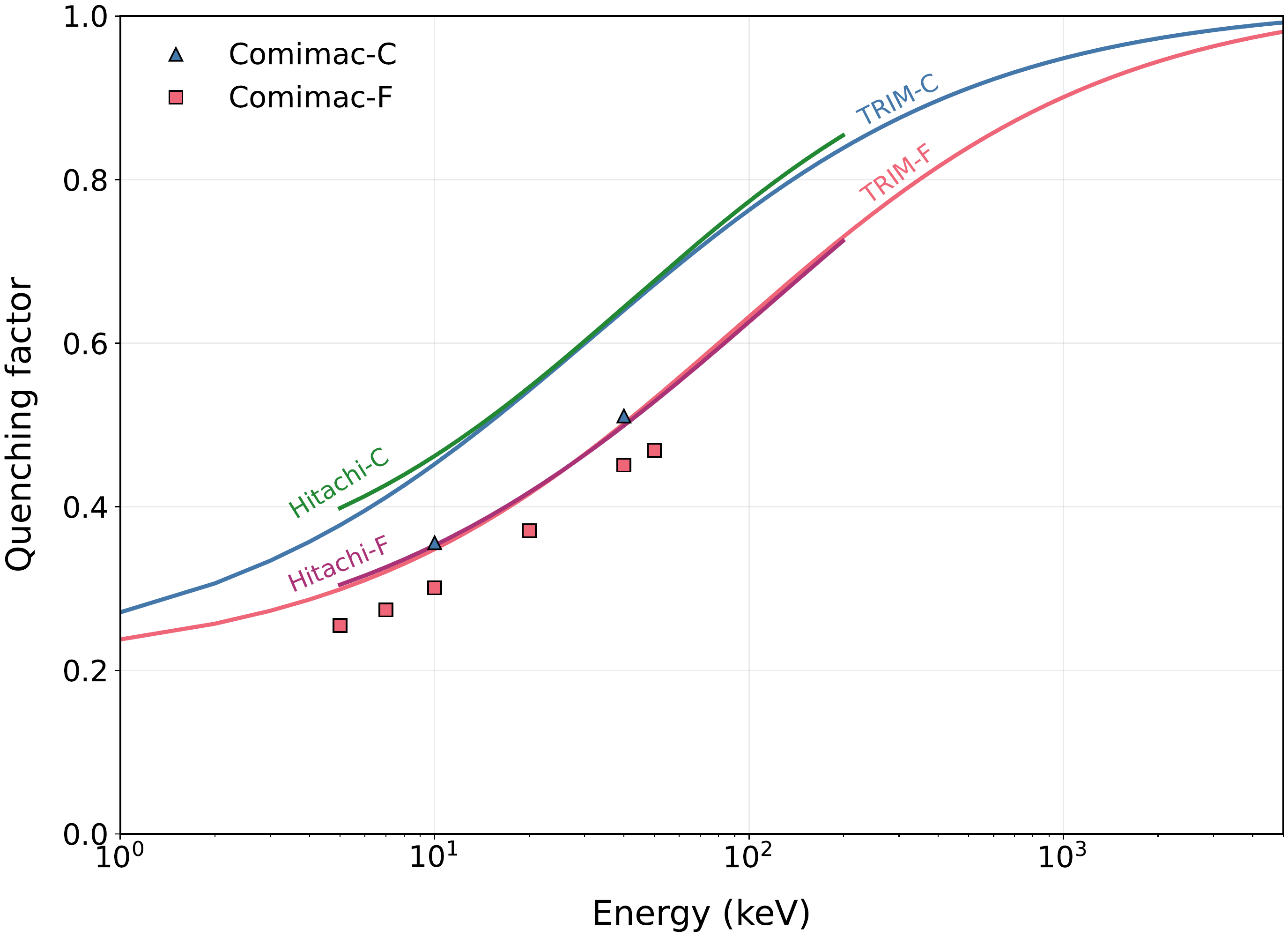}}
\caption{Quenching factor of C and F ions in CF$_4$ estimated using TRIM and the methods found in Ref.~\protect{\cite{Hitachi2008}} (labelled `TRIM-C/F' and `Hitachi-C/F'). Experimental QF values reported in Ref.~\protect{\cite{Guillaudin2012}} for C and F ions in CF$_4$ are also shown (`Comimac-C/F').}
\label{Fig:qf_values}
\end{figure}

Prior to such QF measurements, we will use the quenching factors estimated from TRIM to reconstruct NR energies in the Migdal search. This is a reasonably well established technique where such measurements are unavailable, with TRIM providing reasonable values -- particularly for ion energies in the tens of keV. Other QF estimates for CF$_4$ can be found in \cite{Hitachi2008}, which combines methods from Ref.~\cite{Lindhard1963} and energy loss calculations from SRIM to calculate the quenching factor in molecular gases. Ref.~\cite{Guillaudin2012} also includes a QF measurement for C and F ions in 50~Torr CF$_4$ gas. Figure~\ref{Fig:qf_values} depicts all of these data.


\section{Neutron beam}
\label{S:NeutronSource}

\subsection{Facility and generators}
\label{SS:Facility}

\begin{figure}[t]
\centerline{\includegraphics[width=\linewidth]{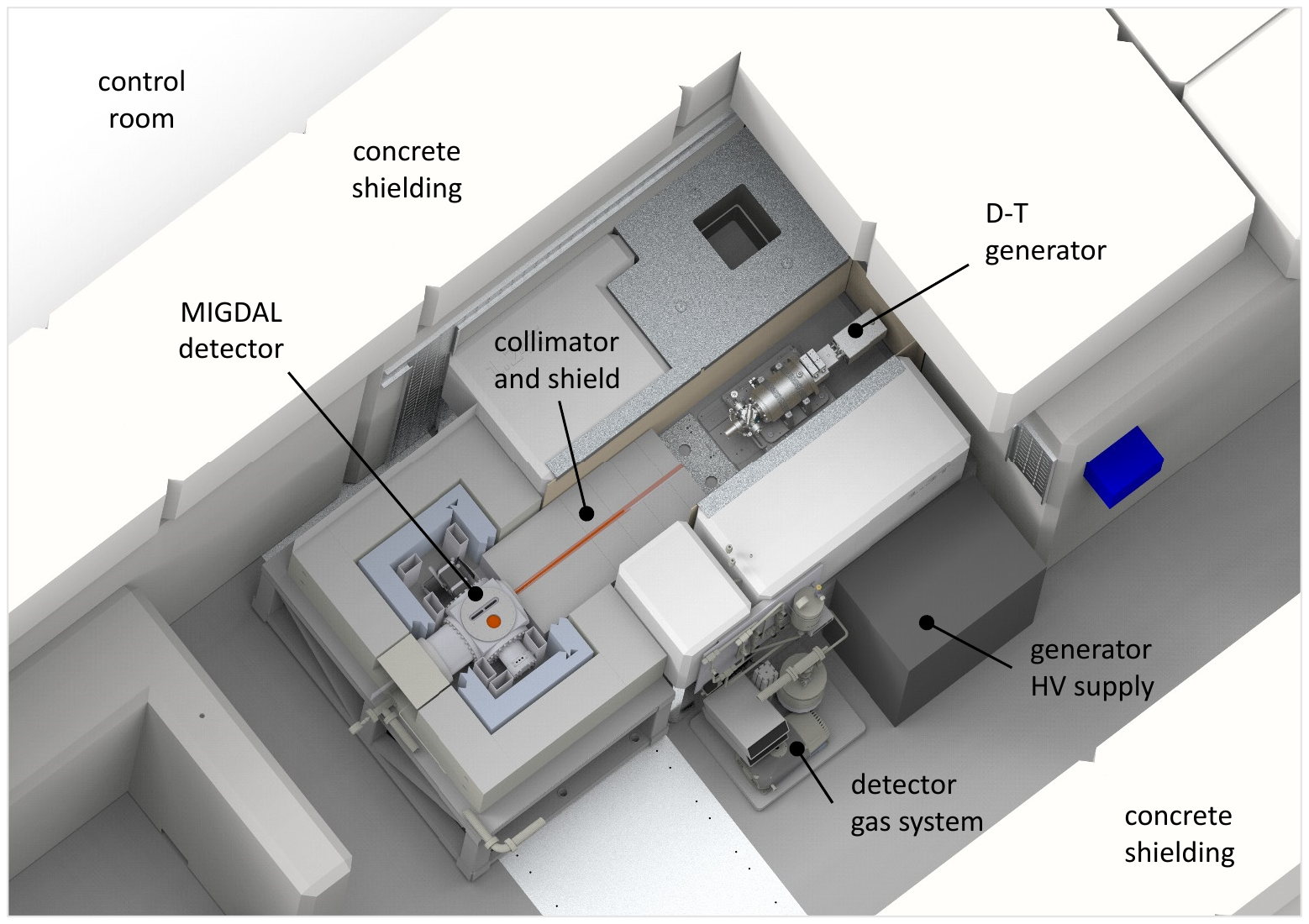}}
\caption{Rendered view of the experimental set up at the NILE facility with the MIGDAL detector deployed in front of the D-T generator within the neutron bunker (some of the upper shielding removed for the purpose of display). The long copper-made D-T collimator can be seen encased in the multi-layer shield. The control room is located outside of the concrete wall.}
\label{Fig:Bunker}
\end{figure}

The MIGDAL experiment will be hosted at the ISIS Neutron and Muon Source facility of the Rutherford Appleton Laboratory (UK). The choice of the specific neutron sources is driven by two factors. Firstly, the suppressed cross section of the Migdal effect requires the use of intense neutron beams. Secondly, monoenergetic sources are preferred for an initial measurement in order to control systematic uncertainties from the neutron energy spectrum. For these reasons MIGDAL will use commercial deuterium-deuterium (D-D) and deuterium-tritium (D-T) fusion generators at the Neutron Irradiation Laboratory for Electronics (NILE) at ISIS, as depicted in Fig.~\ref{Fig:Bunker}.

Initially, the experiment will use a D-D generator from Adelphi Technology Inc. Nominally, this source produces monoenergetic neutrons at around 2.45~MeV, emitted iso\-tropically at a rate of 10$^9$~n/s. However, it should be  noted that 2.45~MeV is the kinetic energy of neutrons emitted in the fusion reaction for the hypothetical case when the reactants are at rest: the precise neutron energy is given by the kinematics of the reactants and, for the real case of a D beam accelerated onto a D target, it is a function of the emission angle and of the energy of the incoming ion -- i.e., of the applied high voltage~\cite{IAEA2012}. At 100~kV operating voltage, appropriate for the Adelphi device, we expect the neutron energy to vary from around 2.15~MeV at 180$^\circ$ (backward direction) to 2.8~MeV at 0$^\circ$ (forward direction). The energy is $\approx$2.47~MeV at 90$^\circ$, where the MIGDAL experiment is placed. The flux per unit solid angle has also a marked dependence on the emission angle~\cite{Aylon2018}. Considering this angular distribution together with the slowing down of the deuterium ions in the target, the result is a broadening of the neutron energy spectrum with respect to the ideal (monoenergetic) case, as described in Ref.~\cite{Waltz2016}. In that reference the broadening is shown to reach a minimum of around 2\% (FWHM) at 90$^\circ$ and a maximum at 0$^\circ$. While the flux per unit angle is lowest in the perpendicular direction (by a factor of $\sim$2) due to the geometry of this generator, the experiment can be placed closer in this orientation, partially compensating for the lower flux. These reasons motivate the decision to place the MIGDAL experiment at 90$^\circ$ for the D-D generator setup.

In a second stage of the experiment we will employ a D-T generator (also from Adelphi) that emits $\sim$14~MeV neutrons with a yield of 10$^{10}$~n/s. This device also uses a microwave plasma ion source and a high voltage to induce the fusion reaction on a target, and can produce neutrons either continuously or in pulses with a minimum length of 100~$\mu$s. The D-T generator energy and flux also have an angular dependence~\cite{IAEA2012,Kasesaz2016}, although the flux anisotropy is much less severe with respect to the D-D case: there is only a 3\% difference in angular flux and the source can be well approximated as isotropic. For 100~kV operation, the neutron energy varies from around 13.5~MeV at 180$^\circ$ to 14.7~MeV at 0$^\circ$, and is $\approx$14.1~MeV in the perpendicular direction. Considering the geometry of the D-T generator, it was decided that the experiment would be placed at 0$^\circ$ in the D-T configuration, corresponding to 14.7~MeV neutron energy. This is the deployment shown in Fig.~\ref{Fig:Bunker}. An experimental study of the flux and energy broadening similar to that presented in Ref.~\cite{Rigamonti2018} is under way.

While the D-D generator has a lower yield than the D-T source, the lower neutron energy in the former case implies suppressed levels of secondary radiation from inelastic scattering, which is expected to lead to lower backgrounds in the search for the Migdal effect. In addition, D-D neutrons will allow the Migdal effect to be studied at lower NR energies, approaching the regime of interest for direct dark matter detection. On the other hand, the D-T device will allow the verification of new predictions in a different kinematic regime which indicate that multiple Migdal ionisation becomes significant~\cite{MIGDALth}.

In addition to neutrons from nuclear fusion, the D-D and D-T sources will also emit an intense flux of secondary radiation due to inelastic interactions and radiative capture in the generator material, plus an X-ray component caused by bremsstrahlung of electrons from the generator plasma. Due to the small mean free path of charged particles in this energy regime, this secondary radiation will consist mostly of $\gamma$-rays. The photon flux will be measured \textit{in-situ} after commissioning of the devices. The background from the secondary radiation produced by the generators is assessed in Section~\ref{SS:Backgrounds}.

The NILE facility, shown in Fig.~\ref{Fig:Bunker}, consists of a concrete bunker situated at ISIS and will accommodate both generators; it features interlocked access. An external control room hosts the DAQ and control electronics for the experiment. The thickness of the bunker walls and the distribution of additional concrete blocks around the D-T generator has been determined by radio-protection considerations. Besides this shielding intended for biological safety, experiments at NILE such as MIGDAL must include additional shielding and collimation in order to ensure that the background from secondary radiation is suppressed to acceptable levels. The specific shielding developed for MIGDAL is discussed below.

In a future phase, the experiment may be deployed at the ChipIr instrument at the Second Target Station of ISIS. The ChipIr instrument~\cite{Cazzaniga2018} provides an intense and collimated beam of neutrons with spectrum similar to that produced in the atmosphere by cosmic radiation (5.6$\times$10$^{6}$~n/cm$^2$/s above 10~MeV)~\cite{Chiesa2018}. One of the main applications of this facility is to test the effect of fast neutrons on electronic devices. The ChipIr neutrons are produced by spallation of 800~MeV protons of the ISIS accelerator onto a tantalum-coated tungsten target. The ChipIR spectrum is continuous and extends to much higher energies, and hence it offers the possibility of testing the Migdal effect in another kinematic regime; this setup would bring a new class of backgrounds that have not been studied.

\subsection{Shield and collimator design}
\label{SS:Shield}

The experiment requires a well-defined beam of primary neutrons that passes through the active volume while avoiding any other primary or secondary radiation that would otherwise interact in the OTPC. For this purpose the detector chamber is surrounded by dedicated shielding, with a frontal penetration to accept primary neutrons (termed the `collimator' in this paper), plus another opening downstream to allow the beam to exit the chamber with minimal interactions.

As mentioned in Section~\ref{SS:OTPC}, an air-filled collimator design has been adopted, followed by a 150-$\mu$m thick Al entrance window: Monte Carlo simulations indicate that the secondary radiation produced by neutron interactions in the air outside of the chamber or in the window are not a dominant source of background. Most secondary radiation detected in the OTPC is typically produced in two steps. Inelastic neutron interactions produce $\gamma$-rays in the surrounding materials (mainly shielding and the collimator itself). These photons then interact in the collimator and detector elements (mainly cathode and first GEM) to produce electrons that easily reach the active volume. Other contributions to the total non-NR event rate are subdominant. In particular, events where a photon interacts directly in the active volume are suppressed owing to the low gas pressure.

The shielding and collimator designs are driven by two needs. Firstly, NR vertices in the active volume should lie between 6~mm and 24~mm from the first GEM, in order to $i$) ensure that the Migdal event is largely contained in the active region, and $ii$) minimise diffusion of the drifting electrons. This requirement constrains the beam width in the direction perpendicular to the GEM surface, and implies the suppression of the beam halo in order to avoid neutron interactions in the GEMs and the cathode. Secondly, the total interaction rate in the active volume must approach the camera frame rate ($\sim$90~Hz) to maximise the number of single-interaction frames. Based on these requirements, the design of the shielding and collimator proceeded by minimising the fraction $f_{other}$ of non-NR events in the total interaction rate. At the end of this optimisation procedure it was verified that the resulting designs do not contribute significantly to the background for the Migdal search (cf.~Section~\ref{SS:Backgrounds}).

This optimisation procedure was organised in three stages. First, only the front part of the shielding design was refined, in the absence of any other material surrounding the experiment, and assuming a simple collimator design by default. Second, the collimator design was iterated using the front shielding configuration resulting from the previous step. Finally, the lateral and posterior parts of the shielding were included and optimised, accounting for the effect of the bunker and the concrete material within.

The above procedure relies extensively on Monte Carlo simulations using two frameworks. Initially, the BDSim software~\cite{Nevay2020} was used for an initial assessment of the thickness of each shielding layer, based on the predicted radiation fluxes entering the detector chamber. BDSim is built on top of GEANT4, customised to study radiation fluxes caused by the interaction of particle beams with surrounding materials. A dedicated GEANT4 simulation of the experiment and its environment was then developed to complete the remaining stages of the optimisation.

The layout of the front part of the shielding depends on the neutron source considered. For the D-D generator, it consists of a layer of borated high-density polyethylene (BHDPE) to efficiently moderate and subsequently capture neutrons, followed by a layer of lead to attenuate $\gamma$-rays from inelastic scattering and radiative capture. The optimal thickness of each layer was found to be 30~cm and 10~cm, respectively. The shielding for the D-T generator (Fig.~\ref{Fig:Bunker}) will also use BHDPE and lead, but is preceded by a layer of iron to efficiently moderate primary neutrons down to $\sim$5~MeV via inelastic scattering. In this case, the optimal thickness of the iron, BHDPE and lead layers was found to be 70~cm, 20~cm and 10~cm, respectively.

The collimator consists of a penetration (hereafter the beam `tunnel') in the front shield, surrounded by walls which fulfill the following purposes: they moderate primary neutrons that lie outside the acceptance of the active volume whilst avoiding part of the shielding as they propagate through the beam tunnel; and they attenuate $\gamma$-rays from the shielding that might enter the beam tunnel and reach the detector chamber. The entire collimator setup will be embedded in a rectangular passage through the shielding with 3$\times$9~cm$^{2}$ cross section.

The D-D collimator walls consist of an inner layer of BHDPE for neutron moderation, surrounded by a thin layer of lead for $\gamma$-ray attenuation. This arrangement is convenient because primary neutrons that reach the collimator lead through the beam tunnel can be moderated by the innermost BHDPE layer, and therefore the probability of producing inelastic scattering in that metal is suppressed. In turn, the D-T collimator walls are made from a single layer of copper only, which provides both efficient fast-neutron moderation via inelastic scattering and $\gamma$-ray attenuation (while iron could also be considered for this purpose, copper was preferred for engineering reasons). Note that since neutrons enter the collimator walls at grazing incidence, even thin BHDPE or Cu layers (for the D-D and D-T generators, respectively) are able to provide sufficient moderation.

\begin{figure}[ht]
\centerline{\includegraphics[width=0.9\linewidth]{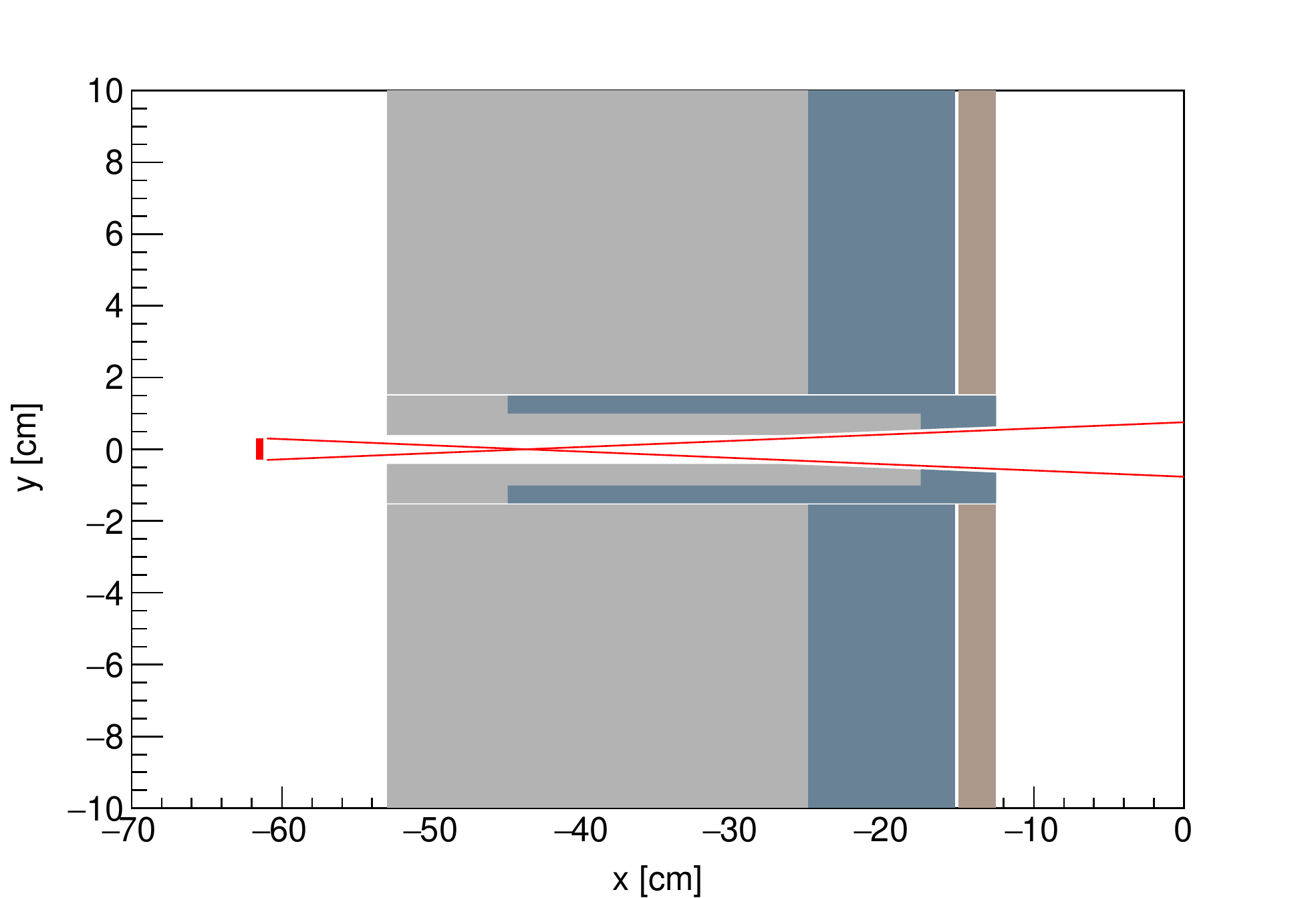}}
\caption{Longitudinal section of the front shield and collimator designed for the D-D source, illustrating the double-trapezoid collimator concept. The materials are borated polyethylene (light grey) and lead (blue grey). 
The approximate size of the D-D generator target is also shown, along with the lines that define the envelope of the beam passing through the collimator (red).}
\label{Fig:DoubleTrapezoid}
\end{figure}

The longitudinal section of the beam tunnel in the camera ($x,y$) plane is shown in Fig.~\ref{Fig:DoubleTrapezoid}. In this plane, the first part of the tunnel has constant width, changing to a linear taper after 35~cm. This design, hereafter referred to as the double-trapezoid configuration, has been devised to account for the fact that primary neutrons are emitted by a small but extended source. For the D-D generator, Monte Carlo simulations show that the double-trapezoid configuration improves $f_{other}$ by $\sim$25\% compared to the simpler fully-tapered design which is optimal for a point source. Similar results were obtained for the D-T generator.

The width of the beam tunnel entrance is determined by the size of the region where neutrons are emitted inside the generator, while the exit aperture is set by the required size of the beam in the active volume. The remaining dimensions of the tunnel (the intersection between the two parts of the tunnel, and the slope of the divergent walls) are adjusted in order to ensure that the walls of the last part of the collimator lie outside the line-of-sight of the primary neutrons (see Fig.~\ref{Fig:DoubleTrapezoid}), thereby suppressing the production of secondary radiation from inelastic scattering near the active volume.

The section of the beam entering the active volume has diffuse edges, and its size is given as the full width at half maximum of the neutron distribution in the transverse plane. For the D-D and D-T generators, the size of the beam section is predicted to be 1.4$\times$9.0~cm$^2$ and 1.3$\times$9.2~cm$^2$, respectively. The size of the beam halo is defined, for each axis, as the width of the region that contains 99\% of the total neutron flux entering the active volume. The corresponding predictions are 1.8$\times$9.0~cm$^2$ and 1.5$\times$9.2~cm$^2$ for the D-D and D-T generators, respectively. While neutrons entering the active volume are mostly monoenergetic, the spectrum features a low-energy tail due to elastic and inelastic interactions in the front shielding and the collimator. This population of degraded energy represents only $\sim$3\% and $\sim$1\% of the total neutron flux for the D-D and D-T generators. We have also confirmed that the fraction of `backward neutrons' entering the OTPC after scattering in the shielding is small: only $\sim$1\% of the flux above 100~keV neutron energy with the D-D generator, and lower still for D-T neutrons.

Despite this being a leading consideration for our material selection, neutron capture in the front shield and collimator will activate materials and produce $\gamma$-rays that may enter the active volume. For the main elements present in these structures (H, C, Fe, Cu and Pb), neutron capture produces either stable nuclides or radioisotopes where $\gamma$-ray emission either does not occur (e.g.~$^{209}$Pb) or is not important (e.g.~$^{54}$Fe and $^{63}$Cu). For $^{65}$Cu (30.8\% of natural abundance), the probability of $\gamma$-ray emission from radiative capture is non-negligible, but still modest (9.2\%).

Gamma-rays produced by inelastic neutron interaction in the generator head will also reach the active volume through the beam tunnel. This is the main contribution to the total $\gamma$-ray flux entering the active volume for the D-T source, while this flux is subdominant for the D-D source due to the lower energy and different configuration of the generator head. The photon rate entering the active volume for the D-T generator is predicted to be approximately 600~Hz, corresponding to $\sim$70\% of the total $\gamma$-ray flux. This rate is $\sim$0.1\% of that of primary D-T neutrons entering the detector. The low-energy component of this photon flux (below 200~keV) is expected to be the dominant source of background for the D-T experiment, as detailed in Section~\ref{SS:Backgrounds}. A thin (2~mm) layer of lead placed between the generator head and the collimator is predicted to attenuate these low-energy photons by 58\% while allowing 82\% of primary neutrons to pass through without scattering. A combination of 1.3~mm of lead followed by 1~mm of tin to absorb the K-shell X-rays from the first layer is able to increase the photon attenuation to 69\% while keeping the same fraction of unscattered primary neutrons.

\section{Track simulations}
\label{S:TrackSimulations}


A dedicated `end-to-end' simulation of the detector has been created to study tracks produced in the OTPC. This comprises three main parts, which model the journey of ionisation electrons from production in the active region to collection by the ITO strips. These simulation elements cover: the active volume, where primary tracks are created and the ionisation drifts towards the entrance to the first GEM; the GEM system, including the two GEMs and the transfer region between them; and the induction gap, where a current is induced on the ITO strips. These three components are detailed below.

\subsection{Primary tracks}
\label{SS:Topologies}

The first stage of the simulation is the production of nuclear and electron recoil tracks in the active volume with appropriate ionisation. Two pieces of software are used to generate the tracks: Degrad~\cite{Degrad}, a Fortran program which produces sites of ionisation for electron tracks with energies above the detection threshold; and TRIM~\cite{SRIM}, a Visual Basic program which is used to simulate the path of nuclear recoils through a medium. TRIM not only details the energy loss due to electronic processes but, crucially, it includes information about the generated secondaries in its detailed collision output.\footnote{TRIM is operated in the fast `Ion Distribution and Quick Calculation of Damage' mode. We utilise the EXYZ.TXT output file, which gives details of all collisions of the primary ion even without the `Full Cascade' mode; where the species of the secondary nuclear recoil is ambiguous, the COLLISION.TXT file can be used to determine it. Once the species and energies of the secondary nuclear recoils have been ascertained, other primary tracks of appropriate energy can be rotated into place and used as secondaries. Since each primary track has details of its own secondary collisions, the process can be repeated recursively until all recoils have been accounted for throughout the whole chain. This is more efficient than using the full cascade mode.} By locating the secondary NR sites and ascertaining the target atom species, subsequent secondary tracks can be `stitched' into place at the appropriate angles by considering the kinematics of elastic scattering. Ignoring inelastic scattering in these determinations is expected to produce angular deviations of $<$5$^{\circ}$, which is not resolvable by our detector after diffusion.

TRIM does not provide the ionisation deposits along NR tracks, but the tabulated electronic energy loss can be used to approximate this. The energy lost due to electronic processes along each step in a track is divided by the W-value of the gas to give the mean number of ionisation electrons expected per step. The number actually generated is re-sampled from a Poisson distribution with a Fano factor of 0.2 appropriate for CF$_4$~\cite{Anderson1992} and released uniformly along each (small) step. We confirmed that the QF calculated for these tracks agrees with that returned by TRIM.

\begin{figure*}[htb]
\centering
\includegraphics[width=0.45\linewidth]{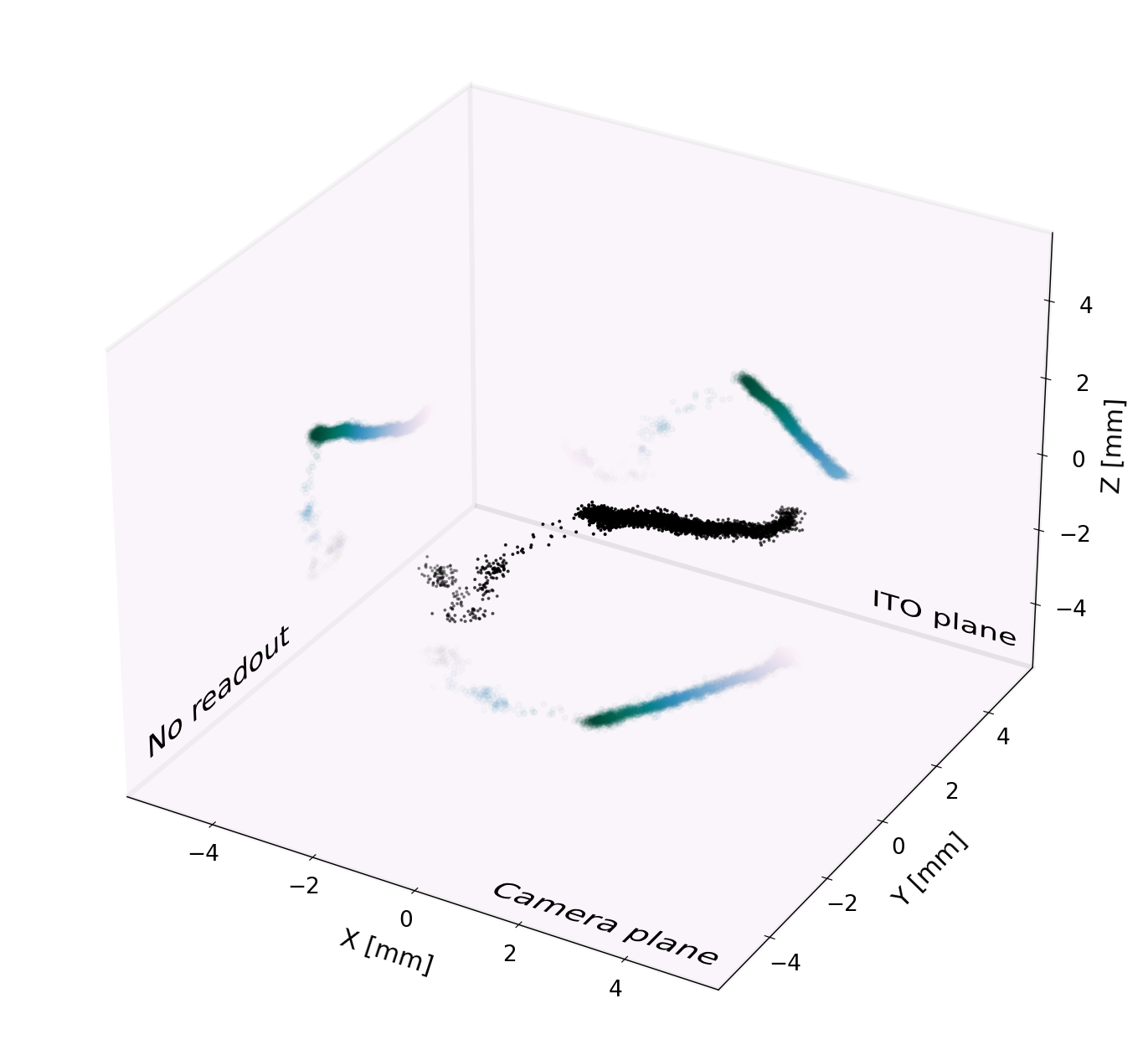}\quad
\includegraphics[width=0.45\linewidth]{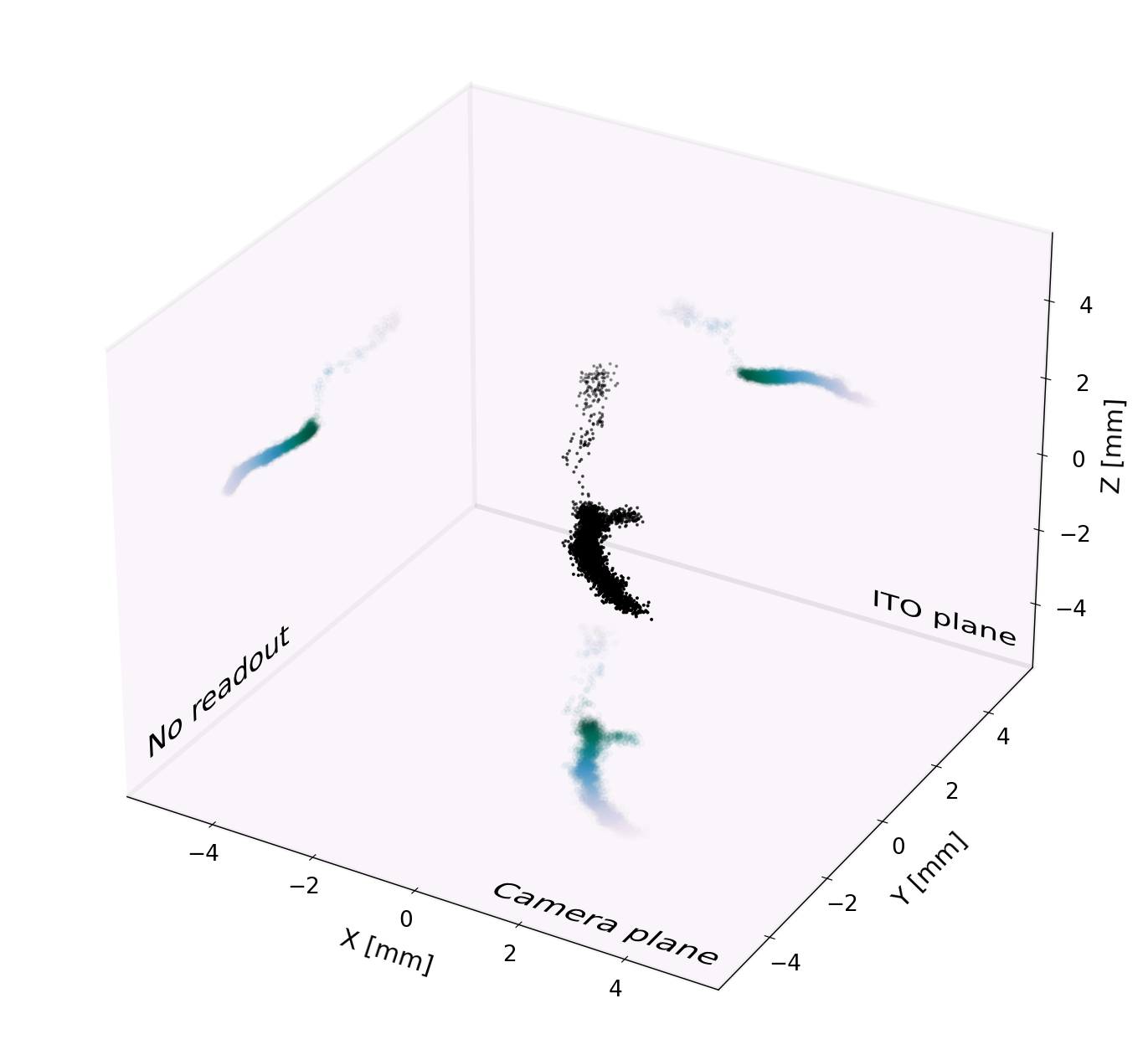}
\caption{Two 3D representations of `Migdal-like' events after 15~mm of drift in the $z$-direction. The $x\!-\!y$ and $x\!-\!z$ projections are labelled according to their mode of acquisition ($y\!-\!z$ is not recorded). The left event contains a 7.5~keV electron, with a 5~keV electron on the right panel. Both are paired with a 150~keV fluorine recoil (denser track). The right image shows a clear fork-like topology towards the beginning of the NR track due to the production of a 14~keV secondary (carbon) recoil (see Section~\protect{\ref{SSS:Secondaries}}). The position of the Migdal vertex is (0,0,0).}
\label{Fig:ExampleMigdalTracks} 
\end{figure*}

In order to produce Migdal-like events, an electron track is overlaid on a nuclear recoil track with a common vertex. The initial NR track direction is obtained either from Monte Carlo or from our analytical calculations (cf.~\ref{A:NeutronScattering}), while the electron is emitted isotropically. Although calculations in helium show that the emission is preferentially in the direction opposite to the nuclear recoil~\cite{Pindzola2014}, we conservatively assume that the Migdal electron is emitted isotropically, which implies some fraction of electrons will be aligned with, and therefore hidden by, the NR track. These `Migdal electrons' are generated with Degrad, specifically from X-ray interactions yielding the appropriate (photo-)electron energy plus the accompanying atomic deexcitation which is also released at the vertex.  Figure~\ref{Fig:ExampleMigdalTracks} shows two examples of `Migdal-like' events. In each event, ionisation electrons are represented by black points in the 3D image and by blue points in the projections onto 2D planes.

\subsection{Electron transport in gas}
\label{SS:Transport}

\begin{figure*}[htb]
\centering
\includegraphics[width=0.43\linewidth]{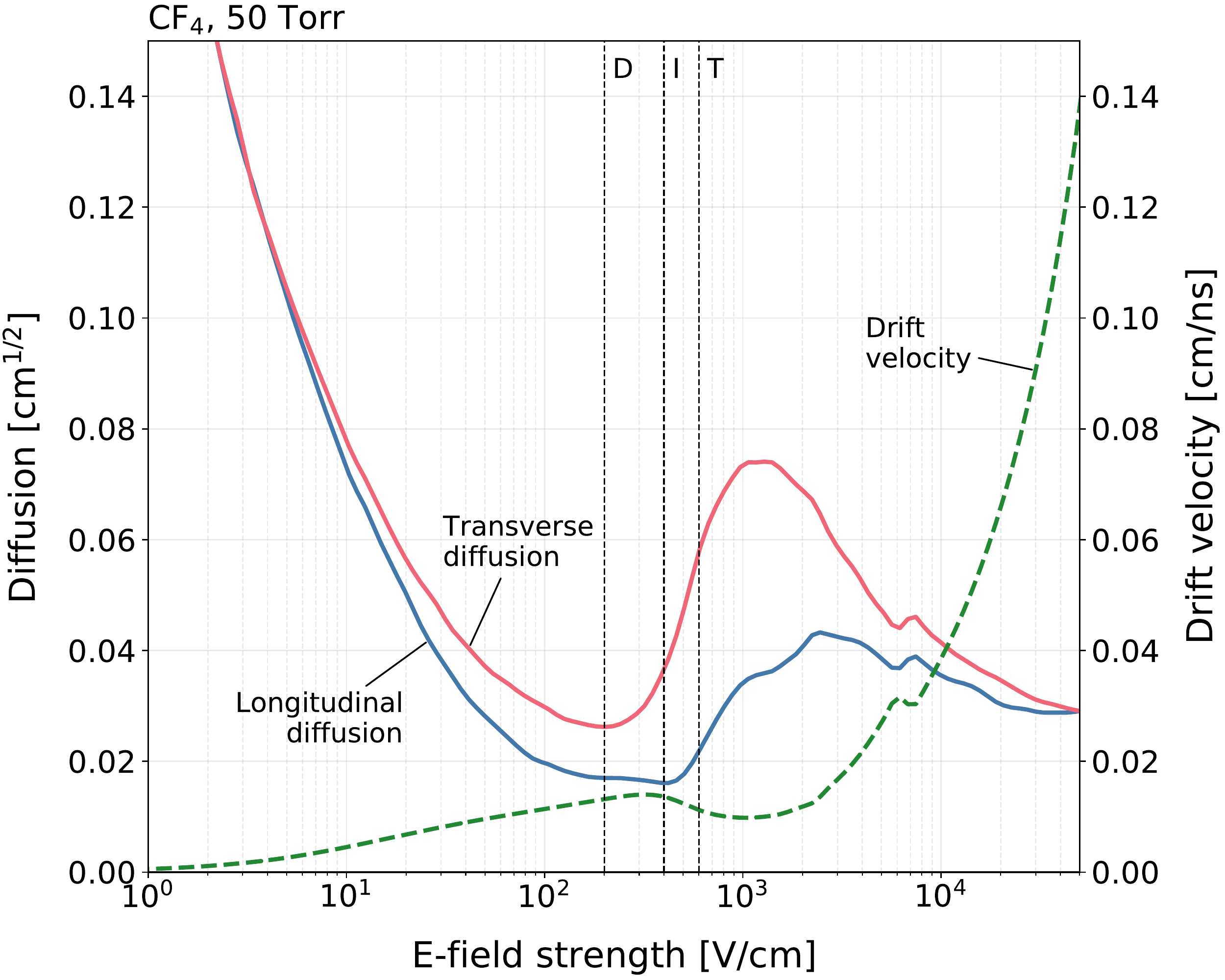}\qquad
\includegraphics[width=0.43\linewidth]{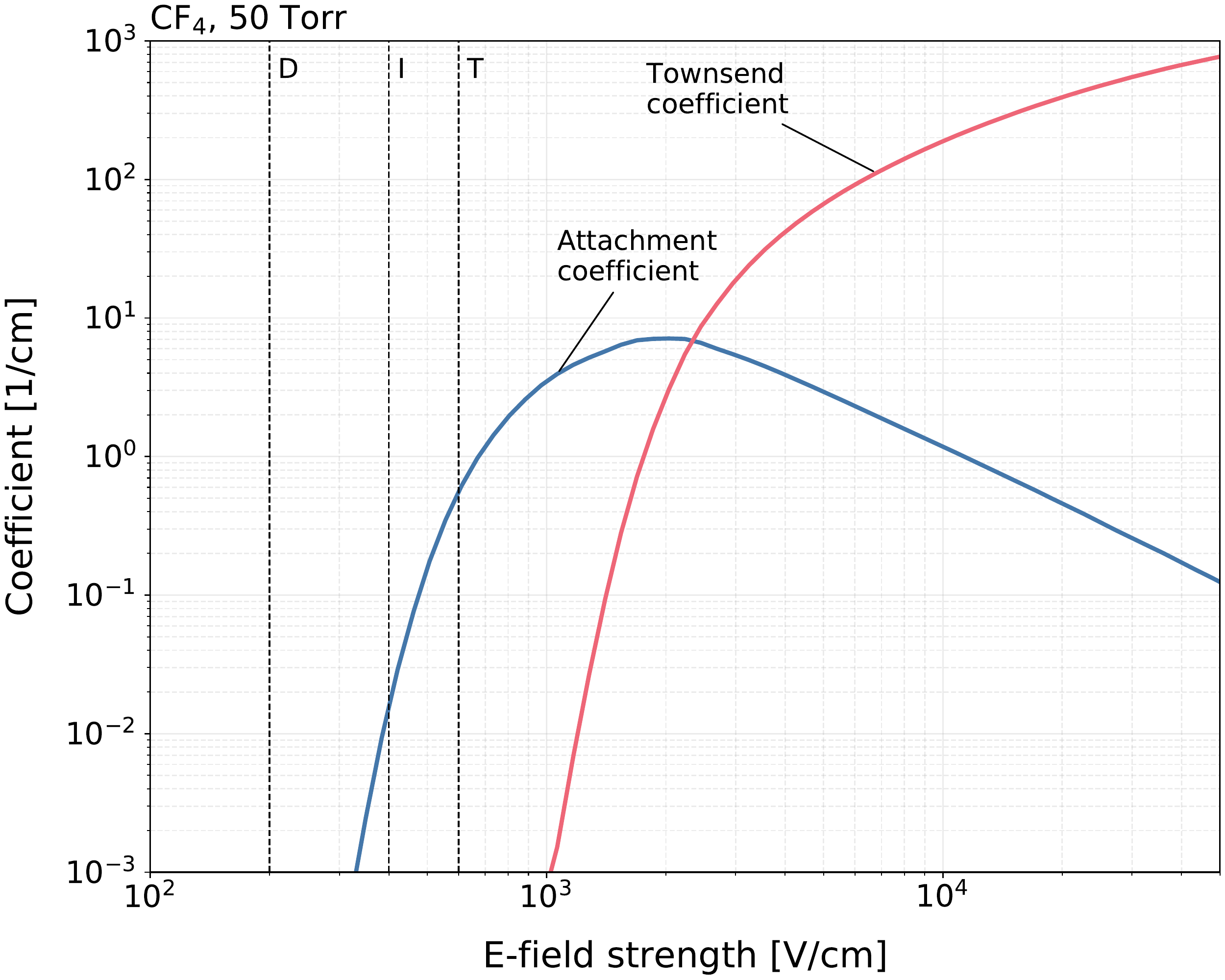}
\caption{Electron transport properties of CF$_4$ at 50~Torr. Left -- Drift velocity and diffusion. Right -- Attachment and Townsend coefficients. Nominal fields in the drift (D), transfer (T) and induction (I) regions are indicated.}
\label{Fig:GEM1}
\end{figure*}

The ionisation electrons produced in the active volume drift under the influence of an electric field towards the anode; avalanche multiplication takes place upon entering the GEM stack. The transport of electrons in the gas is modelled using the Garfield++~\cite{Veenhof:1993hz,Veenhof:1998tt,Garfield} framework, which relies on  Magboltz~\cite{Biagi1999,Magboltz} for the estimation of the transport properties in the gas mixture. The electron drift velocity, longitudinal and transverse diffusion in CF$_4$ at 50~Torr provided by Magboltz are given in Fig.~\ref{Fig:GEM1} (left) as a function of electric field. Figure~\ref{Fig:GEM1} (right) shows the corresponding Townsend and attachment coefficients. 

In an effort to optimise the simulation speed while retaining the highest possible precision, different approaches are used for the different regions. In the drift region diffusion is modelled through a Gaussian smearing. Within the GEMs, where the avalanche amplification takes place, and in the transfer region between the two GEMs, the highest level of detail is obtained by exploiting the microscopic tracking capability of Garfield++. In this, each electron is followed between individual collisions with the atoms of the gas. Finally, in the induction region, where a large number of electrons is present, the Garfield Monte-Carlo integration technique is applied. In this technique, the transport parameters are integrated over 100 collisions, and subsequently the longitudinal and transverse diffusion is sampled from the expected distribution.

\subsection{The Gas Electron Multipliers}
\label{SS:GEMs}

Electron multiplication in the detector is realised as a two-stage process, whereby the electrons produced in the avalanche in the first GEM subsequently initiate new avalanches in the second GEM. Thus, the operating voltages of the two GEMs need to be tuned both for overall gain and for gain balance between the two GEMs. The electric field strength in the transfer region between the two GEMs also needs to be chosen to improve the transparency of the system, in terms of the fraction of electrons produced in the first GEM that enter the holes of the second GEM. At the same time, any attachment of electrons in the CF$_4$ needs to be minimised.

As mentioned above, within the two GEMs and in the transfer region between them the electron transport is modelled using the microscopic tracking capabilities of Garfield++. These provide the highest level of detail but result in substantial computational cost, given the large number of electrons being microscopically tracked, which prohibits the production of the number of events required for more detailed studies. Thus, an ``event library'' approach is pursued, where the library consists of thousands of single electron events. In each case one electron is released at a height of 50~$\mathrm{\mu}$m above the first GEM with a random position within the primitive cell of the GEM structure. The electron and its associated avalanche are microscopically tracked through the system of the two GEMs, and the spatial and timing information of the produced electrons arriving at 20~$\mathrm{\mu}$m below the bottom plane of the second GEM are stored, along with the starting position of the initial electron. For the simulation of large numbers of events, the ionisation electrons are transported within the drift region until they reach a distance of 50~$\mathrm{\mu}$m above the first GEM. For each ionisation electron, its position is projected onto the primitive cell and the spatially closest electron from the event library is identified. The stored outcome for this fully simulated library event is retrieved, and the positions of the resulting electrons (if any) are projected onto the cell of the ionisation electron. This procedure allows for the benefits of microscopic tracking of electrons in the `end-to-end' simulation, while keeping the computational cost for events consisting of hundreds or thousands of ionisation electrons at manageable levels. For the setup used in the following results, the mean charge gain after both GEMs is 0.6$\times$10$^5$.

\subsection{ITO strips}
\label{SS:ITO}

Charge-readout signals are simulated with Gmsh~\cite{gmsh}, Elmer~\cite{Elmer} and Garfield++. Gmsh and Elmer are used to define the geometry of the simulation space and to calculate electrostatic fields, respectively, while Garfield++ is used to simulate the electron drift and calculate the resulting ITO strip signals.

The simulation geometry is a 2D slice through 30~ITO strips, corresponding to a width along $x$ of 25~mm. The total depth is 11.1~mm, starting at the exit of the second GEM and ending 5~mm past the ITO plate (this latter region helps to more accurately model electric-field lines around the ITO strips). A uniform potential of $-$80~V is considered at the second GEM plane, which results in the nominal 400~V/cm electric field in the induction gap. The strips are charged to $+1$~V (individually, for the signal weighting fields), the bottom is grounded, and the sides are periodic.

\begin{figure*}[ht]
    \centering
    \includegraphics[width=0.45\linewidth]{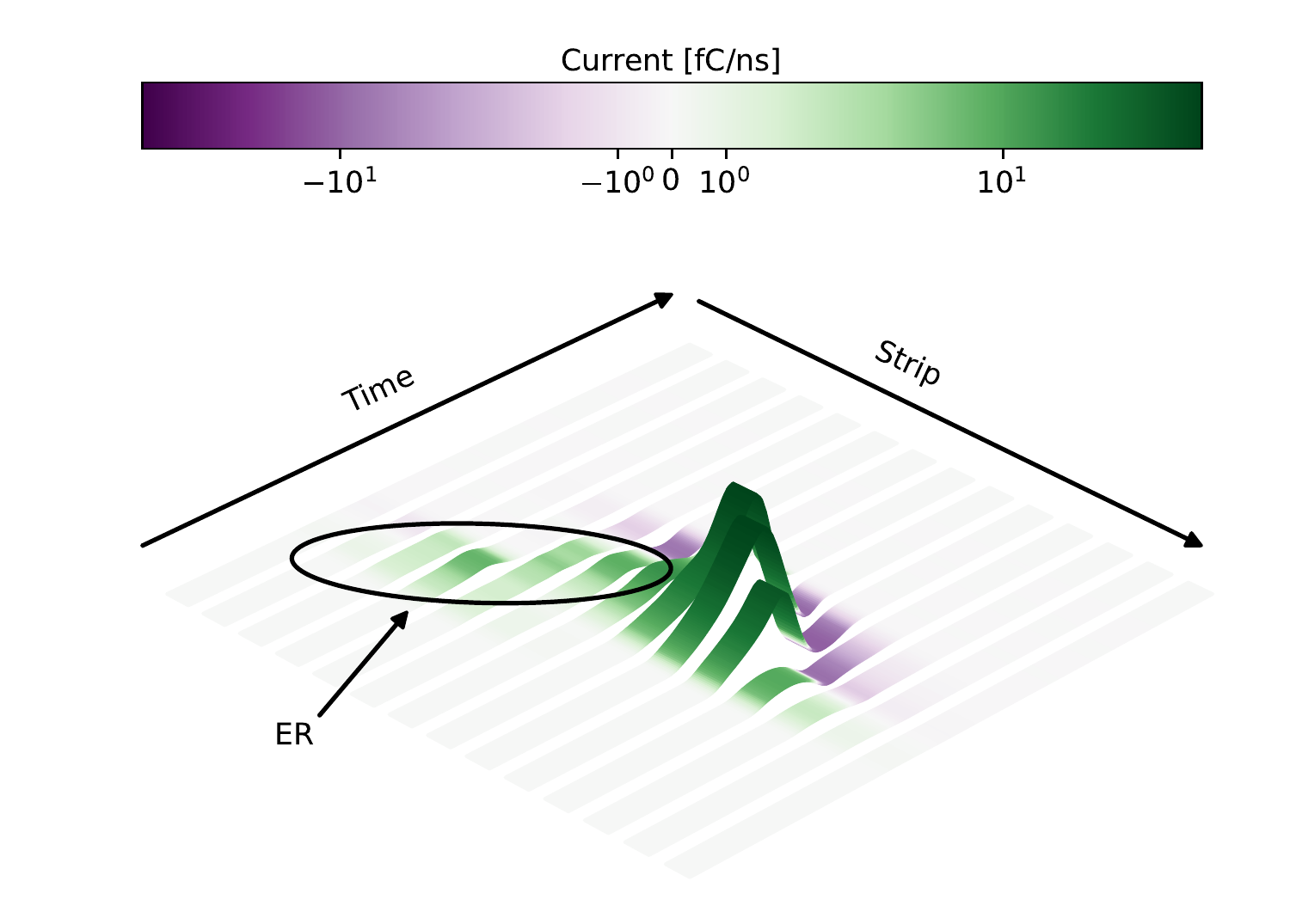}\qquad
    \includegraphics[width=0.45\linewidth]{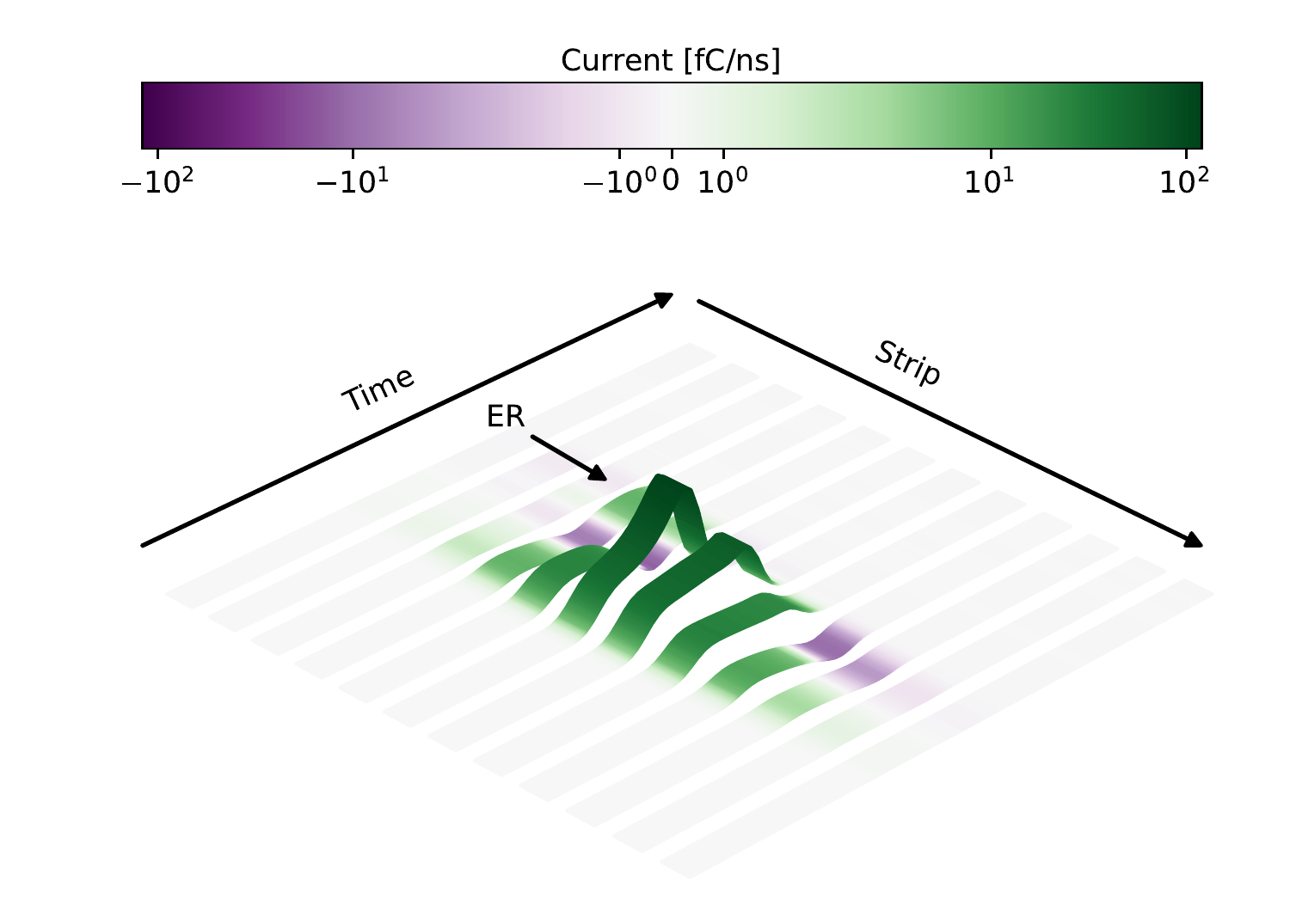}
    \caption{The ITO-strip current signals produced by the two example events shown in Fig.~\protect{\ref{Fig:ExampleMigdalTracks}} prior to folding in the electronics response. The current induced on each strip is represented by the height of the strip and its colour. In each plot the location of the electronic recoil is marked with ``ER''.}
    \label{Fig:ExampleITOSignals}
\end{figure*}

The currents induced on the ITO strips by the drifting electrons leaving the second GEM are convolved with the response of the electronics, including cross-talk, as discussed in Section~\ref{SS:Electronics}, to generate the expected voltage signals on each strip. The current signals prior to this convolution are shown in Fig.~\ref{Fig:ExampleITOSignals}, for the two benchmark events shown previously in Fig.~\ref{Fig:ExampleMigdalTracks}.

\subsection{Electronics response}
\label{SS:Electronics}

The ITO strips exhibit not only significant series resistance (on the order of 6~k$\Omega$/m), but also significant capacitance between strips (65~pF/m). Thus, any simulation of the response of the detector must include not only the amplifier and digitiser, but also the ITO strips themselves. This was implemented by discretising them into 5 segments, and the flat-cable between the ITO and the amplifier into 3 segments, with the resistance and capacitance and each segment modelled with a finite-element method. The resulting electrical circuit was simulated using Cadence Spectre~\cite{Cadence}.

\begin{figure}[ht]
\centerline{\includegraphics[width=0.85\linewidth]{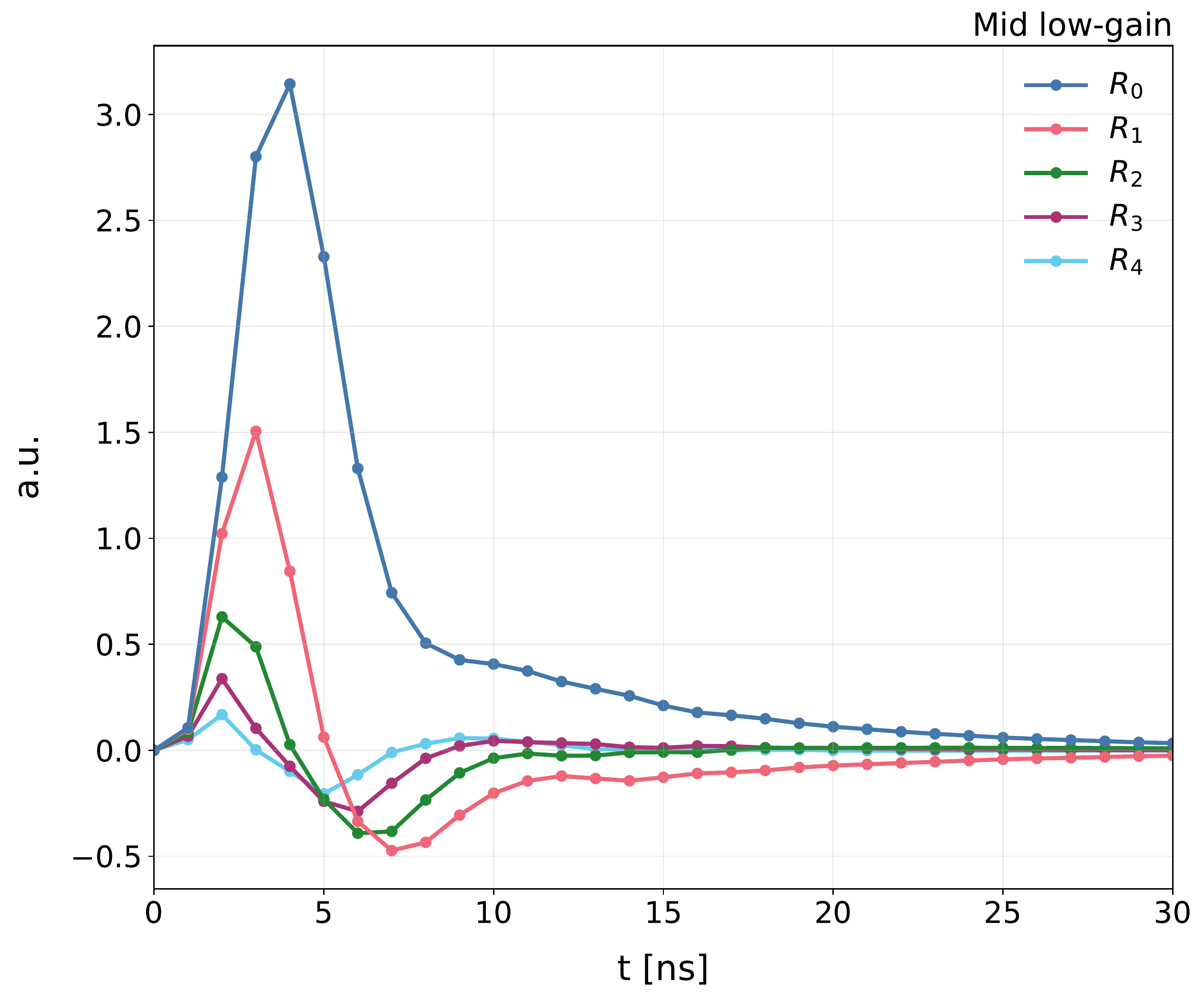}}
\caption{Simulated electronics response to charge deposited in the middle of an ITO strip. The signal depends not only on the charge deposited on the strip of interest, but also that induced on adjacent strips (up to four strips either side).}
\label{Fig:ElectronicsResponse}
\end{figure}

The simulated response to charge deposited at the midpoint along a strip, and how this manifests on different channels, is shown in Fig.~\ref{Fig:ElectronicsResponse}. The plot highlights a non-negligible amount of coupling between strips; therefore, to accurately model the digitised signal from simulated events the output from the electrical model described above was used in Garfield++ to account for both the shaping of the signals and the lateral smearing due to coupling between channels. We plan to deconvolve the electronics response function from the measured waveforms such that the current signals shown previously in Fig.~\ref{Fig:ExampleITOSignals} will resemble the data used for further analysis.

\subsection{The camera image}
\label{SS:Imaging}

\begin{figure*}[ht]
    \centering
    \includegraphics[width=0.38\linewidth]{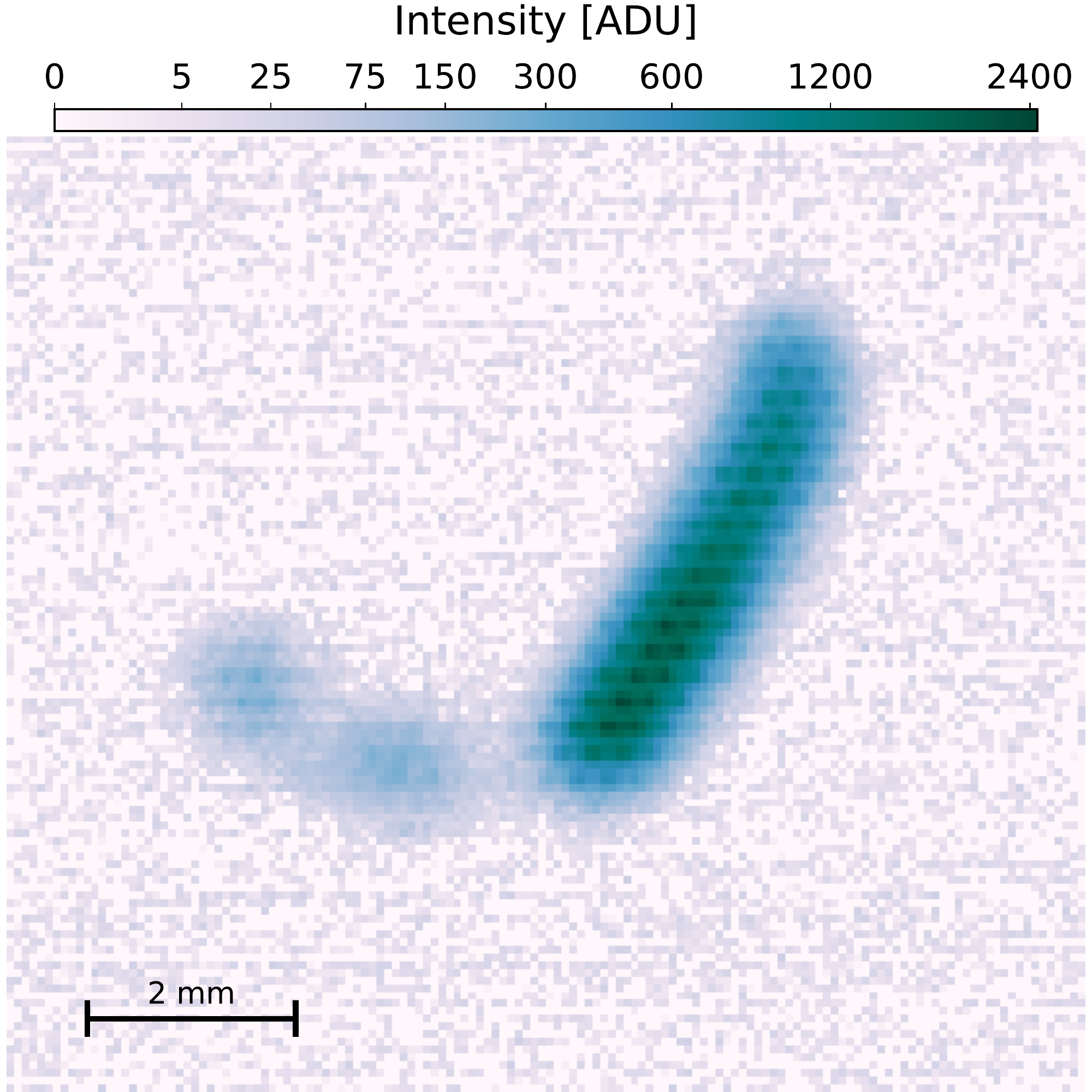}\qquad
    \includegraphics[width=0.38\linewidth]{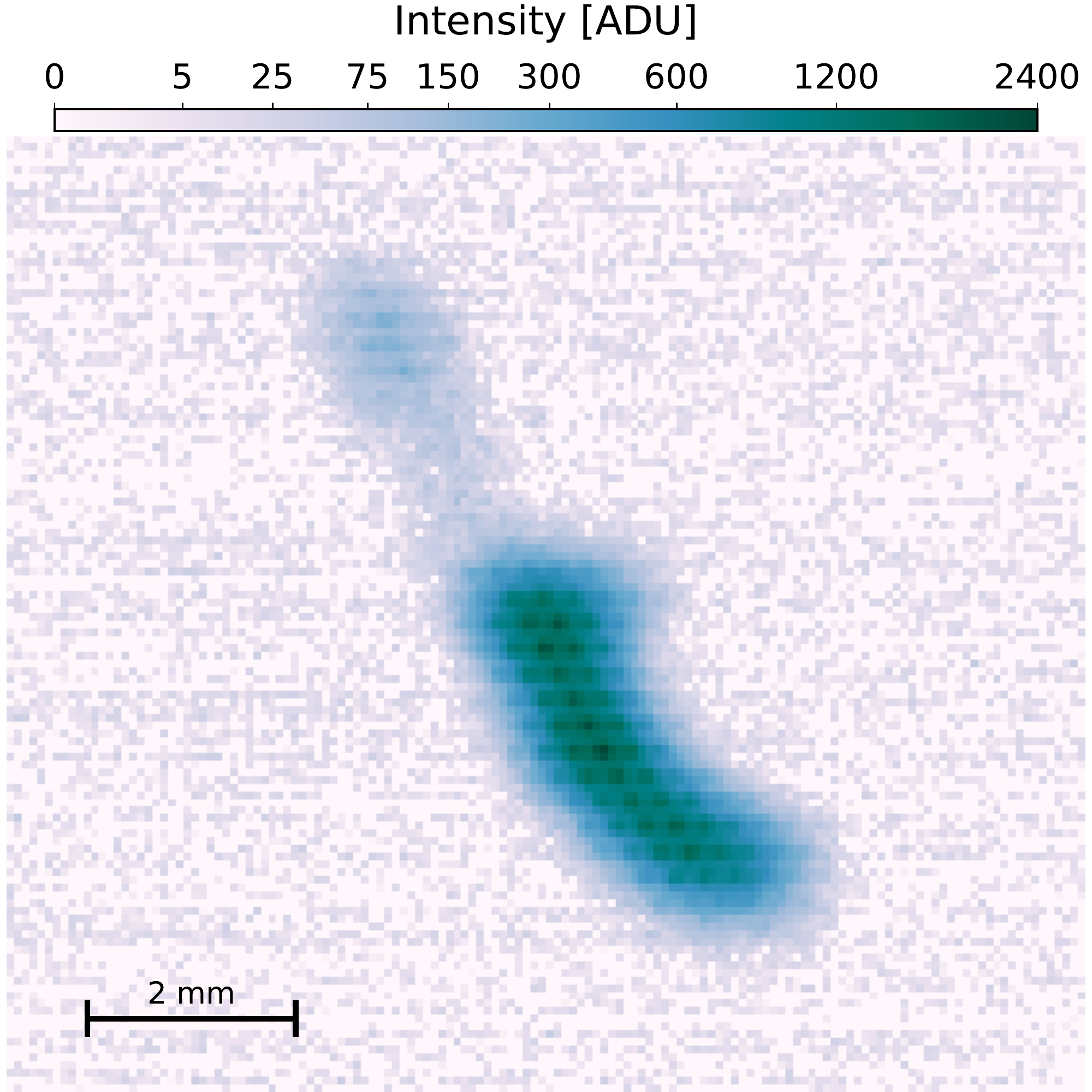}
    \caption{Simulated camera images of the two benchmark Migdal events shown in Fig.~\protect{\ref{Fig:ExampleMigdalTracks}}; both NR and ER tracks are visible in each case, with the latter corresponding to 7.5~keV and 5.0~keV Migdal electrons on the left and right, respectively. Both images have a realistic sample of camera noise added; the average (master) dark frame has been subtracted.}
    \label{Fig:MigdalCameraImage}
\end{figure*}

Electrons exiting the second GEM stage are used to generate the camera images of simulated tracks. The photon yield calculated in Section~\ref{SSS:Camera} is assumed to map onto a geometrically perfect image, which is binned using the parameters of the CMOS camera described in Section~\ref{SS:OpticalReadout}. Noise is added to each pixel based on measurements with this camera model for the proposed mode of operation; the overall signal-to-noise ratio matches that obtained in preliminary tests. Simulated images of the two Migdal events mentioned above are shown in Fig.~\ref{Fig:MigdalCameraImage}.

\subsection{Track reconstruction}
\label{SS:Reconstruction}

Tracks are reconstructed in 3D by combining data from the ITO readout, the camera image and the PMT -- initially matched by the MiDAS data reduction software.

The ITO signals are processed with a 2D deconvolution method similar to that used in Ref.~\cite{Adams_2018}, using the response of the electronics and cross-talk shown in Fig.~\ref{Fig:ElectronicsResponse}, to obtain the induced current on each strip. A second 2D deconvolution is performed using the mean response of a single electron in the induction gap, obtained from simulation, to find the charge per nanosecond on each strip. This provides information about the arrival time of each electron and, using the expected drift velocity, it can be used to estimate the extent of the track in the $z$ direction.

\begin{figure}[htb]
    \centering
    \includegraphics[width=0.90\linewidth]{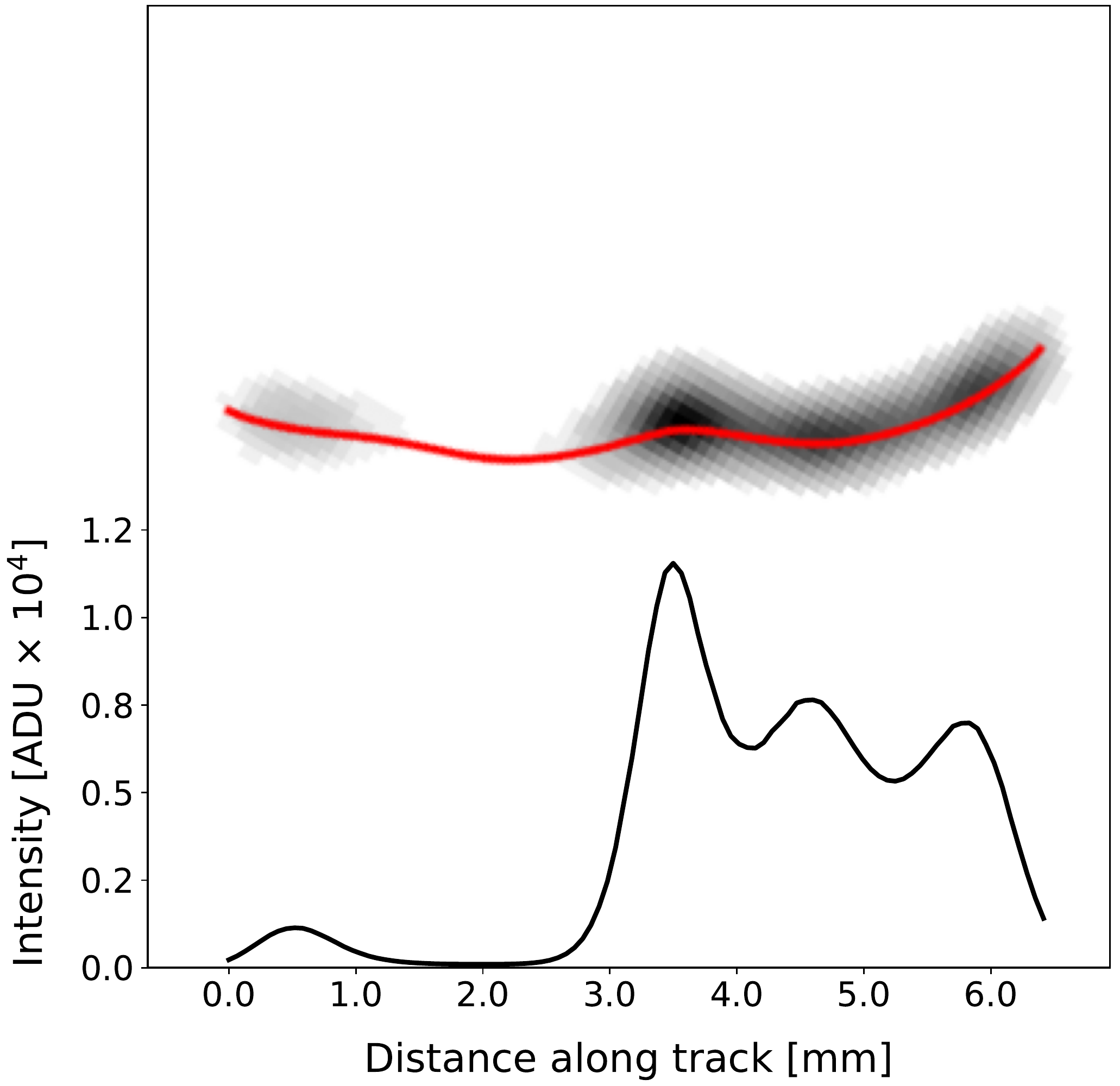}
    \caption{Reconstructed track for the 5~keV Migdal electron event shown in Fig.~\protect{\ref{Fig:MigdalCameraImage}} (right), after applying deconvolution and track-finding algorithms to the camera image; the resultant track ridge is shown in red. The plot in the lower part of the frame shows the intensity integrated transversally along the tracks (rather than a straight projection) and is meant to illustrate the very different energy loss rates along NR and ER tracks.}
    \label{Fig:ImageAnalysis}
\end{figure}

The camera images also undergo several steps of processing to reconstruct the 2D track information. First, a low-pass filter is applied to remove the GEM hole pattern from the images (which is visible in Fig.~\ref{Fig:MigdalCameraImage}). The resulting image is deconvolved using the Richardson-Lucy algorithm~\cite{RichardsonLucy:Lucy,RichardsonLucy:Richardson} with a 2D Gaussian point spread function the width of which is estimated based on the average diffusion at that depth determined from the S1-S2 time delay. The deconvolved images are input to a track finding algorithm~\cite{Steger1998} that extracts the detailed 2D track properties, such as the range, energy loss rate and initial direction of the particle. The result of this procedure applied to the simulated Migdal events are shown in Fig.~\ref{Fig:ImageAnalysis}.\footnote{A grayscale version of \protect{Fig.~\ref{Fig:ImageAnalysis}} was analysed using ImageJ with plugin Ridge Detection~\protect{\cite{RidgeDetection}}.} 

\section{Sensitivity}
\label{S:Sensitivity}

Having described the detector, neutron beam and simulation framework in some detail, we can now present a more realistic calculation of the signal and background rates in the fiducial volume. We begin by setting out the expected number of neutron-induced nuclear recoils and expected Migdal event rates, and then discuss in detail the background topologies which could affect their measurement. From the expected signal and background rates an indicative discovery sensitivity is presented.

\subsection{Signal acceptance}
\label{SS:Acceptance}

We expect to operate the experiment for several consecutive days with high duty cycle, probably limited by the recovery from GEM discharges, progressive loss of gas purity, and the regular calibrations for gain monitoring. We will term each few-day long data-taking period at constant operating conditions (e.g.~pressure) a `run'.

To maximise duty cycle we will operate the neutron generators in continuous mode; technical breaks are not deemed necessary by the manufacturer over such periods. Data can be acquired at the maximum camera speed and transferred from the DAQ computer to the offline PB-scale storage system with no dead time envisaged.

The detector stability will be regularly monitored by gas-gain measurements using the external $^{55}$Fe source, and checking the synchronisation between the camera and the digitiser systems will rely on LED signals. These calibrations are planned to total 30~minutes every 6~hours, reducing the duty cycle to 92\%.

Replenishing the ageing gas and re-biasing the OTPC will take around 30~minutes, and we assume (conservatively) that this will be performed every 6~hours; the above calibrations will be split before and after the gas replenishing. This will bring the duty cycle down to 84\%.

The total number of camera frames recorded in a 5~calendar day run may approach 40~million (including calibrations), taking up 44~TB of disk storage.

We take the interaction rates in CF$_4$ from Table~\ref{Tab:nrates}. The fraction of frames with only one NR interaction in the full active region is $\approx$1/3 for both generators; we remove frames with more than one interaction ($\approx$15\%) -- in many such cases the DAQ will have triggered only once, and there may be scope for event confusion. In addition, in the first instance the Migdal analysis will consider only NR tracks fully contained within the smaller 8$\times$8~cm$^2$ fiducial area. With the above considerations, the 5-day run with 50~Torr of CF$_4$ will accumulate 1.8 and 2.3 million single NR track images for analysis with the D-D and D-T generators, respectively.

The Migdal probabilities for tracks above threshold fully contained in the fiducial volume and with ER and NR energies within our ROI are 3.26$\times$10$^{-5}$ and 8.42$\times$10$^{-5}$ for the two generators, as set out on Section~\ref{S:SignalsBackgrounds}; these calculations follow Ref.~\cite{MIGDALth}.

We must now fold in the efficiency for detecting the Migdal topology in analysis, including factors such as the highly-variable electron track shape, energy resolution and other detector effects in the various subsystems. A detailed study of this topic lies beyond the scope of this paper; instead, using the simulation framework described in the previous section, we undertook a blind data challenge to evaluate this efficiency for the worst case scenario: isotropically-emitted electrons at the nominal 5-keV threshold and fluorine NR tracks also near their energy threshold. This dataset included a few hundred events including both Migdal topologies as well as bare nuclear recoils, in approximately similar numbers; this exercise returned zero false positives, and an average of 75\% of all Migdal events were correctly identified by the analysers inspecting simulated camera images alone (no ITO or PMT information). We apply this efficiency to the above Migdal event rates to arrive at a Migdal detection rate of 8.9~events per calendar day for D-D neutrons and 29.3~events/day with the D-T generator, or around 44 and 147 events in a 5-day run, respectively. Note that the numbers given in Table~\ref{Tab:nrates} do not include this detection efficiency, single-track frame probability, or the operational duty cycle.

\subsection{Backgrounds}
\label{SS:Backgrounds}

We discuss potential backgrounds to this measurement with reference to Table~\ref{Tab:bk1}, which lists the estimated number of counts per million neutron-induced recoil tracks in CF$_4$ gas at 50~Torr. Background counts are considered in the previously-defined ROI, including the 5--15~keV electron energy range; we consider electron vertices located up to 3~mm from the NR track origin. Signal rates in the same ROI are also given for comparison. To highlight effects which we expect to find in the data but that do not necessarily translate into background counts, we list also the count rate integrated for electron energies above 0.5~keV; most of these will contribute to the track `penumbra', but some may cause distinct interactions clearly resolved from the NR track (e.g.~occupying single pixels in the GEM image). In both cases we assume that the electron detection efficiency is 100\%.

In the following it will be confirmed that the sensitivity of the experiment relies significantly on the long photon attenuation lengths provided by the low pressure gas -- as illustrated in Fig.~\ref{Fig:PhotonsInCF4} -- and this mitigates against several backgrounds. A potential major challenge of a different nature comes from the stochastic nature of the atomic cascades which may develop from the primary recoil; in particular, a secondary recoil track may be spawned close to the origin of the initial track to create a fork-like structure which may be confused with a Migdal vertex in some situations. This background topology is the most complex and deserves a longer discussion at the end of this section.

\begin{table*}[htb]
\caption{Number of background and signal events {\em per million} neutron-induced recoil tracks for D-D (2.47~MeV) and D-T (14.7~MeV) neutrons incident on CF$_4$ gas at 50~Torr. Data are given for $\sim$100~keV nuclear recoil threshold and $>$0.5~keV and 5--15~keV electron energies, with the electron vertex located up to 3~mm from the NR track origin. An entry of ``0'' indicates that the process cannot occur in the ROI, while ``$\approx$0'' denotes a negligible rate of $\ll$0.01 events per million recoils. Individual background components and topologies are discussed in the text. Signal rates are those from Table~\protect{\ref{Tab:nrates}} for contained tracks above threshold, normalised per million signal-inducing events.}
\vspace{5pt}
\centering
{\small
\begin{tabular}{l p{5.8cm} | c c c c}
\hline
\multirow{2}{*}{Component} & \multirow{2}{*}{Topology} 
    & \multicolumn{2}{c}{D-D neutrons} & \multicolumn{2}{c}{D-T neutrons} \\
& & $>$0.5 & 5--15~keV & $>$0.5 & 5--15~keV \\
\hline
Recoil-induced $\delta$-rays    & Delta electron from NR track origin 
    & $\approx$0 & 0 & 541,000 & 0 \\
Particle-Induced X-ray Emission (PIXE) &&&&&\\
\qquad X-ray emission       & Photoelectron near NR track origin
    & 1.8 & 0 & 365 & 0 \\
\qquad Auger electrons      & Auger electron from NR track origin
    & 19.6 & 0 & 42,000 & 0 \\
Bremsstrahlung processes$^\dagger$ &&&&&\\
\qquad Quasi-Free Electron Br. (QFEB) & Photoelectron near NR track origin
    & 112 & $\approx$0 & 288 & $\approx$0 \\
\qquad Secondary Electron Br. (SEB)   & Photoelectron near NR track origin
    & 115 & $\approx$0 & 279 & $\approx$0 \\
\qquad Atomic Br. (AB)              & Photoelectron near NR track origin 
    & 70 & $\approx$0 & 171 & $\approx$0 \\
\qquad Nuclear Br. (NB)             & Photoelectron near NR track origin 
    & $\approx$0 & $\approx$0 & 0.013 & $\approx$0 \\
Neutron inelastic $\gamma$-rays & Compton electron near NR track origin
    & 1.6 & 0.47 & 0.86 & 0.25 \\
Random track coincidences &&&&&\\
\qquad External $\gamma$- and X-rays & Photo-/Compton electron near NR track 
    & $\approx$0 & $\approx$0 & $\approx$0 &  $\approx$0 \\
\qquad Trace radioisotopes (gas)         & Electron from decay near NR track origin
    & 0.2 & 0.01 & 0.03 & $\approx$0 \\
\qquad Neutron activation (gas)           & Electron from decay near NR track origin
    & 0 & 0 & $\approx$0 & $\approx$0 \\
\qquad Muon-induced $\delta$-rays & Delta electron near NR track origin
    & $\approx$0 & $\approx$0 & $\approx$0 & $\approx$0 \\
Secondary nuclear recoil fork & NR track fork near track origin
    & -- & $\approx$1 & -- & $\approx$1 \\
\hline
Total background    & Sum of the above components 
    & & 1.5 & & 1.3 \\
\hline
Migdal signal       & Migdal electron from NR track origin 
    & & 32.6 & & 84.2 \\
\hline
\end{tabular}
}
\label{Tab:bk1}
\begin{flushleft}
    {\footnotesize 
    $^\dagger$ These processes were evaluated at the endpoint of the nuclear recoil spectra.
    }   
\end{flushleft}
\end{table*}

The first potential background listed in Table~\ref{Tab:bk1} is that from $\delta$-electrons produced by the recoiling nucleus; this contribution was obtained from a GEANT4 simulation of carbon- and fluorine-induced ionisation for the D-D and D-T recoil spectra. The maximum energy transfer to an electron at rest in a binary collision is far below the 5~keV threshold. For example, a 4~MeV carbon ion (near the endpoint of the D-T spectrum) transfers a maximum of 742~eV to a $\delta$ electron; these energies are modified in collisions with bound electrons, but this is a small effect for light elements. In conclusion, although this process does produce a significant number of low-energy electrons which contribute to broaden the NR track, it does not become a concerning background to the Migdal measurement.

Particle-induced X-ray emission (PIXE) may result in either X-ray fluorescence (which may be absorbed near the NR track) or the emission of Auger electrons.\footnote{Catalogues of atomic transition energies and yields can be found in Refs.~\protect{\cite{Krause1979,Cochlan1973,NISTxrays}}.} For light elements such carbon and fluorine, Auger yields far exceed X-ray fluorescence yields. The GEANT4 PIXE models \cite{Bakr2018,Mantero2010} were used to obtain the particle yields. These were validated for these elements by scaling the proton cross sections in Ref.~\cite{Paul1989} for the effective charge of the heavier projectiles and found to be in good agreement. Auger electron energies have a maximum of 655~eV for CF$_4$~\cite{Larkins1987} and so PIXE does not pose a significant background -- although it should be noted that this is not necessary the case for heavier elements with atomic shell energies in the electron region of interest.

In collisions of recoiling ions with CF$_4$ molecules several Bremsstrahlung processes may occur. These photons can release a $\sim$keV electron near the NR track origin (tail), mimicking a Migdal event. Such processes include Quasi-Free Electron Brems\-strahlung (QFEB)~\cite{Yamadera1981}, Secondary Electron Brems\-strahlung (SEB)~\cite{Folkmann1974}, Nuclear Brems\-strahlung (NB)~\cite{Alder1956}, and Atomic Brems\-strahlung (AB) (also known as polarizational radiation)~\cite{Amusia1992}. In the QFEB process an electron of a target atom is scattered by the Coulomb field of the incident ion and emits bremsstrahlung radiation. In SEB X-rays are produced by ionised electrons in the Coulomb field of the target nuclei. Nuclear bremsstrahlung is produced when a NR is accelerated in the Coulomb field of the target nucleus.\footnote{An identically-named process with interest for dark matter searches~\protect{\cite{Kouvaris2017}} refers instead to the photon emission caused by the electric dipole created between a recoiling nucleus and the atomic electrons; this phenomenon is rarer than the Migdal effect and is not considered here.} In the AB process, a bound electron of a target atom is excited to a continuum state by the incident NR and, returning to its original bound state, a photon is emitted. These Bremsstrahlung processes have been extensively studied in the context of PIXE as they constitute backgrounds for elemental analysis, and the corresponding cross sections have been calculated and confirmed experimentally. In general, the cross sections scale with the charge and the velocity of the ion and with the atomic number of the target atom~\cite{Ishii1984,Ishii1987}. We have calculated a conservative upper limit on the number of electrons ejected per million carbon recoils with energies at the endpoint of the D-D and D-T recoil spectra; this involved the scaling the available cross sections for protons~\cite{Ishii1987,Ishii1988}, and it took into account the photon absorption efficiency in the vicinity of NR tracks. Owing to the low velocity of NRs compared to that of the protons of the same energy, the probability of X-ray emission in the energy range of interest is extremely low in both the D-D and the D-T experiments.

A potentially relevant background occurs when a neutron undergoes inelastic scattering in the active gas volume and a deexcitation $\gamma$-ray interacts near the NR vertex from the same interaction. This type of event dominates the background budget in higher pressure experiments~\cite{Nakamura:2020kex}. While the mean photon interaction length is very large at 50~Torr, the Compton scattering of the 110~keV and 197~keV photons from $^{19}$F($n$,$n'$) is of some concern: their energy is just right to produce Compton electrons in the ROI -- cf.~Fig.~\ref{Fig:PhotonsInCF4}. The calculation in Table~\ref{Tab:bk1} uses GEANT4 to simulate the production and interaction of these and similar $\gamma$-rays in the 5--15~keV range. The corresponding background rate is found to be just below 1~event per million NR tracks -- we predict this to be a leading source of background in our experiment, as shown in Table~\ref{Tab:bk1}.

A different class of background (with several contributors) occurs when an NR track and an unrelated ER track are accidentally recorded in coincidence. In this instance the event topology is similar to that of signal if the electron is emitted sufficiently close to the NR vertex. These electrons can be caused by $\gamma$-rays and X-rays produced in different parts of the experiment and the neutron generator, that reach the active volume. The predicted rate of coincidence events has been assessed by calculating the photon spectrum entering the active volume with GEANT4, and then convolving the corresponding spectrum with the photon probability to yield an electron in the 5--15~keV window. We assumed that tracks with a separation of 3~mm in camera images can be well distinguished, while the time resolution is set to the maximum drift time along the OTPC (230~ns). The total rate of accidental coincidences is found to be $\ll$0.01~events per million NR with either generator. In most of these events the NR and the photon are created by two different neutrons (85\% and 77\% for D-D and D-T generators, respectively), while the remainder comes from coincidences where the photon is produced in an earlier interaction of the same neutron that produces the NR. The volumes where the photon originate differ between the D-D and D-T experiment configurations. In the former case, this is dominated by deexcitation X-rays following photoelectric effect in the cathode (70\% of the total), while in the latter case the most significant are $\gamma$-rays from neutron inelastic scattering occurring in the generator material placed in the line-of-sight of the active volume (80\% of the total).

The previous paragraph discussed accidental coincidences where the origin of the electron can be traced back to a neutron produced in D-D or D-T fusion. However, the electric field used to accelerate deuterium ions in these devices also creates a current of free electrons flowing in the opposite direction, which produce a significant field of bremsstrahlung photons when stopping. For the D-D experiment configuration, such bremsstrahlung photons are produced at a position displaced from the collimator axis, and therefore they are completely stopped by the shielding before they reach the active volume. For the D-T experiment configuration, the rate of coincident events from bremsstrahlung photons has been conservatively estimated using GEANT4 simulations, and we find that this contribution is also negligible. This plus the preceding contribution correspond to the entry for `External $\gamma$- and X-rays' in Table~\ref{Tab:bk1}. It has also been confirmed that $\gamma$-rays from the decay of radioisotopes produced by neutron activation only represents a small contribution to accidental coincidence events (1\% and 2\% for D-D and D-T neutrons, respectively). 

An additional class of accidental background involves the $\beta$-decay of trace radioisotopes in the OTPC gas, with low-energy electrons randomly co-locating with the origin of a NR track. We considered $^{14}$C and $^{39}$Ar in CF$_4$-based gas mixtures, both assumed at typical atmospheric concentrations, as well as 1~mBq of the $^{222}$Rn daughters $^{214}$Pb and $^{214}$Bi, giving a total activity of a few hundred decays per day. This translates into a negligible coincidence rate in the electron ROI.

Neutron scattering may produce $\beta$ emitters without additional charged particles, namely via ($n$,$2n$) or ($n$,$\gamma$) reactions. If the daughter nucleus decays quickly enough, a two-track topology similar to that of the signal may be observed. The most relevant case is $^{19}$F($n$,$2n$)$^{18}$F, where the daughter is produced at a rate of 0.28~nuclei/s with the D-T generator (the reaction is below threshold with D-D) and has a half-life of 110~minutes. This specific decay leads to a background rate well below 1~event per million NR tracks, and other processes such as radiative neutron capture on $^{19}$F are even more unlikely. This background contribution is therefore also negligible.

Finally, atmospheric muons cross the detector at a rate of $\sim$1~per second; although these have too low an energy loss rate to be observable, approximately 2\% cause $\delta$-ray electrons in the ROI which could contribute accidental coincidences; this also translates to a negligible background to the Migdal search.

Next we discuss the more complex issue of atomic cascades in nuclear recoil tracks and the potential for fork-like events involving secondary recoils to be confused with the Migdal signal topology.

\subsubsection{Secondary nuclear recoils}

\label{SSS:Secondaries} 

A primary nuclear recoil created in neutron scattering collides along its path with neighbouring atoms, inducing secondary recoils and initiating an atom cascade. In a binary collision the secondaries may acquire enough kinetic energy to produce a prominent fork-like topology, potentially mimicking a Migdal event. In order to calculate the frequency of such events, we used the TRIM fast calculations to generate 10$^6$ primary recoils from D-D neutrons, with C and F in the correct proportion. Selected events had secondary recoils created within the first 1~mm from the primary NR vertex and energies between 11 and 26~keV for carbon and 13.5 and 32~keV for fluorine, both mapping to the 5--15~keV$_{\mathrm{ee}}$ energy range of the accepted Migdal electrons. Some 10,000 events met these criteria. The relative rates of such events are summarised in Table~\ref{Tab:recrates} for selected primary recoil energies.

\begin{table}[htb]
\caption{Relative rates of events with secondary recoils, expressed per million carbon and fluorine primary ions, created within 1~mm from the vertex in 50~Torr CF$_4$, when the energy of the secondary track is in the range 11.0--26~keV (C) and 13.5--32~keV (F), both corresponding to 5--15~keV$_{\mathrm{ee}}$. These can be further discriminated from electron tracks as explained in the text.}
\vspace{5pt}
\setlength{\tabcolsep}{4pt}
\renewcommand{\arraystretch}{1.2}
\centering
{\small
\begin{tabular}{l | r r }
\hline
Primary ion & \multicolumn{2}{c}{Secondary ion} \\
\hline
Fluorine & Fluorine & Carbon \\
\hline
\qquad 500~keV & 22,310 & 4,800  \\
\qquad 400 &  26,840    & 5,930  \\
\qquad 300 &  36,640    & 7,640  \\
\qquad 200 &  56,130    & 1,263  \\
\qquad 170 &  67,040    & 1,418  \\ 
\hline
Carbon & Fluorine & Carbon \\
\hline
\qquad 500~keV & 6,250  & 1,210 \\
\qquad 400 &  7,950     & 1,610 \\
\qquad 300 &  11,380    & 2,310 \\
\qquad 200 &  17,310    & 3,700 \\
\qquad 130 &  26,120    & 5,770 \\
\hline
\end{tabular}
}
\label{Tab:recrates}
\end{table}

\begin{figure}[t]
\centerline{\includegraphics[width=0.93\linewidth]{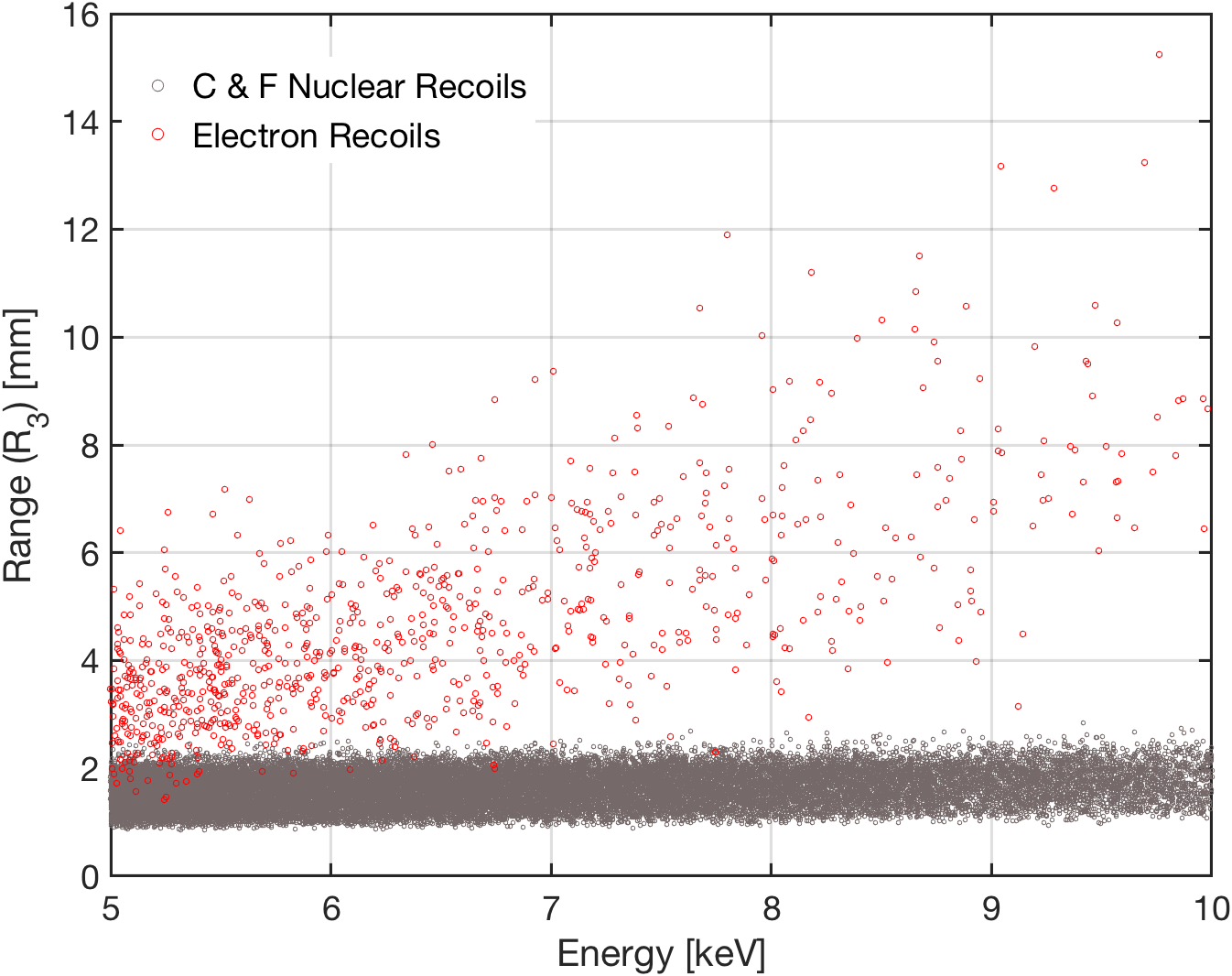}}
\caption{3D range as a function of reconstructed energy (keV$_{\mathrm{ee}}$), showing separation of low-energy electrons (wide band) from carbon and fluorine recoils (narrow band) in CF$_4$ at 50~Torr.}
\label{Fig:r2}
\end{figure}

Next we assess how many of those 10,000 events might be misinterpreted as genuine Migdal interactions -- and hence determine the background rate from this source expressed per million primary NR tracks. We simulated more fully further populations of 10,000 C and F recoils plus 1,000 electrons all with true energies in the range 4.2--12~keV$_{\mathrm{ee}}$: these energies contribute to the 5--10~keV$_{\mathrm{ee}}$ ROI due to the non-zero energy resolution. In these simulations the NR tracks are distributed in energy like those induced by the primary NR spectrum from D-D neutrons, while the electron spectrum follows that expected for the Migdal electrons. For all tracks we follow the simulation steps described in Section~\ref{S:TrackSimulations}. The resulting deconvolved 2D camera images were used to measure a 2D track range using the method described in Ref.~\cite{Phan:2015pda}. This was combined with the $z$ component of the track to calculate an approximate 3D range ($R_3$), which is plotted in Fig.~\ref{Fig:r2} as a function of reconstructed energy.

\begin{figure}[htb]
\centerline{\includegraphics[width=0.93\linewidth]{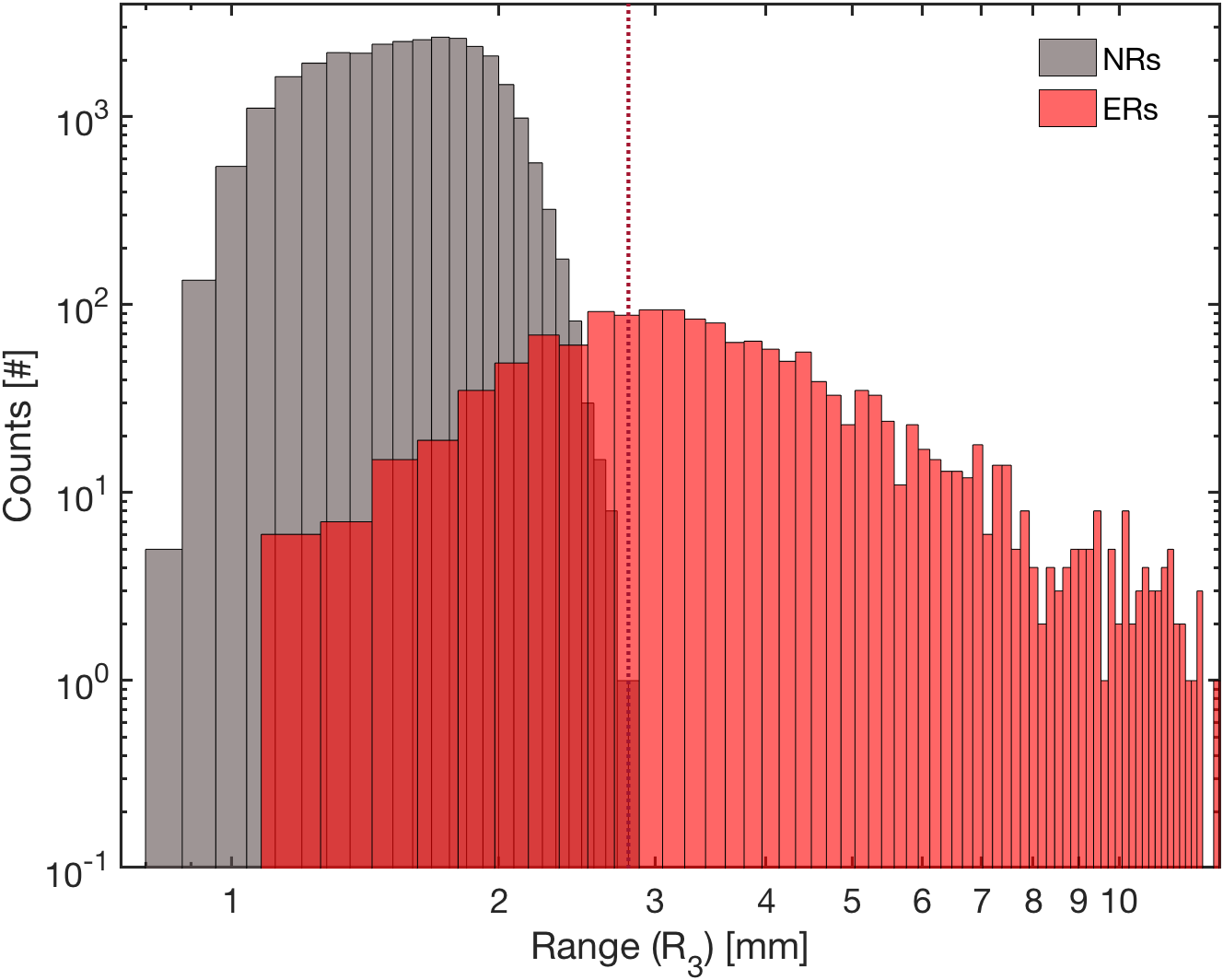}}
\caption{Distribution of 3D range in the 5--10~keV$_{\mathrm{ee}}$ ROI shown in Fig.~\protect{\ref{Fig:r2}}. A cut at $R_3$ = 2.80~mm gives a leakage of 1 background events per million NR tracks, with a corresponding acceptance of ER signals of 87\% (see text for details).}
\label{Fig:skew_gauss_back}
\end{figure}

The average leakage of NR events into the wide ER band was estimated from Fig.~\ref{Fig:skew_gauss_back}, which shows the 3D range distribution of NR and ER tracks across the full ROI. This estimate is derived by simply projecting the events in Fig.~\ref{Fig:r2} onto the $R_3$ axis; a more careful analysis, which corrects for the rise and larger scatter of the NR band at higher energies, would provide a larger acceptance of the ERs. We find that 1~NR event lies above $R_3$ = 2.80~mm, with a corresponding acceptance of ER signals of 87\%. We note that the 25\% of false-negative events found in the data challenge mentioned in Section~\ref{SS:Acceptance} are likely to have a strong overlap with events removed by the $R_3$ cut, and hence we do not consider this as an additional source of inefficiency. In summary, this calculation motivates a total of $\mathcal{O}$(1)~event per million NR tracks in the background tally of Table~\ref{Tab:bk1}.

Since the probability for secondary-recoil generation within the ROI increases towards the end of the track (cf.~Table~\ref{Tab:recrates}), it is worth considering whether `backward neutrons' -- e.g.~back-scattered by the shielding or emitted in ($n$,$2n$) reactions -- could contribute significantly to this background by creating tracks forking near the presumed vertex. As mentioned in Section~\ref{S:NeutronSource}, the backward neutron flux is of order 1\% for neutron energies above 100~keV, so this would be a small contribution to this background even assuming no head-tail discrimination at all.

In conclusion, the number of background events in the signal ROI from the extensive number of sources considered is small -- just over 1~event per million primary NR tracks -- yielding very high signal-to-background ratios for both generators with CF$_4$ at 50~Torr. The dominant background contributions come from the interaction of $\gamma$-ray from inelastic neutron scattering near the NR track vertex, which we assessed realistically via detailed simulations, plus a comparable contribution from secondary nuclear recoils mimicking Migdal electrons; the latter calculation has higher uncertainly as it depends significantly on the quality of the track analyses. Finally, it should be noted that there are various processes producing backgrounds below the ER threshold which may come into the ROI for heavier elements: detailed assessments will be required for other gas mixtures.

\subsection{Sensitivity}
\label{SS:Sensitivity}

From the knowledge of the signal and background rates and their uncertainties we may calculate the expected (median) discovery significance for the D-D and D-T experiments as a function of exposure to the neutron beams: this quantifies how well one can reject the background-only hypothesis assuming the nominal signal hypothesis (the Migdal cross sections calculated in Ref.~\cite{MIGDALth}).

Sophisticated statistical analysis techniques will be employed for robust signal estimation, but a straightforward calculation applicable to counting experiments will suffice here, this one derived from a Profile Likelihood Ratio (PLR) test using the Asimov dataset approximation~\cite{Cowan1998}. We consider initially an uncertainty of 50\% on the background rate, assumed to be Gaussian distributed. This reflects our ability to determine this rate using ancillary measurements, complemented by Monte Carlo where needed: for example, by measuring the rates of lone ER and NR tracks to assess random coincidence backgrounds, or analyse ER vertex locations as a function of distance to the NR vertex to confirm photon interaction rates near the NR track.

From the signal side, no theory uncertainty has been considered in this calculation, and neither did we evaluate the uncertainty on our Migdal event acceptance, as this depends critically on the details of the track analysis. We recognise that the latter may be significant, recalling that, as indicated in Table~\ref{Tab:nrates}, variations in track threshold of 1~mm yield factor of 2 differences in the signal yield. However, given the large signal-to-background ratios expected this will not hinder the measurement significantly.

Under these assumptions we conclude that both experiments have excellent discovery potential, achieving 5$\sigma$ median significance in less than one day of operation: 20~hours for D-D and 4.4~hours for D-T. A more pessimistic scenario, where we halve the signal rate and increase the background uncertainty to 70\%, yields a 5$\sigma$ discovery in a little over 7~calendar days for the D-D experiment, while in the D-T case only 7~hours are required to reach that significance.

In summary, we expect to make conclusive detections with both generators with a 5-day run in the baseline scenario, even assuming the restrictive parameters considered here: analysis of camera frames with single interactions fully contained within a restricted fiducial region, and imposing a conservatively high electron threshold of 5~keV.

\section{Migdal in other elements}
\label{S:Gases}

Measuring Migdal probabilities for different elements, with their particular electron configurations, will be important to study fully this effect -- and indeed critical should the measured rates disagree with the theoretical predictions. In future stages of the project we plan to explore CF$_4$-based mixtures with noble gases, from He to Xe, which include leading elements used in direct dark matter detection. Krypton may be considered too despite its radioactive isotopes, which do not pose an insurmountable challenge. Other elements of interest, e.g.~Si and Ge used in cryogenic bolometers, also form gaseous compounds, e.g.~as nonpolar tetrafluorides which have similar electron transport properties to CF$_4$; however, the dominant contribution of fluorine to the Migdal rate (due to stoichiometry) must be overcome. Other promising group-IV compounds include mono-silane (SiH$_4$) and germane (GeH$_4$), which are expected to have similar properties to methane, a well-known quencher. It must be noted that the Migdal effect may be subtly different in dense liquids and solid materials, but our measurements in atomic and molecular species will provide a sound basis for further study. Table~\ref{Tab:elements} lists neutron cross sections and calculated Migdal probabilities for these elements~\cite{MIGDALth}. With the exception of He, the probabilities for emission of a hard Migdal electron are comparable across the elements considered. Without undertaking a more detailed study for each element, this provides some initial confidence that the signal rate in other gases should be comparable to the CF$_4$ rate, which we have found to be eminently measurable.

\begin{table*}[t]
\caption{Neutron scattering cross sections (mb) at 2.47~MeV and 14.7~MeV from ENDF/B-VIII.0~\cite{Brown2018} -- total ($\sigma_0$) and bare-recoil processes ($\sigma_s$) are shown for the leading isotope; a weighted average is given where natural abundance is indicated; also given are the Migdal probabilities for the full neutron-induced NR spectrum for each element, integrated down to zero NR threshold for electron detection thresholds of 0.5~keV and 5~keV~\protect{\cite{MIGDALth}}.}
\vspace{5pt}
\centering
{\small
\setlength{\tabcolsep}{5pt}
\renewcommand{\arraystretch}{1.3}
\begin{tabular}{r | r r r r | r r r r }
\hline
& \multicolumn{4}{c|}{2.47~MeV (D-D)} & \multicolumn{4}{c}{14.7~MeV (D-T)} \\
\hline
                    & $\sigma_0$, mb      & $\sigma_s$, mb    & P($>$0.5~keV)   & P($>$5~keV) 
                    & $\sigma_0$, mb      & $\sigma_s$, mb    & P($>$0.5~keV)   & P($>$5~keV) 
                    \\
\hline
$^4$He     & 3,239 & 3,239 & $2.98\!\times\!10^{-3}$ & $4.29\!\times\!10^{-7}$ 
           & 1,017 & 1,017 & $9.01\!\times\!10^{-2}$ & $2.48\!\times\!10^{-6}$ \\ 
$^{12}$C   & 1,613 & 1,613 & $6.01\!\times\!10^{-3}$ & $1.45\!\times\!10^{-5}$ 
           & 1,379 & 1,321 & $2.15\!\times\!10^{-2}$ & $4.09\!\times\!10^{-5}$ \\ 
$^{19}$F   & 3,038 & 3,038 & $2.81\!\times\!10^{-3}$ & $2.01\!\times\!10^{-5}$ 
           & 1,786 & 1,272 & $9.95\!\times\!10^{-3}$ & $6.50\!\times\!10^{-5}$ \\ 
$^{nat}$Ne & 2,474 & 2,465 & $2.62\!\times\!10^{-3}$ & $2.32\!\times\!10^{-5}$ 
           & 1,677 & 1,055 & $8.50\!\times\!10^{-3}$ & $6.89\!\times\!10^{-5}$ \\ 
$^{nat}$Si & 3,111 & 3,111 & $2.39\!\times\!10^{-3}$ & $2.87\!\times\!10^{-5}$ 
           & 1,725 & 1,150 & $1.10\!\times\!10^{-2}$ & $1.25\!\times\!10^{-4}$ \\ 
$^{40}$Ar  & 5,050 & 5,050 & $2.18\!\times\!10^{-3}$ & $2.92\!\times\!10^{-5}$ 
           & 2,818 & 2,754 & $6.85\!\times\!10^{-3}$ & $8.94\!\times\!10^{-5}$ \\
$^{nat}$Ge & 3,401 & 3,401 & $1.64\!\times\!10^{-3}$ & $2.46\!\times\!10^{-5}$ 
           & 3,227 & 3,130 & $5.47\!\times\!10^{-3}$ & $8.12\!\times\!10^{-5}$ \\
$^{nat}$Kr & 3,825 & 3,825 & $1.56\!\times\!10^{-3}$ & $2.37\!\times\!10^{-5}$ 
           & 3,741 & 3,717 & $4.65\!\times\!10^{-3}$ & $7.03\!\times\!10^{-5}$ \\
$^{nat}$Xe & 5,760 & 5,760 & $7.31\!\times\!10^{-4}$ & $1.55\!\times\!10^{-5}$ 
           & 4,871 & 4,861 & $2.80\!\times\!10^{-3}$ & $5.95\!\times\!10^{-5}$ \\
\hline
\end{tabular}
}
\label{Tab:elements}
\end{table*}

Here we discuss briefly the noble element mixtures with CF$_4$, assuming that the Migdal probabilities for F and C will have been measured previously within some small uncertainty -- CF$_4$ is likely always needed for its visible luminescence spectrum. We conclude with a brief discussion on gases involving the group-IV elements Si and Ge.

\subsection{Binary mixtures with noble elements}

\begin{figure}[t]
\centerline{\includegraphics[width=1.\linewidth]{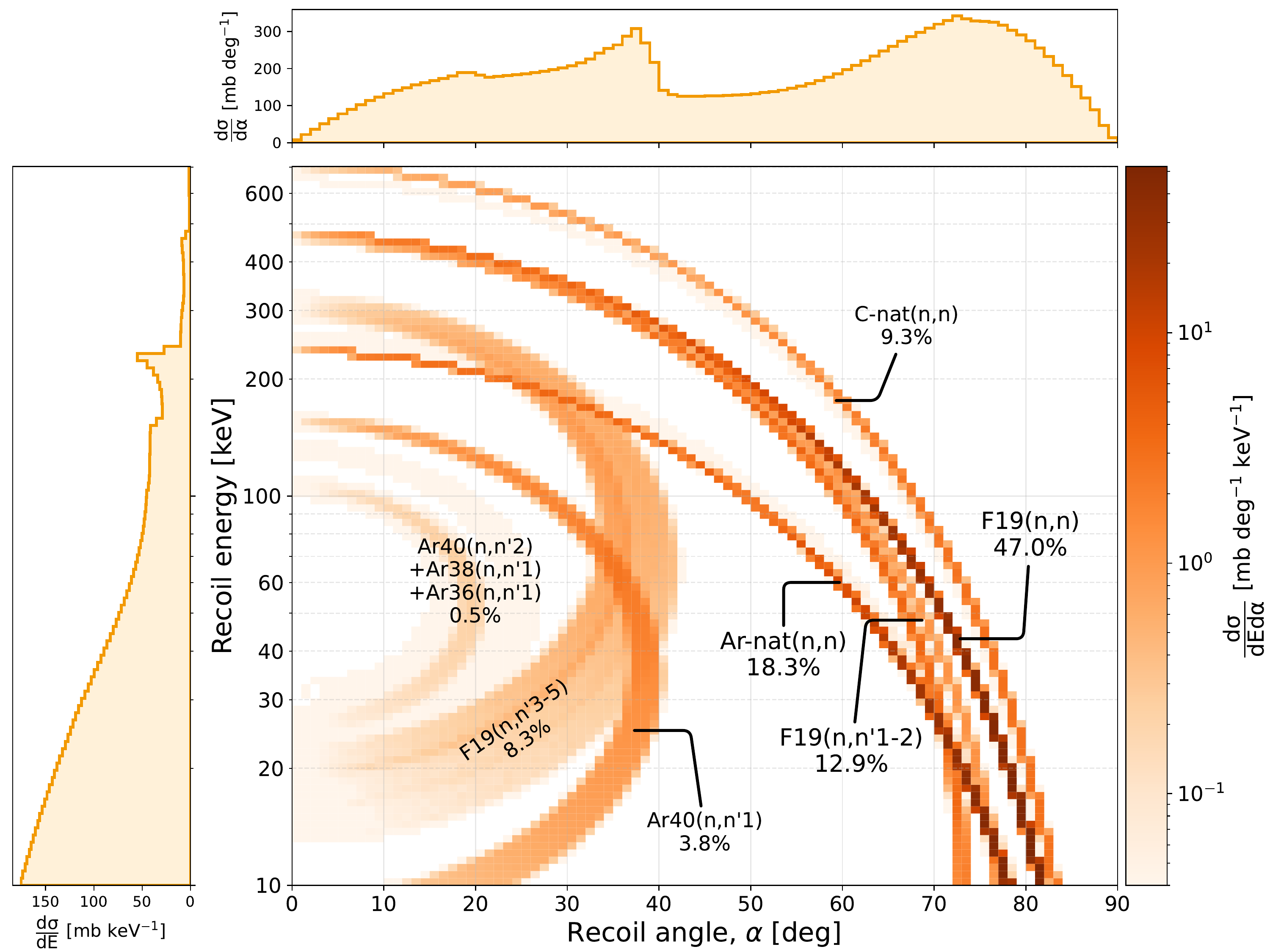}}
\caption{Doubly-differential cross sections for D-D neutron scattering from a 50\% Ar/CF$_4$ mixture, calculated with GEANT4. The fraction of each signal-inducing scattering processes is indicated; the recoil energy spectrum and recoil angle projections are also shown.}
\label{Fig:ArCF4}
\end{figure}

Different experimental approaches may be needed depending on the mass of the noble element. Since Ne and F have similar atomic weights and predicted Migdal rates, the measurement must rely on subtracting the CF$_4$ contribution across the NR spectrum; this may be challenging due to the abundance of fluorine in any viable binary mixture. If instead the noble species is either lighter or heavier than fluorine, one may be able to utilise NR spectral information to improve the search sensitivity for that particular element. Further, an analysis using the correlated NR energy-angle information (cf.~\ref{A:NeutronScattering}) may be able to identify particular regions of that 2D parameter space where the Migdal yield from the noble species is most distinct from that of CF$_4$. One such example is shown in Fig.~\ref{Fig:ArCF4} for a 50\% Ar/CF$_4$ mixture.

Mixture-specific backgrounds need to be evaluated in each case. In general, K-shell energies of the elements reach the 5~keV electron threshold when $Z\!\ge\,$20, which may render the detection of the Migdal effect in elements heavier than Ar inherently more difficult at our high NR energies. Some backgrounds are closely related to the atomic physics, and we note that both the Migdal probability and the photon interaction probability both follow the photoelectric cross section~\cite{Liu2020}, so we expect some backgrounds to remain in proportion to the signal. Other backgrounds are driven by the nuclear physics instead, and here the complexity of heavier nuclei and the richer isotopic composition of the heavier elements makes neutron scattering more complex.

However, heavier species may bring some benefits. It has been mentioned previously that the Auger emission in light elements such as C and F does not protrude out of the NR-track penumbra, and hence it will not interfere with either the signal or the background in a measurement with pure CF$_4$. However, in Ar there is an interesting effect: the dominant Migdal electron with energy above 5~keV comes from the $n=1$ shell, and the KLL Auger yield is high ($\sim$90\%) and relatively energetic itself ($\sim$2.6~keV~\cite{Puttner2020}) -- hence it is expected that Migdal events will contain not one, but two visible electrons. Even in Xe, where the highest Migdal probability for electron emission above 5~keV comes from the $n=2$ level, LMM Auger emission dominates the ensuing deexcitation, with energy $\sim$3--4~keV~\cite{Kovalik1998}. Such Auger Electron Spectroscopy can in fact assist with the identification of the Migdal effect in some elements. It should be noted that multiple Migdal ionisation must also be considered in this context, as this will lead to more complex deexcitation signatures than in pure CF$_4$.

An alternative strategy to explore the heavier elements is to operate at much higher pressures and forego the direct identification of the Migdal electron, in favour of detecting the accompanying atomic X-ray fluorescence in the OTPC -- as proposed in Ref.~\cite{Nakamura:2020kex}. In this case the search is for the (sub-dominant) $n=1$ electron, where X-ray emission dominates the ensuing atomic deexcitation. We may later explore this avenue, after a full reevaluation of ER backgrounds which will be much more severe in this regime.

At the other end, He is essentially transparent to photons in this regime, and the collisional energy losses for electrons and ions are also very small in helium: hence, `backfilling’ a low CF$_4$ partial pressure with He to near ambient pressure does not affect the particle interaction properties of the mixture significantly, and the measurement can proceed much as that in low-pressure gas. Increasing the helium pressure will also help to overcome the lower Migdal probability of He relative to other elements (cf.~Table~\ref{Tab:elements}). Other systems are well placed to measure the Migdal effect in this element~\cite{Amaro2022}.

In each case a significant number of technical parameters must be considered in the search for an optimal total pressure and relative gas composition. Firstly, operating conditions must yield long enough ER and NR tracks to allow identification of the Migdal topology and associated energies. From DEGRAD and SRIM simulations we conclude that, at constant total pressure, the electron range increases with the relative molar fraction of He, Ne, Ar; it is reasonably constant with Kr concentration; and it decreases somewhat for Xe. The noble-element NR tracks increase with concentration, but here we note that the D-D neutron endpoint energy does not reach the nominal 4~mm threshold for Kr and Xe recoils at 50~Torr; in both cases the noble-element NR track will be $\lesssim$1~mm.

Secondly, the transport properties of the mixture must be favourable; in the noble elements the dominance of elastic scattering at low energies will slow down electron transport and increase diffusion compared to molecular species, the latter presenting additional energy loss mechanisms that cool the drifting electrons; controlling diffusion argues for higher pressures, in tension with seeking long tracks.

Next, the luminescence properties of these mixtures (spectrum and yield) must be considered, along with our ability to detect this light in the camera and PMT systems. The scintillation mechanisms of the noble gases~\cite{AprileBook} are quite distinct from those of CF$_4$~\cite{vanSprang1978,Washida1983}, and they interact with one another in various ways (e.g.~photoabsorption and photoionisation, excimer quenching, Penning transfer), leading to a variety of outcomes depending on the actual mixture. In all cases CF$_4$ is still likely to be needed for its visible-wavelength scintillation.

Finally, sufficient GEM gain must be stably achieved to allow identification of the Migdal topology. In this context the addition of quencher gases deserves consideration.

In conclusion, detecting the Migdal effect in the noble elements will bring various challenges, and detailed studies are needed to identify suitable conditions for each measurement -- alongside dedicated background simulations for each case. Fortunately, there is a growing body of work related to these mixtures, e.g.~He/CF$_4$~\cite{Fraga2003b,Baracchini2000a,Baracchini2000b}, Ne/CF$_4$~\cite{Coimbra2012,Fujiwara2018}, Ar/CF$_4$~\cite{Fraga2003b,Margato2004,Fujiwara2016b,Snyder2018}, Kr/CF$_4$~\cite{Fujiwara2017} and Xe/CF$_4$~\cite{Ovchinnikov2013,Henriques2019}. 

\subsection{Group-IV compounds}

Few compounds of Si and Ge exist that are gaseous at room temperature. Two exceptions are the tetrafluorides (SiF$_4$, GeF$_4$) and the tetrahydrides (SiH$_4$ and GeH$_4$), the latter more commonly known as silane and germane. Unfortunately, some of these are highly toxic, and all are environmentally damaging, though the quantities involved would be small. The tetrafluorides are nonpolar compounds of the very electronegative fluorine (Si and Ge are somewhat less electronegative than C), and we expect some commonalities with CF$_4$ for electron transport and other properties that would make them good detector media. As in CF$_4$~\cite{Christophorou1996}, electronic excited states are mostly dissociative, and dissociative electron attachment plays a key role in determining several properties in these compounds~\cite{Bjarnason2013}. In turn, the nonpolar tetrahydrides have low overall electronegativity and we expect similar properties to those of methane, a well-known quencher gas: lower gain, low diffusion, no significant electron attachment, but also the absence of useful scintillation. These four gases could yield reasonable Migdal signals in Si or Ge, especially the tetrahydrides as the absolute Migdal rates from hydrogen would be very small; in the tetrafluorides, the high atomic number of Ge compensates in part for the low stoichiometry in GeF$_4$, as shown in Table~\ref{Tab:elements}.

The fluorescence of SiF$_4$ and GeF$_4$ has much in common with that observed in CF$_4$~\cite{Biehl1997,Boyle1998}, but we are not aware of the precise emission spectra and total yields in response to energetic particles or from electroluminescence; hence, measurements are needed to determine if an additional partial pressure of CF$_4$ is required to provide sufficient sensitivity for the PMT and camera systems. This is likely to be the case for GeF$_4$ since emission lies mostly in the UV region~\cite{Boyle1998}. If CF$_4$ is anyway needed, one might consider silane and germane instead: although these fluoresce partly in the visible region~\cite{Suto1986,Mitchell1987}, their quantum yields are low (this is also the case for methane). Therefore, addition of CF$_4$ or other scintillating gas would be likely needed.

\section{Conclusion}
\label{S:Conclusion}

In this article we detailed the design and expected performance of the MIGDAL experiment, a new project aiming to measure the atomic Migdal effect in nuclear scattering induced by fast neutrons. This phenomenon is regularly invoked by direct dark matter detection collaborations to extend the reach of their experiments to lower particle masses -- but there has been, to date, no experimental confirmation of this effect in nuclear scattering.

The MIGDAL detector employs an optical time projection chamber operating with low-pressure CF$_4$ gas, which will image particle tracks initiated by an intense neutron beam. We show that the Migdal topology, consisting of an electron and a nuclear recoil sharing the same vertex, should be readily identifiable above the nominal 5~keV (4~mm track length) electron threshold. We evaluated the performance of the design through detailed calculations of signal and background and simulated detector response, and demonstrate that it should be able to detect 8.9 (29.3) Migdal events per calendar day with the D-D (D-T) neutron generators available to us -- resulting in a conclusive detection of this effect in fluorine.

The initial experiment in pure CF$_4$ will be extended to other CF$_4$-based mixtures, especially those including the key elements of interest for dark matter searches. Such a programme will study systematically how the Migdal rates vary with atomic/molecular species in the distinct energy regimes probed by the two neutron generators. Although the MIGDAL experiment will not reach the lowest nuclear recoil energies where this effect brings a significant increase in sensitivity to dark matter experiments, it is important to establish that the quantum mechanical calculations are correct in the higher-energy regime -- where the dipole approximation breaks down and multiple ionisations are possible (and in some cases dominant) -- and probe for observable differences between emission from atoms and molecules.

Other measurements are being pursued elsewhere using gaseous, liquid and solid detector media. The uniqueness of our approach is the direct imaging and measurement of the Migdal electron track, which has the potential for a very clear identification of this topology with low systematic uncertainty -- most other efforts are aiming to detect only the combined energy from the Migdal electron (plus its binding energy) along with the nuclear recoil energy, while yet other experiments are targeting the resolved detection of X-ray fluorescence accompanying Migdal emission. The low density of our active detector medium means that a low background measurement is achievable in a surface laboratory.

At the time of writing the MIGDAL experiment is entering commissioning and a first deployment at the NILE neutron facility is expected soon.


\section*{Data access statement}
No experimental datasets were generated by this research; all simulated data were produced using standard software packages as described in the text.

\section*{Acknowledgements}

\begin{sloppypar}
This work has been supported by the UKRI's Science \& Technology Facilities Council through the Xenon Futures R\&D programme (awards ST/T005823/1, ST/T005882/1, ST/V001833/1, ST/V001876/1), Consolidated Grants (ST/S000739/1, ST/T000759/1), CM's Ernest Rutherford Fellowship (ST/N004663/1) and TM's PhD scholarship (ST/T505894/1); by the U.S. Department of Energy, Office of Science, Office of High Energy Physics, under Award Number DE-SC0022357; by the Portuguese Foundation for Science and Technology (FCT) under award number PTDC/FIS-PAR/2831/2020; and by the European Union’s Horizon 2020 research and innovation programme under the Marie Sk\l{}odowska-Curie grant agreement No. 841261 (DarkSphere) and No. 101026519 (GaGARin). ELA acknowledges the support from Spanish grant CA3/RSUE/2021-00827, funded by Ministerio de Universidades, Plan de Recuperacion, Transformacion y Resiliencia, and Universidad Autonoma de Madrid. We are grateful to the Particle Physics Department at RAL for significant additional support which made this project possible. Thanks are also due to the CERN RD51 collaboration for their support through Common Project funds, hardware tests and training, and useful discussions. We would also like to thank the ISIS facility for technical assistance and for hosting this experiment. We thank Master's students E.~Brookes (RHUL) and L.~Shanahan (Imperial), as well as summer students M.~Lau and I.~Andreou (Imperial), E.~Zammit-Lonardelli, M.~Collier, C.~Jolly and M.~Handley (RAL) for their contributions. Thanks are also due to D.~Parker (Imperial HEP Electronics Workshop). We are grateful to S.~Biagi (U.~Liverpool) for his assistance with Degrad; and to P.~Cox, M.~Dolan and H.~Quiney (U.~Melbourne) for discussions on the atomic theory underlying the Migdal effect. 

For the purpose of open access, the authors have applied a Creative Commons Attribution (CC BY) licence to any Author Accepted Manuscript version arising from this submission.
\end{sloppypar}


\FloatBarrier

\def\appendixname{Appendix\!}
\appendix

\section{Nuclear recoils in neutron scattering}
\label{A:NeutronScattering}

A description of neutron scattering kinematics focused on the parameters of the recoiling nucleus is useful for Migdal signal calculations, to validate Monte Carlo tools and simulations, and for analysing experimental data. The equations may be readily derived from the classical theory of binary collisions (e.g.~\cite{goldstein1980}), but they usually appear in the literature with a focus on the neutron parameters rather than those of the residual nucleus. These relations are required to relate the neutron cross sections tabulated in the neutron data libraries to the properties of the recoiling nucleus.

In this appendix we describe kinematic relations for the four processes likely to be of interest in the search for Migdal events; these feature only neutral products in the final state, potentially yielding a clean Migdal `vertex': elastic ($n,n$) and inelastic $(n,n^\prime)$ scattering, radiative capture ($n$,$\gamma$) and ($n$,$2n$) threshold reactions. We present below formulas for non-relativistic neutron scattering solved for nuclear recoil energy ($E_r$) and nuclear recoil angle ($\alpha$) in the laboratory and centre-of-mass frames, expressed as a function of neutron scattering angle in the laboratory frame ($\theta$). Both angles are measured with respect to the incident neutron direction. Relations between angular cross sections for the outgoing particles are also given.

In the following we consider D-D and D-T neutrons with kinetic energies of 2.47~MeV and 14.7~MeV, respectively, noting that both the mean energy and the spectral width depend somewhat on the operating parameters of the generator as well as on its orientation as discussed previously. The non-relativistic approximation is sensible up to D-T neutron energies. The target nuclei considered below include C and F but also Ar, as an example of a heavier species with a richer nuclear excitation spectrum.

\medskip
\subsection*{Elastic scattering}

For elastic scattering the energy of the nuclear recoil as a function of neutron scattering angle $\theta$ in the laboratory frame is given by:
\begin{eqnarray}
\label{Er,el}
E_r &=& \frac{E_n}{2}\frac{4mM}{(m+M)^2} \times \\*
    & & \left[1-\cos\theta\sqrt{1-\left(\frac{m}{M}\sin\theta\right)^2}
    +\frac{m}{M}\sin^2\theta\right]\,, \nonumber
\end{eqnarray}
where $E_n$ is the energy of the incident neutron, and $m$ and $M$ are the neutron and atomic masses, respectively. For D-T neutrons, the maximum recoil energies are 4.21~MeV for $^{12}$C, 2.82~MeV for $^{19}$F and 1.41~MeV for $^{40}$Ar. For a heavy target, when the laboratory and centre-of-mass frames approximately coincide, this reduces to:
\begin{equation}
E_r \approx \frac{E_n}{2}\frac{4mM}{(m+M)^2}
\left(1-\cos\theta\right)\,.
\end{equation}
This approximation can be poor for light targets, deviating up to 19\% for $^{12}$C. It is worth noting that the endpoint of the recoil spectrum given by these equations is $\sim$0.5\% smaller than that obtained using the relativistic calculation for D-T neutrons.

The recoil energy may be written as a function of recoil angle $\alpha$ in the (exact) form:
\begin{equation}
E_r = E_n \frac{4mM}{(m+M)^2}\cos^2\alpha\,.
\end{equation}

The relation between the neutron and recoil angles in the laboratory frame is given by:
\begin{equation}
\cos 2\alpha = \frac{m}{M} \sin^2 \theta
-\cos\theta\sqrt{1-\left(\frac{m}{M}\sin\theta\right)^2 }\,.
\label{alpha}
\end{equation}
In the heavy target approximation this reduces to:
\begin{equation}
\cos\alpha \approx \sqrt{\frac{1-\cos\theta}{2}}\,.
\end{equation}
The inverse relation to (\ref{alpha}) may be written as:
\begin{equation}
\cos\theta = \frac{m/M-\cos2\alpha} {\sqrt{(m/M+1)^2-4m/M\cos^2\alpha}}\,.
\label{theta}
\end{equation}

\medskip
\subsection*{Inelastic scattering}

For $(n,n^\prime)$ inelastic scattering reactions, with the nucleus excited to energy level $\epsilon$, the energy of the {\em excited nucleus} before deexcitation is given by:
\begin{eqnarray}
E^\star_r &=& \!\!\frac{E_n}{2}\frac{4mM}{(m+M)^2} \times \\
&& \!\!\left[ 1 - \cos\theta\sqrt{1-\!\left(\frac{m}{M}\sin\theta \right)^2
\!- \!\frac{\epsilon}{E_n}\left(1+\frac{m}{M}\right)} \right. \nonumber \\
&& \!\! + \left. \frac{m}{M}\sin^2\theta-\frac{1}{2}\frac{\epsilon}{E_n}\left(1+\frac{m}{M}\right) \right]\,, \nonumber
\label{Er,inel}
\end{eqnarray}
where the nuclear recoil mass has been approximated by the ground state value. For zero excitation energy this reduces to the elastic scattering formula (\ref{Er,el}).

For D-T neutrons the main excited state in $^{12}$C has $\epsilon=4.439$~MeV. The heavier $^{19}$F nucleus has various levels up to 5~MeV, with important ones at 0.197~MeV and 1.554~MeV. In $^{40}$Ar, scattering via the `continuum' of highly excited states dominates by a large factor for D-T neutrons, whereas the first level at 1.461~MeV is the only important one for D-D neutrons. Upon decay to the ground state with emission of a single $\gamma$-ray, the final energy of the nuclear recoil becomes:
\begin{equation}
E_r = E^\star_r + \frac{1}{2}\frac{\epsilon^2}{Mc^2}
- \sqrt{\frac{2E^\star_r}{Mc^2}}\epsilon\cos{\beta}\,,
\label{gammarecoil}
\end{equation}
where $\beta$ is the angle between the excited recoil direction and the emitted photon. This may be expressed in terms of the photon angles relative to the beam direction:
\begin{equation}
\cos{\beta} = \cos\alpha^\star\cos\psi+\sin\alpha^\star\sin\psi\cos\phi \,,
\end{equation}
where $\alpha^\star$ is the recoil angle before deexcitation, and $\psi$ and $\phi$ are the $\gamma$-ray angles projected onto and out of the scattering plane, respectively. More generally, deexcitation of the higher energy levels occurs via the emission of several $\gamma$-rays, which may be accounted for separately, including any angular correlations and other anisotropies.

For D-T neutrons incident on the targets of interest, the second term in (\ref{gammarecoil}) is essentially negligible (2\% correction to $E_r$ for forward scattering), but the recoil energy smearing introduced by the $\gamma$-ray emission is significant for low neutron scattering angles, when $E^\star_r$ is small: the amplitude of the third term represents $\approx$25\% of the recoil energy at zero scattering angle for the leading excited states in all targets, decreasing to a few percent above $\theta=45{^\mathrm{o}}$ for the heavier targets; for the lighter $^{12}$C it is still around 7\% at that angle -- so there is appreciable energy smearing in this case.

A key feature of inelastic scattering is the existence of a maximum recoil angle  $\alpha_{0}\!<\!90^{\rm o}$ given by:
\begin{equation}
\cos\alpha_{0}= \sqrt{\frac{\epsilon}{E_n}\left(1+\frac{m}{M}\right)}\,,
\label{alpha_max}
\end{equation}
and $E_r(\alpha)$ is in fact double-valued below that angle. Using this definition we may express the recoil energy as a function of recoil angle in the following way:
\begin{eqnarray}
\label{Er,inel_alpha}
E^\star_r &=& \!\!\!\frac{E_n}{2}\frac{4mM}{(m+M)^2} \times \\*
  & & \!\!\!\cos^2\alpha \! \left[1 \pm \sqrt{1\! -\! \left(\frac{\cos\alpha_{0}}{\cos\alpha}\right)^2}\!
    - \!\frac{1}{2}\left(\frac{\cos\alpha_{0}}{\cos\alpha}\right)^2\right]. \nonumber
\end{eqnarray}
It is interesting to note, both from this equation and from (\ref{Er,inel}), that there is a (small) minimum recoil energy that occurs for zero neutron angle and zero recoil angle (taking the minus sign).

To obtain an explicit relation between the neutron and recoil angles, equation (\ref{theta}) may be used as a reasonable approximation for inelastic scattering, valid to $\sim$2\% for the targets of interest, and the inverse relation (\ref{alpha}) is equally suitable up to $\alpha_0$. An accurate expression may be obtained by introducing a kinematic factor $\gamma$, defined as the ratio of velocities of the centre-of-mass frame to that of the outgoing neutron in that frame. This takes on a particularly simple form for elastic scattering: $\gamma=m/M$; for inelastic scattering via energy level $\epsilon$, we have instead:
\begin{equation}
\gamma=\frac{m/M}{\sqrt{1-{\epsilon}/{E_n}(1+{m}/{M})}} 
= \frac{m/M}{\sin\alpha_0} \,.
\label{gamma}
\end{equation}
This expression is accurate when $M$ represents the mass of the excited nucleus, rather than that of the ground state. So, (\ref{theta}) may be used for inelastic scattering by replacing the $m/M$ factor by the full expression for $\gamma$ (several of the above equations for inelastic scattering are often found in the literature using this $\gamma$ factor).

This calculation of the recoil angle ignored the momentum imparted by $\gamma$-ray emission following the binary collision. This smearing effect is significant for light targets with energetic transitions, but probably not resolvable within the experimental resolution in most cases. For each $\gamma$-ray, the final recoil angle is given by:
\begin{equation}
\cos\alpha = \frac{\sqrt{2Mc^2E^\star}\cos\alpha^\star-\epsilon\cos\psi}
{\sqrt{2Mc^2E_r}}\,.
\label{inelastic_gammas}
\end{equation}
Maximal deflection occurs for perpendicular emission in the scattering plane.
For the $\epsilon=4.4$~MeV state in $^{12}$C it introduces a maximum smearing of $\pm3^\mathrm{o}$ near $\alpha_0$.

In conclusion, the inelastic scattering via particular nuclear levels can be calculated accurately from neutron cross section data given in the data libraries. Derivation of the Migdal emission for these cases may be important as they may produce recognisable features in the data owing to their distinct angular distribution ($\alpha_0<90^\mathrm{o}$). However, this becomes more challenging for isotopes exhibiting a large density of states or for scattering off the continuum of unresolved levels at higher energies -- or indeed when many isotopes are present at natural abundance. In these instances it is sensible to employ Monte Carlo tools such as GEANT4 to calculate these recoils.

\medskip
\subsection*{Radiative capture}

The capture of MeV neutrons produces sizeable recoil energies, but the cross section is very small (cf.~Table~\ref{Tab:nxs}). We extend our discussion to this case since it is likely that some capture events will be recorded in a Migdal search dataset clustering at recoil angles $\alpha \simeq 0$. The recoil energy of the product nucleus is obtained from momentum conservation:
\begin{equation}
E^\star_r = E_n\frac{m}{M'^*} \approx E_n\frac{m}{M'} \approx\frac{E_n}{A+1} \,,
\label{capturerecoil}
\end{equation}
where $M'^*$ is the mass of the excited compound nucleus, $M'$ is its ground state mass, and $A$ is the atomic mass of the target species. This amounts to $E_r \sim 1$~MeV for the $^{13}$C recoil from D-T neutron capture on $^{12}$C.

The subsequent $\gamma$-ray emission smears both the recoil energy and angle, as above.
Typically, this involves several $\gamma$-rays even for thermal neutron capture (e.g.~6 distinct energies from $^{13}$C deexcitation, and 168 from $^{20}$F~\cite{Reedy2002}), and a larger number still for the case of fast neutrons. The excitation energy of the compound nucleus is equal to $(M+m-M')c^2+E_n$; therefore, for D-T neutrons a total of $\sim$20~MeV will appear in the $\gamma$-ray cascade. The recoil energy change induced by a single $\gamma$-ray may be calculated from (\ref{gammarecoil}) using the product mass $M'$.

The smearing of recoil angle around $\alpha=0$ may be calculated from (\ref{inelastic_gammas}), replacing $M$ by $M'$. The maximal smearing becomes: 
\begin{equation}
\sin\alpha_{max} = \frac{\epsilon}{\sqrt{2M'c^2E_r}}\,.
\label{capture}
\end{equation}
In this instance the full excitation energy should be used in the calculation as this limiting case corresponds to all photons being emitted in the same direction. For D-D neutrons this defines a cone with 6--7$^\mathrm{o}$ half-angle around the beam direction for the species of interest.

\medskip
\subsection*{Threshold reactions}

Threshold reactions are important for D-T neutrons, and in particular ($n$,$2n$) reactions can produce detectable Migdal events; in some instances these reactions involve also $\gamma$-ray emission. The threshold for ($n$,$3n$) reactions lies above D-T energies for all targets. Other important threshold reactions for D-T neutrons include ($n$,$\alpha$) and ($n$,$n\alpha$), followed by ($n$,$p$) and ($n$,$np$).

It is not possible to express the kinematic parameters of the 3-body final state in closed form, and there may exist angular correlations between the two indistinguishable neutrons in this case. Hence, data on the properties of the residual nucleus are scarce (although they are important in some applications related to radiation damage). Energy-differential cross sections can be found in the neutron data libraries (File~6), with the residual nucleus considered to be emitted isotropically in the laboratory frame.

\medskip
\subsection*{Angular cross sections}

Differential angular cross sections for the various neutron reactions can be retrieved from the nuclear reaction data libraries in the ENDF~6 format~\cite{ENDF6} (File~4 for 2-body reactions, File~6 for more complex reactions requiring energy-angle correlation); these are typically expressed in terms of centre-of-mass frame coordinates. It should be noted that some plotting tools automatically convert these to the laboratory frame (although this may not be clearly indicated). It is convenient to use the centre-of-mass frame for the purpose of parameterising the angular cross sections or for calculating the Migdal signal, and converting to the laboratory frame is then necessary to relate to the experimental observables.

Elastic and inelastic angular cross sections are often parameterised by Legendre polynomials of order $L$ valid in the centre-of-mass frame, with coefficients $\alpha_l$ given at fixed incident energy $E_n$ in the laboratory frame:
\begin{equation}
\left(\frac{d\sigma}{d\Omega_1}\right)_{\!c}=\frac{\sigma_0}{2\pi}
\sum_{l=0}^L\frac{2l+1}{2}{\alpha_lP_{l}(\cos\theta_c)}\,,
\label{legendre}
\end{equation}
where $\theta_c$ is the neutron scattering angle in the centre-of-mass frame and $\sigma_0$ is the total cross section for the process at the relevant energy.

To convert the angular cross section to the laboratory frame we require the relationship between the neutron scattering angles in the two frames: 
\begin{equation}
\cos\theta 
= \frac{\gamma+\cos\theta_c}{\sqrt{1+\gamma^2+2\gamma\cos\theta_c}}\,.
\label{thetaconv}
\end{equation}
The lab-frame cross section is then given by:
\begin{eqnarray}
\label{frames_neutron}
\left(\frac{d\sigma}{d\Omega_1}\right)_{\!l} &=&
\left| \frac{d\cos\theta_c}{d\cos\theta} \right|
\left(\frac{d\sigma}{d\Omega_1}\right)_{\!c} \\
&=& \frac{(1+\gamma^2+2\gamma\cos\theta_c)^{3/2}}{1+\gamma\cos\theta_c}
\left(\frac{d\sigma}{d\Omega_1}\right)_{\!c} \nonumber \\
&=& \frac{(\gamma\cos\theta+\sqrt{1-\gamma^2\sin^2\theta})^2}
{\sqrt{1-\gamma^2\sin^2\theta}}
\left(\frac{d\sigma}{d\Omega_1}\right)_{\!c}\!; \nonumber
\end{eqnarray}
these equations are valid for both elastic and inelastic scattering using the appropriate $\gamma$ factor.

Similarly, we may obtain the nuclear recoil cross section using the centre-of-mass neutron cross section, as this applies to both outgoing particles (inverting the sign of $\cos\theta_c$ in the case of the recoil). For elastic scattering, the equivalent relation to (\ref{thetaconv}) for the recoil angle is simply:
\begin{equation}
\alpha = \frac{\pi-\theta_c}{2}\,.
\label{alpha_el}
\end{equation}
The corresponding differential cross section for the recoil solid angle in the laboratory frame can be calculated:
\begin{eqnarray}
\left(\frac{d\sigma}{d\Omega_2}\right)_{\!l} &=& \label{dsdalpha_el}
\left|\frac{d\cos\theta_c}{d\cos\alpha}\right|
\left(\frac{d\sigma}{d\Omega_2}\right)_{\!c} \\
&=& 4\sin\frac{\theta_c}{2}\left(\frac{d\sigma}{d\Omega_2}\right)_{\!c}
\nonumber \\ 
&=& 4\cos\alpha \left(\frac{d\sigma}{d\Omega_2}\right)_{\!c}\,.\nonumber
\end{eqnarray}

For inelastic scattering the recoil angle relation is similar in form to that for the outgoing neutron in (\ref{thetaconv}):
\begin{equation}
\cos\alpha = \frac{\gamma_r - \cos\theta_c}
{\sqrt{1+\gamma_r^2-2\gamma_r\cos\theta_c}} \,,
\label{alpha_thetac}
\end{equation}
where we defined a kinematic factor $\gamma_r$ for the recoil:
\begin{equation}
\gamma_r = \frac{M}{m}\gamma = \sin \alpha_0^{-1} \,.
\label{gamma_r}
\end{equation}

From this we derive how the recoil cross section transforms between frames:
\begin{equation}
\left(\frac{d\sigma}{d\Omega_2}\right)_{\!l} =
\frac{\left(1+\gamma_r^{2}-2\gamma_r\cos\theta_c\right)^{3/2}}
{|1-\gamma_r\cos\theta_c|} 
\left(\frac{d\sigma}{d\Omega_2}\right)_{\!c}\,.
\label{frames_recoil}
\end{equation}
These equations reduce to (\ref{alpha_el}) and (\ref{dsdalpha_el}) for the case of elastic scattering, when $\alpha_0=90^{\mathrm o}$ and $\gamma_r=1$. This cross section has a singularity at $\alpha_0$ due to the frame transformation. It should be noted also that, below that angle, two centre-of-mass angles contribute to the cross section at each $\alpha$, calculated by inverting (\ref{alpha_thetac}):
\begin{equation}
\cos\theta_c = \gamma_r\sin^2\alpha\pm
\cos\alpha\sqrt{1-\gamma_r^2\sin^2\alpha} \,.
\label{thetac_alpha}
\end{equation}

So far in this section we have treated inelastic scattering as a binary collision, applying momentarily before the excited residual nucleus deexcites with emission of one or more $\gamma$-rays. Although this does not change the outgoing neutron, the photon momentum may smear the recoil angle and energy appreciably -- notably for $^{12}$C, a light species emitting an energetic $\gamma$-ray. To calculate the full angular cross section the angular distribution of the emitted photons must be considered too, and these are not generally isotropic (File~14 of the data libraries).

\begin{figure}
\centerline{\includegraphics[width=0.9\linewidth]{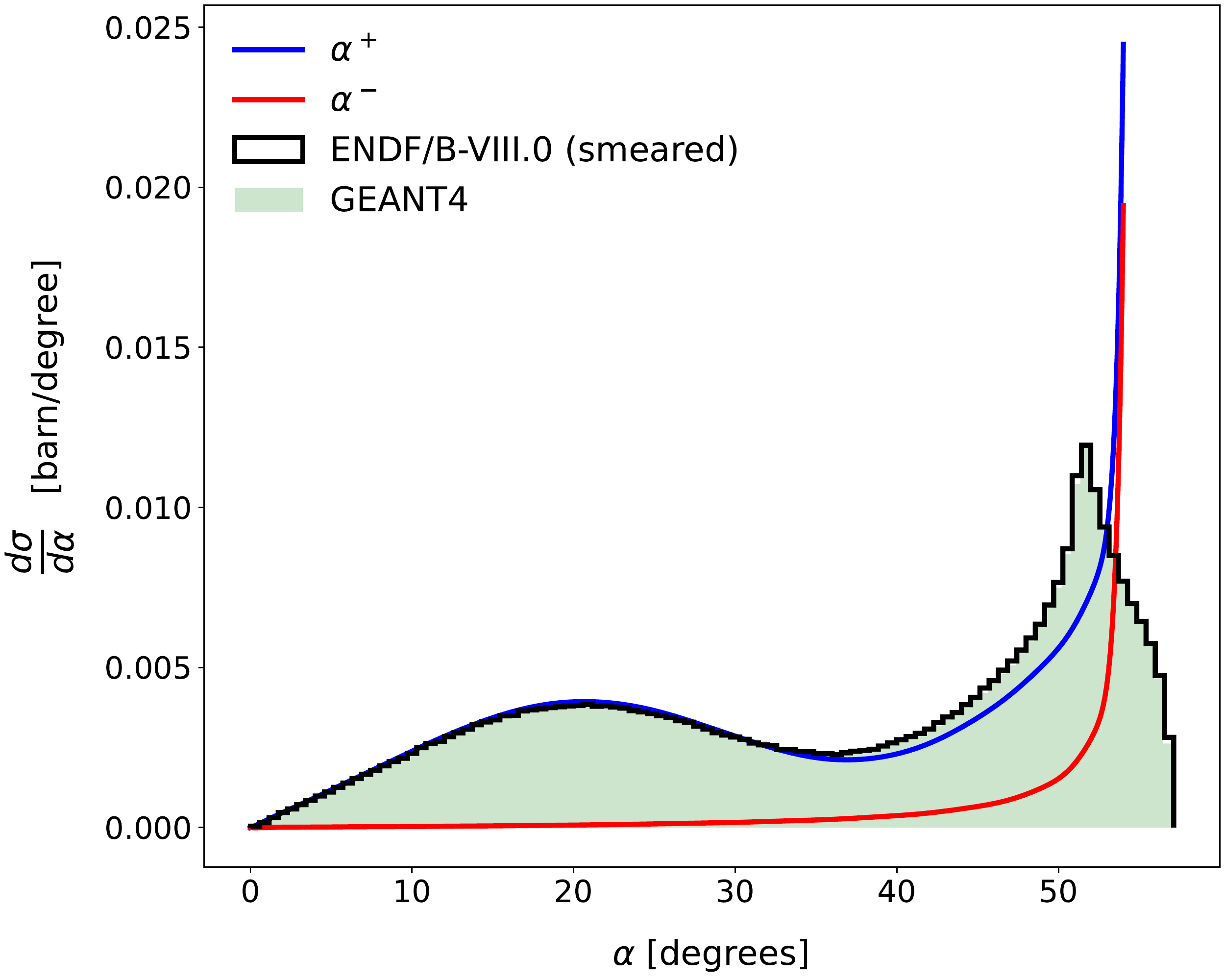}}
\caption{Differential cross section for the nuclear recoil angle ($\alpha$) for inelastic scattering of D-T neutrons via the first excited state of $^{12}$C at $\epsilon$=4.4~MeV. The two centre-of-mass angle contributions are shown separately ($\alpha^+$ and $\alpha^-$ in blue and red, respectively), derived from ENDF/B-VIII.0 data; these are combined and smeared according to (\protect{\ref{gammarecoil}}), adding to the thick black histogram. The shaded green histogram shows the same process modelled with GEANT4, which is in good agreement.}
\label{fig:C12inelastic} 
\end{figure}

Figure~\ref{fig:C12inelastic} illustrates the various steps involved in the calculation of the angular cross section for $^{12}$C recoils from D-T neutron scattering via the first excited state with $\epsilon=4.4$~MeV. This includes the following steps: calculation of the neutron cross section from the Legendre polynomial parameterisation obtained from ENDF File~4 using (\ref{legendre}); evaluation at the two centre-of-mass angles contributing to each recoil angle $\alpha$ using (\ref{thetac_alpha}); conversion of the cross section to the laboratory frame using (\ref{frames_recoil}); smearing for $\gamma$-ray emission with angular distribution obtained from ENDF File~14. This is overlaid on a GEANT4 simulation of the same process.\footnote{Note that the ENDF libraries do not provide all necessary particle correlations to satisfy energy and momentum conservation in all cases, and the GEANT4 NeutronHP models adjust the final state of some reactions through the emission of artificial low-energy $\gamma$-rays~\protect{\cite{G4ADG}}. This is problematic for some Migdal simulations, and a compilation flag may be used to suppress this adjustment. However, note that this may affect some nuclear recoil distributions (the effect is negligible for the case shown.)} To circumvent the singularity in the angular cross section, it may be more expedient to differentiate (\ref{Er,inel}) to convert (\ref{frames_neutron}) to an energy-differential cross-section, which is continuous. The scattering angle $\alpha$ can then be obtained by inverting  (\ref{Er,inel_alpha}).

\bibliography{MigdalBibFile}

\end{document}